\documentclass[sn-mathphys-num]{sn-jnl}

\usepackage{float}
\usepackage{graphicx}
\usepackage{multirow}
\usepackage{amsmath,amssymb,amsfonts}
\usepackage{amsthm}
\usepackage{mathrsfs}
\usepackage[title]{appendix}
\usepackage{xcolor}
\usepackage{textcomp}
\usepackage{manyfoot}
\usepackage{booktabs}
\usepackage{algorithm}
\usepackage{algorithmicx}
\usepackage{algorithm}
\usepackage{listings}
\usepackage{algpseudocode}
\usepackage{enumitem}

\newtheorem{procedure}{Procedure}

\begin{document}

\title[Article Title]{FinBERT-BiLSTM: A Deep Learning Model for Predicting Volatile Cryptocurrency Market Prices Using Market Sentiment Dynamics}

\author*[1]{\fnm{Mabsur Fatin} \sur{Bin Hossain}}\email{mabsurfatinbin-2019317809@cs.du.ac.bd}
\equalcont{These authors contributed equally to this work.}

\author[1]{\fnm{Lubna Zahan} \sur{Lamia}}\email{lubnazahan-2019417808@cs.du.ac.bd}
\equalcont{These authors contributed equally to this work.}

\author[1]{\fnm{Md Mahmudur} \sur{Rahman}}\email{mahmudur@cse.du.ac.bd}

\author[1]{\fnm{Md Mosaddek} \sur{Khan}}\email{mosaddek@du.ac.bd}

\affil[1]{\orgdiv{Department of Computer Science and Engineering}, \orgname{University of Dhaka}, \orgaddress{\city{Dhaka}, \country{Bangladesh}}}

\abstract{
Time series forecasting is a key tool in financial markets, helping to predict asset prices and guide investment decisions. In highly volatile markets, such as cryptocurrencies like Bitcoin (BTC) and Ethereum (ETH), forecasting becomes more difficult due to extreme price fluctuations driven by market sentiment, technological changes, and regulatory shifts. 
Traditionally, forecasting relied on statistical methods, but as markets became more complex, deep learning models like LSTM, Bi-LSTM, and the newer FinBERT-LSTM emerged to capture intricate patterns. Building upon recent advancements and addressing the volatility inherent in cryptocurrency markets, we propose a hybrid model that combines Bidirectional Long Short-Term Memory (Bi-LSTM) networks with FinBERT to enhance forecasting accuracy for these assets. This approach fills a key gap in forecasting volatile financial markets by blending advanced time series models with sentiment analysis, offering valuable insights for investors and analysts navigating unpredictable markets.
}


\maketitle

\section{Introduction}\label{sec1}

Time series forecasting has long been a critical tool in financial analysis, applied to predict price movements in assets such as stocks and gold. As markets evolve, accurate forecasting becomes increasingly important for investors navigating complex and unpredictable market dynamics. Even in stock and gold markets, forecasting has never been straightforward due to the complex interplay of numerous factors \cite{dhingra} \cite{Alameer2019}. This makes accurate forecasting indispensable yet increasingly difficult, especially as financial landscapes continue to evolve at a rapid pace.

In recent years, cryptocurrencies have unlocked vast opportunities within the global financial ecosystem, offering innovative investment options and driving economic growth \cite{mittal2020cryptocurrency}. As the first decentralized cryptocurrency, Bitcoin gained popularity for facilitating online transactions without intermediaries, standing in sharp contrast to traditional financial systems \cite{nakamoto2008bitcoin}. Ethereum (ETH), with the second-largest market capitalization, follows closely behind \cite{hu2021transaction}. While the potential for profitability makes Bitcoin (BTC) and Ethereum (ETH) attractive to forward-looking investors, cryptocurrencies are notorious for their extreme price volatility. This volatility stems from a complex interplay of factors, including internal elements such as transaction costs, reward systems, mining difficulty, and coin circulation; external influences like market trends and speculation; macro-financial factors including stock markets, exchange rates, gold prices, and interest rates; and political factors such as legalization, restrictions, and regulatory changes \cite{sovbetov2018factors}. Understanding these dynamics is crucial for investors seeking to capitalize on the rapid fluctuations characteristic of cryptocurrency markets.

Cheah and Fry \cite{cheah2015speculative} explored the speculative nature of Bitcoin markets, highlighting substantial deviations from fundamental values that contribute to its pronounced volatility. Corbet et al. \cite{Corbet2018} investigated the dynamic relationships between cryptocurrencies and traditional financial assets, revealing that cryptocurrencies like BTC remain relatively isolated from traditional markets, with distinct volatility patterns that may offer diversification benefits for investors with short-term horizons. Further studies, including those by Baur et al.\cite{baur2018bitcoin} and Urquhart \cite{urquhart2016inefficiency}, underscore the inefficiencies and high volatility in cryptocurrency markets. Baur et al. emphasized the time-varying linkages driven by external shocks, while Urquhart suggested that Bitcoin may be evolving toward greater market efficiency. These characteristics highlight the speculative nature of the cryptocurrency markets, presenting significant challenges for accurate price prediction and risk management. Nevertheless, they also create opportunities to develop advanced trading strategies that capitalize on rapid price fluctuations.

Price prediction in cryptocurrency markets, especially for assets like Bitcoin and Ethereum, has been a focal point of research. Early approaches to financial forecasting relied heavily on traditional statistical models and time-series analysis. Techniques such as ARIMA (AutoRegressive Integrated Moving Average) and GARCH (Generalized Autoregressive Conditional Heteroskedasticity) have been extensively used to model financial time series data. Katsiampa \cite{katsiampa2017volatility} employed GARCH models to estimate Bitcoin's volatility, demonstrating their effectiveness in capturing short-term price movements. However, these models often struggle to account for the non-linear dynamics inherent in cryptocurrency markets. Bouri et al. \cite{bouri2017returnvolatility} used asymmetric GARCH models to analyze the return-volatility relationship in Bitcoin around the 2013 price crash, emphasizing the limitations of these models in fully capturing the evolving complexities of cryptocurrency markets.

As machine learning techniques advanced, models like LSTM and GRU emerged. These are specialized kinds of Recurrent Neural Networks (RNNs) that address the vanishing gradient problem common in traditional RNNs. Studies, such as by Rizwan et al. \cite{rizwan2019bitcoin}, demonstrated that LSTM and GRU models significantly outperform traditional methods like ARIMA in predicting Bitcoin prices, providing higher accuracy and better capturing the non-linear patterns in cryptocurrency markets. While traditional LSTMs marked a significant advancement in time series analysis, they are limited by their unidirectional nature, processing information only in one direction-typically forward in time. Meanwhile, Bidirectional LSTMs (Bi-LSTMs) were developed, enabling the model to consider both past and future data simultaneously. Recent research in 2023 by Seabe et al. \cite{seabe2023forecasting} compared LSTM, GRU, and Bi-LSTM models in predicting the prices of Bitcoin, Ethereum, and Litecoin and found that Bi-LSTM consistently delivered the most accurate results. The study emphasized that the bidirectional flow of information in Bi-LSTMs improves time-series prediction, especially in highly volatile markets like cryptocurrencies.

On the other hand, research has shown that media sentiment, derived from news articles and blog posts, has a significant impact on cryptocurrency prices, with findings indicating that investors tend to overreact to news in the short term, influencing Bitcoin's market fluctuations \cite{karalevicius2018sentiment}. Researchers such as Valencia et al. \cite{valencia2019price} have incorporated media sentiment into cryptocurrency price prediction models. Valencia et al. \cite{valencia2019price} employed machine learning models such as Neural Networks (NN), Support Vector Machines (SVM), and Random Forests (RF) to predict the price movements of Bitcoin, Ethereum, Ripple, and Litecoin, demonstrating that neural networks combined with Twitter and market data outperformed the other models in cryptocurrency prediction. As sentiment became more integrated with these prediction models, in a comparative analysis of ten sentiment classification algorithms, Hartmann et al. \cite{hartmann2019comparing} found Naïve Bayes to be highly effective, outperforming several other methods. The study evaluated five lexicon-based techniques and five machine learning approaches, with Naïve Bayes emerging as one of the top performers, particularly for its efficiency with smaller datasets. Further enhancing sentiment-based classification models, the BERT transformer model significantly outperformed traditional models like Naïve Bayes, Support Vector Machines, and TextCNN, demonstrating superior accuracy and effectiveness in predicting stock market trends due to its bidirectional context understanding \cite{sousabert}. Araci \cite{araci2019finbert} presented FinBERT, a variant of the BERT transformer model specifically fine-tuned for financial sentiment analysis. This model has proven effective in extracting nuanced sentiment from financial texts, offering deeper insights into how market sentiment influences asset prices \cite{arijit2024}.

The integration of LSTM networks with transformer models like FinBERT has emerged as a highly effective approach in financial prediction, particularly for tasks involving sentiment analysis in time series forecasting. This hybrid model combines the strengths of both: LSTM’s ability to capture long-term dependencies in sequential data and FinBERT’s capacity to process contextual information from textual data. Halder et al. \cite{halder2022} explored various deep learning techniques, including MLP, LSTM, and their proposed FinBERT-LSTM model, which integrates financial news sentiment into stock price predictions. Their experiments focused on short-term predictions for the NASDAQ-100 index\footnote{\url{https://www.nasdaq.com}, last accessed on October 27, 2024.} and highlighted the performance of these models. Similarly, Gu et al. \cite{gu2024} employed NASDAQ-100 stock data and trained their model using Benzinga\footnote{\url{https://www.benzinga.com/}, last accessed on October 28, 2024} news articles, comparing the predictive capabilities of different models. In both studies, FinBERT-LSTM gave the best results, outperforming the other models tested. Zeng and Jiang \cite{jiang2023financial} leveraged a publicly available historical financial news dataset from Kaggle\footnote{\url{https://www.kaggle.com}, last accessed on October 27, 2024.}, covering over 800 U.S. companies across 12 years, and combined it with stock market data from the yfinance API\footnote{\url{https://pypi.org/project/yfinance/}, last accessed on October 27, 2024.}. They compared several models, including ARIMA, LSTM, BERT-LSTM, and FinBERT-LSTM, demonstrating that the FinBERT-LSTM model achieved superior performance in both training and testing phases. Recently in 2023, Girsang and Stanley \cite{girsang2023hybrid} collected one year of daily market data and tweets for Solana (SOL) and Ethereum (ETH), spanning from January 1, 2021, to December 31, 2021. The data, sourced from Binance\footnote{\url{https://www.binance.com/en}, last accessed on October 28, 2024}, was complemented with Twitter sentiment data. Their research demonstrated that incorporating sentiment scores extracted using FinBERT significantly enhanced prediction performance compared to models relying solely on traditional time series methods, such as LSTM, GRU, or hybrid models. Despite advancements in financial stock prediction, the application of transformer models combined with LSTM networks has been explored far less in the volatile cryptocurrency markets. While Girsang and Stanley (2023) \cite{girsang2023hybrid} examined the use of these combined models for Ethereum and Solana, they primarily focused on Twitter sentiment rather than news sentiment. This leaves a gap in exploring cryptocurrencies like Bitcoin, one of the most volatile and widely traded assets using these advanced techniques. Both the incorporation of news sentiment and the application of these techniques have been underexplored in cryptocurrency prediction, particularly in contrast to the more common use of social media sentiment.

To address a significant research gap, our study makes the following contributions:

\begin{enumerate}
    \item[\text{i.}] Tailored for the volatile cryptocurrency market, we present a combined model of FinBERT and Bi-LSTM. This model is designed to capture market volatility by incorporating detailed financial news sentiment and tracking temporal price patterns.
    
    \item[\text{ii.}] Specifically designed for the cryptocurrency market, we introduce a distinctive dataset comprising Bitcoin and Ethereum financial news alongside historical price data. This robust foundation facilitates comprehensive market analysis and informed trading simulations.

    \item[\text{iii.}] We extensively evaluate our FinBERT-Bi-LSTM model in comparison with benchmark models: LSTM, FinBERT-LSTM, and Bidirectional LSTM. We analyze their performance through intra-day and one-day-ahead predictions while assessing the influence of sentiment on prediction accuracy. Additionally, we benchmark the profitability of these models using a straightforward trading strategy for one-day-ahead predictions. Lastly, we explore two distinct approaches for forecasting cryptocurrency prices over multiple days, as making accurate multi-day predictions becomes increasingly challenging in the volatile cryptocurrency market.

    \item[\text{iv.}]In terms of results, the FinBERT-Bi-LSTM model demonstrates superior performance, achieving approximately 98\% accuracy for both intra-day and one-day-ahead predictions on Bitcoin (BTC), along with notable profits for the one-day-ahead strategy. For Ethereum (ETH), the model achieves around 97\% accuracy for intra-day and one-day-ahead predictions, generating significant profits as well. In long-term price predictions, the first approach yields about 98\% accuracy for BTC, while the latter approach performs better for ETH with around 95\% accuracy.
\end{enumerate}

In Section~\ref{sec2}, we discuss the foundational models and sentiment analysis techniques. In Section~\ref{sec3}, we formulate the problem by outlining the prediction tasks and a trading strategy for forecasting cryptocurrency prices using historical price data and sentiment analysis. Section~\ref{sec4} presents our proposed FinBERT-Bi-LSTM hybrid model, developed for improved cryptocurrency price forecasting. Sections~\ref{sec5} and \ref{sec6} detail the experimental setup, datasets, and evaluation metrics, along with a comprehensive analysis of the results. Finally, Section~\ref{sec7} offers concluding remarks and discusses future work aimed at enhancing predictive performance through innovative approaches.

\section{Background}
\label{sec2}

In this section, we discuss the key models that form the foundation of our approach to cryptocurrency price prediction, including the Long Short-Term Memory (LSTM) model, the Bidirectional LSTM (Bi-LSTM) model, and the FinBERT model for financial sentiment analysis.

\subsection{LSTM Network}
\begin{figure}[h]
    \centering
    \includegraphics[width=0.65\textwidth]{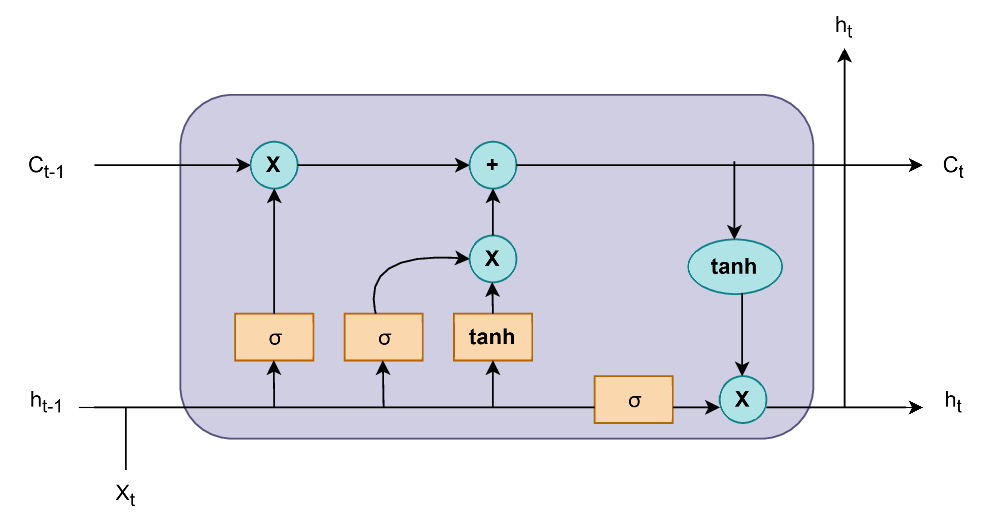}
    \caption{LSTM Cell }
    \label{fig:lstm}
\end{figure}

Long Short-Term Memory (LSTM) networks are designed to retain and learn from important sequential data over long periods due to their gating mechanisms. These gates-\textbf{Input Gate}, \textbf{Forget Gate}, and \textbf{Output Gate}-are essential for controlling the flow of information into, within, and out of the memory cells. Each gate is governed by a sigmoid activation function that outputs a value between 0 and 1, indicating how much information should be passed through. The structure of an LSTM Cell \cite{chevalier2018larnn} is shown in Fig. \ref{fig:lstm}

\paragraph{Forget Gate}
The \textbf{Forget Gate} is responsible for determining which information from the previous hidden state should be discarded. The decision is made by taking the previous hidden state $h_{t-1}$ and the current input $x_t$, and passing them through a sigmoid function. 
\begin{equation}
\label{eq1}
f_t = \sigma (x_t \cdot U_f + h_{t-1} \cdot W_f)
\end{equation}
In equation \ref{eq1}, $U_f$ is the weight matrix for the input $x_t$, and $W_f$ is the weight matrix for the previous hidden state $h_{t-1}$. The output $f_t$ is a value between 0 and 1, where 0 means ``completely forget" and 1 means ``completely retain" the information from the previous cell state $C_{t-1}$. This gate ensures that irrelevant information is removed from the memory, preventing the accumulation of unnecessary details as the LSTM processes longer sequences.

\paragraph{Input Gate}
The \textbf{Input Gate} determines which new information should be added to the memory cell. It consists of two parts:
\begin{itemize}
    \item First, a sigmoid function controls how much of the current input should be written to the cell.
    \begin{equation}
    \label{eq2}
        i_t = \sigma (x_t \cdot U_i + h_{t-1} \cdot W_i)
    \end{equation}
    In equation \ref{eq2}, $i_t$ represents the input gate's activation at time step $t$, $x_t$ is the current input, $h_{t-1}$ is the previous hidden state, $U_i$ is the weight matrix for the input $x_t$, and $W_i$ is the weight matrix for the hidden state $h_{t-1}$.
    
    \item Second, a tanh function creates a candidate value for the memory update, where $\tilde{C}_t$ is the candidate memory update, $U_g$ and $W_g$ are the respective weight matrices for the input $x_t$ and the hidden state $h_{t-1}$, as shown in equation \ref{eqn3}.
    \begin{equation}
    \label{eqn3}
        \tilde{C}_t = \text{tanh}(x_t \cdot U_g + h_{t-1} \cdot W_g)
    \end{equation}
\end{itemize}

The product of $i_t$ and $\tilde{C}_t$ determines how much of the candidate memory update $\tilde{C}_t$ should be added to the cell state, given by the product of $f_t$ and $C_{t-1}$ in equation \ref{eqn4}.

\begin{equation}
\label{eqn4}
    C_t = f_t \cdot C_{t-1} + i_t \cdot \tilde{C}_t
\end{equation}

This mechanism ensures that only relevant information from the current input is stored in the memory, controlling how much new data enters the cell.

\paragraph{Output Gate}
The \textbf{Output Gate} regulates how much information from the cell state should be passed to the next hidden state and used as the output of the LSTM unit. The process unfolds as follows:

First, the previous hidden state $h_{t-1}$ and the current input $x_t$ are multiplied by their respective weight matrices, $W_o$ and $U_o$. These results are then combined and passed through a sigmoid function, yielding the activation of the output gate, $o_t$, as shown in equation \ref{eqn5}.

\begin{equation}
\label{eqn5}
o_t = \sigma (x_t \cdot U_o + h_{t-1} \cdot W_o)
\end{equation}

Next, as shown in equation \ref{eqn6}, the current cell state $C_t$ is passed through a tanh function, and the result is multiplied by $o_t$ to determine the hidden state $h_t$, which becomes the final output.

\begin{equation}
\label{eqn6}
h_t = o_t \cdot \text{tanh}(C_t)
\end{equation}

This mechanism ensures that the model outputs relevant information at each time step, drawing from both the current input and the historical context stored in the memory cell.

\subsection{Bi-LSTM}
\begin{figure}[H]
    \centering
    \includegraphics[width=0.60\textwidth]{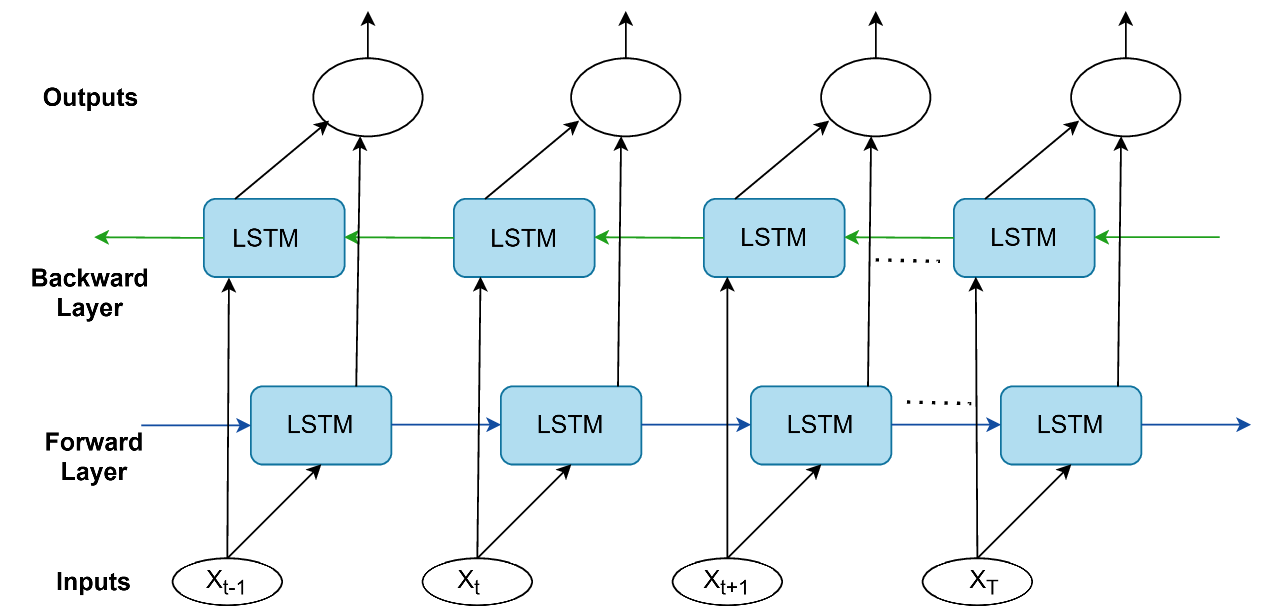}
    \caption{The structure of a bi-directional LSTM (Bi-LSTM)}
    \label{fig:Bilstm}
\end{figure}

Bi-LSTM (Bidirectional Long Short-Term Memory) is a type of Recurrent Neural Network (RNN) designed to process sequential data in both forward and backward directions, capturing both past and future context for more accurate predictions. Traditional RNNs face limitations in retaining long-term dependencies as they only process sequences in a single direction, either forward or backward. LSTMs overcome this limitation by introducing memory cells and gating mechanisms to selectively retain, forget, or update information over time. These internal memory states enable LSTMs to capture long-range dependencies in data \cite{hochreiter1997long}. Bi-LSTM enhances this by adding bidirectional processing, allowing the model to understand the sequence by considering both past and future contexts \cite{graves2005framewise} \cite{schuster1997bidirectional}. It consists of two LSTM layers: one processing the sequence in the forward direction (from the first time step to the last) and another in the backward direction (from the last time step to the first), as shown in Fig. \ref{fig:Bilstm}. 

Each LSTM layer maintains its own hidden states and memory cells, which are updated based on the current input and the previous hidden states. After the forward and backward passes, the hidden states from both directions are combined at each time step, typically through concatenation or another transformation. This approach enables Bi-LSTM to capture richer dependencies. By leveraging both past and future information, Bi-LSTM has proven to be successful in time series prediction, where understanding the full context of data is crucial for making accurate forecasts \cite{ben2021predicting}.

\subsection{FinBERT}

FinBERT, developed by Dogu Araci \cite{araci2019finbert}, is a model built on the BERT (Bidirectional Encoder Representations from Transformers) architecture, which processes input text bidirectionally, meaning it considers both the preceding and following context of each word. FinBERT was specifically fine-tuned for financial text classification, making it particularly suitable for sentiment analysis of cryptocurrency-related news. Financial texts often contain specialized language and complex expressions that general sentiment models struggle to interpret. FinBERT overcomes this by using transfer learning, enabling it to be fine-tuned on financial data to better understand the context and nuances of financial language. The model's strength lies in its ability to leverage pre-trained knowledge from large language corpora and adapt to the financial domain. By further training FinBERT on financial-specific data, the model becomes more effective in handling the unique terminology and sentiment patterns present in financial news and reports. The FinBERT model assigns a sentiment score based on the classification output, where:

\begin{itemize}
    \item \textbf{Positive Sentiment:} FinBERT identifies positive news by assigning a positive sentiment label, indicating a favorable outlook.
    \item \textbf{Neutral Sentiment:} For neutral news, FinBERT assigns a neutral sentiment label, representing no strong positive or negative sentiment.
    \item \textbf{Negative Sentiment:} FinBERT assigns a negative sentiment label to news that indicates an unfavorable outlook.
\end{itemize}

FinBERT also incorporates advanced training techniques to ensure it retains the broad language understanding of BERT while adapting to the financial context. This makes it a valuable tool for improving sentiment analysis in the financial domain, where accurate interpretation of sentiment can enhance decision-making and market predictions.

\section{Problem Formulation}
\label{sec3}
Let \(x = [p_{d-n}, p_{d-n+1}, \dots, p_{d-1}]\) represent the input vector consisting of a window of historical price data of size \(n\), where \(p_i\) denotes the closing price at day \(i\). Additionally, let \(s_i\) denote the sentiment score for day \(i\) obtained through FinBERT. We explore the following prediction tasks.

\subsection{Predicting the Closing Price of Day \(d\)}

The objective is to predict the closing price for day \(d\) under two scenarios:

\subsubsection{Intra-Day Price Predictions}

The goal is to predict the closing price within the same trading day.
\newline

The plain LSTM and Bi-LSTM models predict the closing price for day \(d\) based solely on historical price data, without any sentiment information. The predicted closing price \(\hat{p}_d\) is given by:

\begin{equation}
\hat{p}_d = f(x)
\end{equation}

where \(f(x)\) takes the input vector \(x\), consisting of historical prices.
\newline
\newline

The FinBERT-integrated models incorporate the sentiment score from day \(d\), denoted as \(s_d\), into the LSTM and Bi-LSTM models. The predicted closing price \(\hat{p}_d\) is given by:

\begin{equation}
\hat{p}_d = f(x, s_d)
\end{equation}

where \(f(x, s_d)\) takes the input vector \(x\) and the sentiment score \(s_d\) as inputs.

\subsubsection{One-Day-Ahead Price Predictions}

The goal is to predict the closing price for the following trading day.
\newline

For the plain LSTM and Bi-LSTM models, the prediction for day \(d\) is made using the same historical price data as in intra-day predictions. The predicted closing price \(\hat{p}_d\) is given by:

\begin{equation}
\hat{p}_d = f(x)
\end{equation}

where \(f(x)\) takes the input vector \(x\), which consists of historical prices.
\newline
\newline
In the FinBERT-integrated models, the sentiment score from day \(d-1\), \(s_{d-1}\), is used as the last available sentiment data, since sentiment data for day \(d\) is not available at the time of prediction. The predicted closing price \(\hat{p}_d\) is given by:

\begin{equation}
\hat{p}_d = f(x, s_{d-1})
\end{equation}

where \(f(x, s_{d-1})\) takes the input vector \(x\) and the sentiment score \(s_{d-1}\).

\subsection{One-Day-Ahead Trading Strategy}

In this setting, a bot makes decisions to buy, sell, or hold cryptocurrencies immediately after observing the closing price of day \(d-1\), making a one-day-ahead prediction for day \(d\), and then acting based on this prediction. The FinBERT-integrated models use the sentiment score from day \(d-1\), as the decision must be made one day in advance. The performance of the models is evaluated by calculating the profit:

\begin{equation}
\text{Profit} = V - C_0 - T
\end{equation}

Where \(V\) is the final portfolio value, including remaining cash and the value of any cryptocurrency held, \(C_0\) is the initial capital, and \(T\) represents the transaction costs incurred during trading.
\newline

In this strategy, we make the following assumptions:

\begin{itemize}
    \item If a decision is made to buy or sell, the transaction is assumed to occur at the same price as the closing price of the day \(d-1\). This assumption is reasonable given that, although cryptocurrency prices can fluctuate rapidly within minutes, the bot can execute the decision quickly due to the simplicity of the strategy, minimizing the impact of price fluctuations.

    \item It is assumed that the required amount of Bitcoin or Ethereum for purchase will always be available on the market, and transactions will not fail due to liquidity constraints or other technical issues. This assumption is reasonable due to the high liquidity of Bitcoin and Ethereum on major exchanges.
\end{itemize}

\subsection{Predicting the Closing Prices for \(m\)-Days Ahead}

The objective is to predict the closing prices for \(m\) days ahead, where \(m > 1\), starting from day \(d\). The plain LSTM/Bi-LSTM models, as well as the FinBERT-LSTM and FinBERT-Bi-LSTM models, are used without sentiment data for future days. The only key difference between the plain LSTM/Bi-LSTM models and FinBERT-integrated models is that the latter incorporates sentiment data during training, capturing the influence of sentiment in shaping the model's understanding of price trends. This setup enables the FinBERT-integrated models to assess and project long-term price trends independently, demonstrating their forecasting power even when sentiment data for future days is unavailable. The predicted closing prices \(\{\hat{p}_d, \dots, \hat{p}_{d+m-1}\}\) are given by:

\begin{equation}
\{\hat{p}_d, \hat{p}_{d+1}, \dots, \hat{p}_{d+m-1}\} = f(x)
\end{equation}

where \(f(x)\) takes the input vector \(x\), consisting of historical prices.

\section{FinBERT-BiLSTM: A Sentiment-Driven Forecasting Model}
\label{sec4}
This section introduces our proposed FinBERT-Bi-LSTM model, designed to address the challenges outlined in the problem formulation. We hypothesize that the bidirectional nature of the FinBERT model will synergize effectively with the Bidirectional LSTM architecture, enhancing its capability to capture complex sentiment dynamics. We describe its application across various prediction tasks using tailored algorithms.

\subsection{Intra-Day Price Prediction using FinBERT-Bi-LSTM}

\begin{algorithm}[H]
\caption{Intra-Day Cryptocurrency Price Prediction}
\label{alg:1}
\textbf{Input:} Historical price data $\mathbf{S}$, News sentiment data $\mathbf{N}$, Hyperparameters: sequence length $n$, learning rate $\alpha$, epochs $T$\\
\textbf{Output:} Predicted prices $\hat{\mathbf{P}}$, Evaluation metrics

\begin{algorithmic}[1]

\State $\mathbf{S} \rightarrow \{\mathbf{S}_{\text{train}}, \mathbf{S}_{\text{test}}\}$, \quad $\mathbf{N} \rightarrow \{\mathbf{N}_{\text{train}}, \mathbf{N}_{\text{test}}\}$
\State $\mathbf{S}_{\text{train}} \rightarrow \{\mathbf{S}_{\text{sub-train}}, \mathbf{S}_{\text{val}}\}$, \quad $\mathbf{N}_{\text{train}} \rightarrow \{\mathbf{N}_{\text{sub-train}}, \mathbf{N}_{\text{val}}\}$

\State Normalize $\mathbf{S}_{\text{sub-train}}$, $\mathbf{S}_{\text{val}}$, $\mathbf{S}_{\text{test}}$

\State $\{\mathbf{X}_{\text{sub-train}}, \mathbf{y}_{\text{sub-train}}\} \gets \text{Seq\_With\_Cur\_Sent}(\mathbf{S}_{\text{sub-train}}, \mathbf{N}_{\text{sub-train}}, n)$
\State $\{\mathbf{X}_{\text{val}}, \mathbf{y}_{\text{val}}\} \gets \text{Seq\_With\_Cur\_Sent}(\mathbf{S}_{\text{val}}, \mathbf{N}_{\text{val}}, n)$
\State $\{\mathbf{X}_{\text{test}}, \mathbf{y}_{\text{test}}\} \gets \text{Seq\_With\_Cur\_Sent}(\mathbf{S}_{\text{test}}, \mathbf{N}_{\text{test}}, n)$

\State Initialize model $f_\theta$ with parameters $\theta$
\State Train $f_\theta$ on $\{\mathbf{X}_{\text{sub-train}}, \mathbf{y}_{\text{sub-train}}\}$ for $T$ epochs using learning rate $\alpha$, with validation on $\{\mathbf{X}_{\text{val}}, \mathbf{y}_{\text{val}}\}$
\State Predict $\hat{\mathbf{P}}$ using $f_\theta(\mathbf{X}_{\text{test}})$
\State Apply inverse normalization to $\hat{\mathbf{P}}$ and $\mathbf{y}_{\text{test}}$

\end{algorithmic}
\end{algorithm}

In Algorithm \ref{alg:1}, lines 1-3 handle the loading of historical price and sentiment data, splitting it into training and test sets. The training set is further divided into training and validation subsets, with each normalized using MinMaxScaler. In lines 4-6, the procedure \texttt{Seq\_With\_Cur\_Sent} generates input sequences by creating historical windows of the previous \( n \) days' prices, with each sequence's target output being the price of the current day across the training, validation, and test subsets. Additionally, each input sequence is enriched with the sentiment score for the current day to support more accurate price prediction. Details of this procedure are discussed in Procedure \ref{proc:seq_with_cur_sent}. Next in line 7, the FinBERT-Bi-LSTM model is constructed based on the chosen architecture and initialized with parameters \(\theta\). In line 8, the model is trained on the training set and validated using the validation data. In line 9, predictions are made on the test set, and in line 10, inverse normalization is applied to return the predicted and actual prices to their original scale.

\subsection{One-Day-Ahead Price Prediction with Trading Strategy Simulation using FinBERT-Bi-LSTM}

\begin{algorithm}[H]
\caption{One-Day-Ahead Price Prediction}
\label{alg:2}
\textbf{Input:} Historical price data $\mathbf{S}$, News sentiment data $\mathbf{N}$, Hyperparameters: sequence length $n$, learning rate $\alpha$, epochs $T$\\
\textbf{Output:} Predicted prices $\hat{\mathbf{P}}$, Evaluation metrics

\begin{algorithmic}[1]

\State $\mathbf{S} \rightarrow \{\mathbf{S}_{\text{train}}, \mathbf{S}_{\text{test}}\}$, \quad $\mathbf{N} \rightarrow \{\mathbf{N}_{\text{train}}, \mathbf{N}_{\text{test}}\}$
\State $\mathbf{S}_{\text{train}} \rightarrow \{\mathbf{S}_{\text{sub-train}}, \mathbf{S}_{\text{val}}\}$, \quad $\mathbf{N}_{\text{train}} \rightarrow \{\mathbf{N}_{\text{sub-train}}, \mathbf{N}_{\text{val}}\}$

\State Normalize $\mathbf{S}_{\text{sub-train}}$, $\mathbf{S}_{\text{val}}$, $\mathbf{S}_{\text{test}}$

\State $\{\mathbf{X}_{\text{sub-train}}, \mathbf{y}_{\text{sub-train}}\} \gets \text{Seq\_With\_Cur\_Sent}(\mathbf{S}_{\text{sub-train}}, \mathbf{N}_{\text{sub-train}}, n)$
\State $\{\mathbf{X}_{\text{val}}, \mathbf{y}_{\text{val}}\} \gets \text{Seq\_With\_Cur\_Sent}(\mathbf{S}_{\text{val}}, \mathbf{N}_{\text{val}}, n)$
\State $\{\mathbf{X}_{\text{test}}, \mathbf{y}_{\text{test}}\} \gets \text{Seq\_With\_Prev\_Sent}(\mathbf{S}_{\text{test}}, \mathbf{N}_{\text{test}}, n)$

\State Initialize model $f_\theta$ with parameters $\theta$
\State Train $f_\theta$ on $\{\mathbf{X}_{\text{sub-train}}, \mathbf{y}_{\text{sub-train}}\}$ for $T$ epochs using learning rate $\alpha$, with validation on $\{\mathbf{X}_{\text{val}}, \mathbf{y}_{\text{val}}\}$
\State Predict $\hat{\mathbf{P}}$ using $f_\theta(\mathbf{X}_{\text{test}})$
\State Apply inverse normalization to $\hat{\mathbf{P}}$ and $\mathbf{y}_{\text{test}}$

\end{algorithmic}
\end{algorithm}

The Algorithm \ref{alg:2}, is almost identical to Algorithm \ref{alg:1}, with the only difference being how the sentiment scores are handled in the test set. For the training and validation phases, as in Algorithm \ref{alg:1}, the sentiment score for the current day is appended to the sequence of prices from the previous \( n \) days using the procedure \texttt{Seq\_With\_Cur\_Sent}. However, for the test set in line 6, the procedure \texttt{Seq\_With\_Prev\_Sent} is used instead, where the sentiment score from the previous day is appended to each input sequence.
In this case, the input sequence is enriched with sentiment information from the prior day to predict the price of the current day. This adjustment reflects the one-day-ahead prediction scenario, where sentiment data for the current day is unavailable during testing. The model must instead rely on the most recent sentiment data from the previous day to predict the price for the current day.

It is worth noting that we also experimented with incorporating the previous day’s sentiment during both the training and validation phases, in addition to testing for one-day-ahead prediction. The results will be presented in Section~\ref{sec6}.

\begin{algorithm}[H]
\caption{Simulating a Trading Strategy Based on One-Day-Ahead Price Predictions}
\label{alg:3}

\textbf{Input:} Actual prices $\mathbf{y}_{\text{test}}$, Predicted prices $\hat{\mathbf{P}}$, Transaction rate $r_t$, Initial capital $C_0$, Buy threshold $T_b$, Sell threshold $T_s$\\
\textbf{Output:} Total portfolio value $V$, Profit

\begin{algorithmic}[1]

\State Initialize capital $C \gets C_0$, cryptocurrency held $H \gets 0$
\For{$i \in (1, \text{len}(\mathbf{y}_{\text{test}}))$}
    \State $y_{\text{prev}} \gets \mathbf{y}_{\text{test}}[i-1]$
    \State $\hat{P}_i \gets \hat{\mathbf{P}}[i]$
    
    \If{$\frac{\hat{P}_i - y_{\text{prev}}}{y_{\text{prev}}} > T_b$ \textbf{and} $C > 0$}
        \State  $H \gets \frac{C}{y_{\text{prev}}} \times (1 - r_t)$
        \State $C \gets 0$
    \ElsIf{$\frac{y_{\text{prev}} - \hat{P}_i}{y_{\text{prev}}} > T_s$ \textbf{and} $H > 0$}
        \State  $C \gets H \times y_{\text{prev}} \times (1 - r_t)$
        \State $H \gets 0$
    \EndIf
\EndFor
\State $V \gets C + H \cdot \mathbf{y}_{\text{test}}[\text{len}(\mathbf{y}_{\text{test}})-1]$
\State $\text{profit} \gets V - C_0$
\State \textbf{Return} $V$, $\text{profit}$

\end{algorithmic}
\end{algorithm}

In Algorithm \ref{alg:3}, lines 2 to 12 simulate the trading strategy by iterating over the test data. In line 3, we retrieve the closing price of the current day, followed by accessing the predicted price for the next day in line 4. From lines 5 to 7, if the predicted price shows an expected increase beyond the buy threshold and we have available capital, the buy condition is triggered. In these steps, we invest the entire available capital into cryptocurrency, acquiring an amount adjusted for the transaction rate. Similarly, in lines 8 to 11, if the predicted price is expected to fall below the sell threshold and we hold cryptocurrency, the sell condition is activated. Here, all held assets are sold, and the capital is recovered as the product of the current price and the cryptocurrency held, adjusted by the transaction rate. This strategy maximizes returns by fully investing during price increases while selling during declines to avoid losses. In line 13, the total portfolio value is computed by summing the remaining capital with the value of any cryptocurrency still held. Lastly, in line 14, the profit is determined by subtracting the initial capital from this final portfolio value.
\subsection{Future m-day Price Prediction using FinBERT-Bi-LSTM}

Next, we explore two distinct approaches for predicting future cryptocurrency prices:

\begin{enumerate}
    \item In the first approach, Maximum Data Training(MDT), the model is trained on all available historical data up to the point immediately before the future testing data. This method does not use a validation set, allowing the model to learn from the maximum amount of data before making predictions.

    \item In the second approach, Validation-Enhanced Training(VET), a validation set is introduced between the training and testing phases. The model is trained and validated, with the version of the model that achieves the lowest validation loss being used to predict future prices, ensuring better generalization.
\end{enumerate}

Both approaches are demonstrated in Algorithm \ref{alg:4} and Algorithm \ref{alg:5}, respectively.

\begin{algorithm}[H]
\caption{Future m-Day Price Prediction using MDT}
\label{alg:4}

\textbf{Input:} Historical price data $\mathbf{S}$, News sentiment data $\mathbf{N}$, Hyperparameters: sequence length $n$, learning rate $\alpha$, epochs $T$, future forecast horizon $m$\\
\textbf{Output:} Predicted future prices $\hat{\mathbf{P}}$, Evaluation metrics

\begin{algorithmic}[1]

\State $\mathbf{S}, \mathbf{N} \rightarrow \{\mathbf{S}_{\text{train}}, \mathbf{S}_{\text{test}}\}, \{\mathbf{N}_{\text{train}}, \mathbf{N}_{\text{test}}\}$

\State $\mathbf{S}_{\text{train}} \gets \text{Normalize}(\mathbf{S}_{\text{train}})$

\State $\{\mathbf{X}_{\text{train}}, \mathbf{y}_{\text{train}}\} \gets \text{Seq\_With\_Cur\_Sent}(\mathbf{S}_{\text{train}}, \mathbf{N}_{\text{train}}, n)$

\State $f_\theta \gets \text{Initialize model with parameters } \theta$

\State $\theta \gets \text{Train } f_\theta \text{ on } \mathbf{X}_{\text{train}}, \mathbf{y}_{\text{train}} \text{ for } T \text{ epochs with learning rate } \alpha$

\State $\hat{\mathbf{P}} \gets \text{Pred\_Next\_M\_Days}(f_\theta, \mathbf{S}_{\text{train}}, n, m)$

\State $\hat{\mathbf{P}} \gets \text{InverseNormalize}(\hat{\mathbf{P}})$

\end{algorithmic}
\end{algorithm}

 In line 1 of Algorithm \ref{alg:4}, the dataset is split into training and test sets, with the test set containing the next m days following the training period. Lines 2-3 prepare the training data by normalizing prices, creating rolling windows of input sequences with their corresponding outputs, and appending sentiment scores to each sequence using the \texttt{Seq\_With\_Cur\_Sent} procedure described in Procedure \ref{proc:seq_with_cur_sent}. Lines 4 and 5 construct the model and train it. In line 6, the procedure \texttt{Pred\_Next\_M\_Days} takes as input the trained model, the training set, the sequence length $n$, and the future forecast horizon $m$ to predict prices for the next $m$ days in the testing set. The procedure \texttt{Pred\_Next\_M\_Days} here uses the training data to make predictions for the testing set, with further details provided in Procedure \ref{proc:pred_next}. At last in line 7, we apply inverse normalization to the predicted prices.

\begin{algorithm}[H]
\caption{Future m-Day Price Prediction using VET}
\label{alg:5}

\textbf{Input:} Historical price data $\mathbf{S}$, News sentiment data $\mathbf{N}$, Hyperparameters: sequence length $n$, learning rate $\alpha$, epochs $T$, future forecast horizon $m$\\
\textbf{Output:} Predicted prices $\hat{\mathbf{P}}$, Evaluation metrics

\begin{algorithmic}[1]

\State $\mathbf{S}, \mathbf{N} \rightarrow \{\mathbf{S}_{\text{train}}, \mathbf{S}_{\text{test}}\}, \{\mathbf{N}_{\text{train}}, \mathbf{N}_{\text{test}}\}$

\State $\mathbf{S}_{\text{train}} \rightarrow \{\mathbf{S}_{\text{subtrain}}, \mathbf{S}_{\text{val}}\}, \mathbf{N}_{\text{train}} \rightarrow \{\mathbf{N}_{\text{subtrain}}, \mathbf{N}_{\text{val}}\}$

\State Normalize $\mathbf{S}_{\text{subtrain}}$, $\mathbf{S}_{\text{val}}$

\State $\{\mathbf{X}_{\text{subtrain}}, \mathbf{y}_{\text{subtrain}}\} \gets \text{Seq\_With\_Cur\_Sent}(\mathbf{S}_{\text{subtrain}}, \mathbf{N}_{\text{subtrain}}, n)$

\State $f_\theta \gets \text{Initialize model with parameters } \theta$
\State $best\_mae \gets \infty$
\For{each epoch $i \in (1, T)$}
    \State Train $f_\theta$ on $\mathbf{X}_{\text{subtrain}}, \mathbf{y}_{\text{subtrain}}$ \text{ on the } $i^{th}$ epoch with learning rate $\alpha$
    \State $MAE \gets \Call{Validate\_Model}{f_\theta, \mathbf{S}_{\text{subtrain}}, \mathbf{S}_{\text{val}} , n, m}$
    \If{$mae < best\_mae$}
    \State $best\_mae \gets mae$
    \State $\theta^* \gets \theta$ \Comment{Store the weights of the best model}
\EndIf

\EndFor

\State $\theta \gets \theta^*$ \Comment{Load the weights of the model with the lowest MAE}

\State $\hat{\mathbf{P}} \gets \text{Pred\_Next\_M\_Days}(f_\theta, \mathbf{S}_{\text{val}}, n)$

\State $\hat{\mathbf{P}} \gets \text{InverseNormalize}(\hat{\mathbf{P}})$

\vspace{1em}
\Function{Validate\_Model}{$f_\theta, \mathbf{S}, \mathbf{V}, n, m$}
    \State $\hat{\mathbf{P}} \gets \text{Pred\_Next\_M\_Days}(f_\theta, \mathbf{S}, n, m)$
    \State Calculate MAE between $\hat{y}$ and $\mathbf{V}$
    \State \Return MAE
\EndFunction

\end{algorithmic}
\end{algorithm}

In lines 1-3 of Algorithm \ref{alg:5}, the dataset is first split into training and test sets, with the test set containing the first m days following the training set. The training set is then further divided into a training subset and a validation set, with the validation set representing the next m days immediately following the training subset. Later on, the training subset and validation set are normalized. As in Algorithm~\ref{alg:4}, line 4 uses the \texttt{Seq\_With\_Cur\_Sent} procedure described in Procedure \ref{proc:seq_with_cur_sent} to create input sequences with corresponding outputs for training.
In Line 5, we create the model with initial parameters. In lines 7-14, a loop iterates over the specified number of epochs, during which the model is trained and then evaluated using a validation function that takes as input the trained model after each epoch, the training subset, the validation subset, the sequence length, and the future forecast horizon. This function, implemented in lines 18-22, calls the \texttt{Pred\_Next\_M\_Days} procedure to predict the next $m$ days in the validation set, using the given training subset as input. After generating predictions, the Mean Absolute Error (MAE) is calculated between the actual validation prices and the predicted prices. The function \texttt{VALIDATE\_MODEL} concludes by returning the MAE, which is used to determine the model with the lowest validation loss. The best model is saved for future use in line 12. In lines 15-16, after the training is complete, the saved model (with the best validation loss) is used to predict the first $m$ days of the test dataset, with the procedure \texttt{Pred\_Next\_M\_Days} taking as input the saved model, the validation subset, the sequence length $n$, and the future forecast horizon $m$. Finally, in line 17, inverse normalization is applied to the predicted prices.

Now, the procedures mentioned in the algorithms are provided in detail.

\noindent\rule{\textwidth}{0.5pt} 

\begin{procedure}\label{proc:seq_with_cur_sent}
\vspace{0.25em} 

Seq\_With\_Cur\_Sent($\mathbf{S}$, $\mathbf{N}$, $n$)
\newline
\vspace{0.25em} 
\noindent\rule{\textwidth}{0.5pt} 

\begin{algorithmic}[1]
    \State Initialize empty sequences $\mathbf{X}$ and targets $\mathbf{y}$
    \For{$i \in (n, \text{len}(\mathbf{S}))$}
        \State $\mathbf{X}[i] \gets \mathbf{S}[i - n : i]$
        \State $\mathbf{y}[i] \gets \mathbf{S}[i]$
        \State $\mathbf{X}[i] \gets \mathbf{X}[i] \cup \mathbf{N}[i]$
    \EndFor
    \State \Return $\mathbf{X}, \mathbf{y}$
\end{algorithmic}
\vspace{0.25em} 
\end{procedure}

\noindent\rule{\textwidth}{0.5pt} 

\begin{procedure}\label{proc:seq_with_prev_sent}
\vspace{0.25em}

Seq\_With\_Prev\_Sent($\mathbf{S}$, $\mathbf{N}$, $n$)
\newline
\vspace{0.25em}
\noindent\rule{\textwidth}{0.5pt} 

\begin{algorithmic}[1]
    \State Initialize empty sequences $\mathbf{X}$ and targets $\mathbf{y}$
    \For{$i \in (n, \text{len}(\mathbf{S}))$}
        \State $\mathbf{X}[i] \gets \mathbf{S}[i - n : i]$
        \State $\mathbf{y}[i] \gets \mathbf{S}[i]$
        \State $\mathbf{X}[i] \gets \mathbf{X}[i] \cup \mathbf{N}[i - 1]$
    \EndFor
    \State \Return $\mathbf{X}, \mathbf{y}$
\end{algorithmic}
\vspace{0.25em}
\end{procedure}

\noindent\rule{\textwidth}{0.5pt} 

\begin{procedure}\label{proc:pred_next}
\vspace{0.25em}

Pred\_Next\_M\_Days($f_\theta$, $\mathbf{S}$, $n$, $m$)
\newline
\vspace{0.25em}
\noindent\rule{\textwidth}{0.5pt} 

\begin{algorithmic}[1]
    \State $t \gets \text{length of } \mathbf{S}$
    \State $\text{input\_seq} \gets \mathbf{S}[t - n : t]$
    \For{$i \in (1, m)$}
        \State $\hat{P}_i \gets f_\theta(\text{input\_seq})$
        \State $\text{input\_seq} \gets \text{input\_seq} \cup \hat{P}_i$
        \State $\text{input\_seq} \gets \text{input\_seq}[1:]$
    \EndFor
    \State \Return $\hat{\mathbf{P}}$
\end{algorithmic}
\vspace{0.25em}
\end{procedure}
\noindent\rule{\textwidth}{0.5pt} 

The procedure \texttt{Seq\_With\_Cur\_Sent} takes as input the price data \( \mathbf{S} \), sentiment data \( \mathbf{N} \), and sequence length \( n \). Inside the procedure, lines 2-6 apply a rolling window of size \( n \) days to generate input sequences and their corresponding outputs. In lines 3-4, for each \( i \)-th day, the model processes a sequence of prices from days \( i-n \) to \( i-1 \) as input to predict the price for day \( i \) as the corresponding output. In line 5, the sentiment score for current day  \( \mathbf{N}[i] \) is appended to the price sequence, enriching the input with relevant sentiment information. The procedure concludes by returning the sequences \( \mathbf{X} \) and targets \( \mathbf{y} \) in line 7.
\newline

The procedure \texttt{Seq\_With\_Prev\_Sent} is similar to \texttt{Seq\_With\_Cur\_Sent}, with the key difference being that in line 5, it integrates the sentiment score from the previous day  \( \mathbf{N}[i-1] \) into the input sequence instead of appending the sentiment score for day \( i \).
\newline

The procedure \texttt{Pred\_Next\_M\_Days} takes as input the model $f_\theta$, price data \( \mathbf{S} \), the sequence length $n$, and the future forecast horizon $m$. The input sequence is initialized with the last \(n\) days of the price data in line 2. In lines 3 to 7, the model iterates for m steps: in line 4, the model is provided with the current input sequence to predict the price for the next day. In line 5, the predicted price is appended to the input sequence, and in line 6, the oldest value is removed. The key difference between this approach and the one in Algorithm \ref{alg:1} and Algorithm \ref{alg:2} is how we handle predictions. Here, we use the predicted value from the previous step as part of the input sequence for the next prediction, while in line 9 of Algorithm \ref{alg:1} and Algorithm \ref{alg:2}, we provided the entire $\mathbf{X}_{test}$ at once for prediction, that is, we are using the actual test values instead of using the predicted values to predict the next one. Finally, the predicted values for the next \(m\) days are returned in line 8, marking the end of the procedure.

\section{Experimental Setting}\label{sec5}
In this section, we detail the experimental setup used to evaluate the performance of our proposed model for cryptocurrency price prediction. We describe the datasets used, including historical price data, financial news and texts, and the preprocessing steps applied to ensure data consistency and relevance. Additionally, the sentiment analysis process, architecture of the models, and key hyperparameters are outlined. The computational environment used to conduct the experiments is also discussed, ensuring transparency and reproducibility in the experimental workflow.

\subsection{Dataset Description}
To forecast the prices of Bitcoin (BTC) and Ethereum (ETH), we compiled distinct datasets encompassing historical price data, financial news articles, and texts. These datasets were carefully selected and preprocessed to capture the intricate dynamics of cryptocurrency markets, ensuring that the model could effectively leverage the rich information available. Below is a detailed description of the data sources and preprocessing steps used in our research.

\subsubsection{Historical Price Data Collection}
The historical price data for Bitcoin (BTC) and Ethereum (ETH) was obtained from Yahoo Finance\footnote{\url{https://finance.yahoo.com/}, last accessed October 28, 2024}, a highly reputable source for reliable financial information. The dataset spans from January 1, 2023, covering a total of 585 days for both Bitcoin and Ethereum. In this study, we focus exclusively on the daily closing price for training. This choice provides a clear and concise summary of the day’s market activity while maintaining a streamlined and simplified model.

To prepare the historical price data for analysis, normalization is a crucial preprocessing step that scales the data to a common range, typically between 0 and 1. This transformation helps the model converge more efficiently during training and ensures uniformity across the input data. In this study, we applied the \texttt{MinMaxScaler} function from the Scikit-Learn library to normalize the financial data.

The formula for normalization is as follows:

\begin{equation}
X_t^n = \frac{X_t - \min(X_t)}{\max(X_t) - \min(X_t)}
\end{equation}

where \(X_t^n\) represents the normalized data point of \(X_t\). To revert the data to its original scale, inverse normalization is applied using the following formula:

\begin{equation}
X_t = X_t^n \times \left[\max(X_t) - \min(X_t)\right] + \min(X_t)
\end{equation}

Here, \(X_t^n\) is the normalized data, and the formula restores the original values by applying the inverse transformation.

\subsubsection{Transaction Rate}
 In this study, the transaction rates for Bitcoin and Ethereum were sourced from CoinRemitter\footnote{\url{https://coinremitter.com/fees}, last accessed October 28, 2024}, a reliable service provider known for offering competitive fees. Specifically, we utilized a transaction rate of 0.0001 BTC for Bitcoin and 0.003 ETH for Ethereum, which includes the gas station optimization feature for Ethereum. The gas station rate was chosen because it reflects a more efficient and cost-effective transaction fee, as traders often use this feature to minimize costs. By incorporating these transaction rates, our simulation ensures a more accurate modeling of the transaction costs that traders might face on both networks.

\subsubsection{News and Financial Text Collection}

To capture the impact of market sentiment on Bitcoin (BTC) and Ethereum (ETH) price movements, we collected news articles and financial texts from reputable cryptocurrency news platforms, including Coindesk\footnote{\url{https://www.binance.com/en}, last accessed on October 28, 2024}, Binance\footnote{\url{https://www.coindesk.com/}, last accessed on October 28, 2024} and other specialized websites focused on the cryptocurrency market. These sources provide up-to-date and comprehensive coverage of events, regulatory changes, and technological developments influencing market sentiment.

The dataset spans 585 days from January 1, 2023, and includes news headlines, article summaries, and key financial reports that provide valuable context for understanding BTC and ETH price movements. This period was selected due to the availability of comprehensive news coverage, providing a rich context for sentiment analysis. Separate datasets were created for BTC and ETH to allow for distinct analysis of each cryptocurrency's market sentiment.

\subsubsection{Dataset Splitting}
The datasets were divided using a hierarchical segmentation approach for intraday predictions, one-day-ahead predictions, and future m-day predictions with Validation-Enhanced Training. In this approach, 85\% of the data was allocated to the training set and 15\% to the test set. Additionally, within the training set, 85\% was used for model training, while the remaining 15\% was reserved for validation. In contrast, future m-day predictions with Maximum Data Training used a simpler data split, with 85\% of the data allocated for training and 15\% for testing, without any further division for validation.

\subsection{Model Architectures and Hyperparameters}
This section outlines the architectures and hyperparameters used in our models, detailing the configurations of LSTM and Bi-LSTM, followed by the integration of FinBERT with both architectures to leverage sentiment analysis for enhanced predictive performance.
\subsubsection{LSTM and FinBERT-LSTM Architecture}
For the LSTM module, we designed a sequential model comprising the following layers: 
(i) an LSTM layer with 50 units and \texttt{tanh} activation; 
(ii) a second LSTM layer with 30 units, also with \texttt{tanh} activation; 
(iii) a third LSTM layer with 20 units and \texttt{tanh} activation; 
(iv) Finally, a dense layer with 1-dimensional output using a \texttt{linear} activation function. This architecture was used for the Bitcoin (BTC) dataset. For the Ethereum (ETH) dataset, the same configuration was used, except for the number of units in each LSTM layer, where 55, 25, and 20 units were used for the first, second, and third LSTM layers, respectively. The model was compiled using \texttt{Mean Squared Error} as the loss function and the \texttt{Adam} optimizer with the chosen learning rate. The FinBERT-LSTM model follows the same architecture, with the only difference being the incorporation of news sentiment data.

\subsubsection{Bi-LSTM and FinBERT-Bi-LSTM Architecture}
For the Bi-LSTM module, we used a similar sequential model, with the following layers: 
(i) a Bidirectional LSTM layer with 55 units and \texttt{tanh} activation; 
(ii) a second Bidirectional LSTM layer with 25 units, also using \texttt{tanh} activation; 
(iii) a third Bidirectional LSTM layer with 20 units and \texttt{tanh} activation; 
(iv) Finally, a dense layer with 1-dimensional output using a \texttt{linear} activation function. This architecture was applied to both Bitcoin (BTC) and Ethereum (ETH) datasets, as this configuration provided consistent performance across both datasets. The model was also compiled using \texttt{Mean Squared Error} as the loss function and the \texttt{Adam} optimizer with the chosen learning rate. The FinBERT-Bi-LSTM model follows the same architecture, incorporating news sentiment data.

\subsubsection{Hyperparameters} The following hyperparameters were used for both the intra-day, one-day ahead, and future prediction experiments for both BTC and ETH:

\begin{itemize} 
    \item \textbf{Epochs:} The number of epochs was set to 100 across all experiments.
     
    \item \textbf{Sequence Length:} For intra-day and one-day ahead predictions, the sequence length was fixed at 10 for both BTC and ETH. For future predictions, the sequence length varied between 10, 30, and 31, adjusted through trial and error to optimize performance.

    \item \textbf{Learning Rate:} For intraday and one-day ahead predictions, a learning rate of 0.02 was used consistently for both BTC and ETH. For future predictions, the learning rate was varied between 0.00075 and 0.02, manually tuned based on model performance.
    
\end{itemize}

In the future prediction context, the sequence length and learning rate were adjusted more flexibly to account for the unique characteristics of the forecasting task. In Validation-Enhanced Training approach, a validation dataset was used to select the optimal values for sequence length and learning rate. In the Maximum Data Training approach, where no validation set was available, the selection was based on model performance during training.

\subsubsection{Forecast Horizon} 
For the future prediction tasks, we set the forecast horizon to \( m = 30 \) days. This choice balances the need for a meaningful prediction window with the inherent volatility of the cryptocurrency market, as extending the forecast too far into the future resulted in erratic outcomes.

\subsection{Sentiment Analysis}

In this study, sentiment analysis has been conducted on the collected financial news articles related to Bitcoin (BTC) and Ethereum (ETH) using FinBERT. 

\begin{figure}[h]
    \centering
    \includegraphics[width=.99\textwidth]{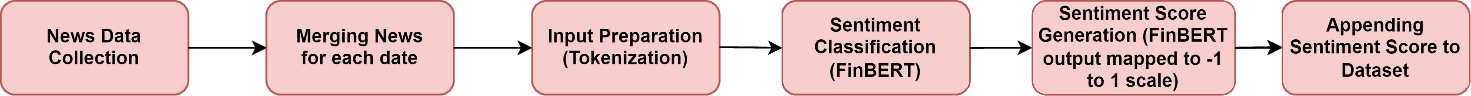}
    \caption{Flowchart of the Sentiment Analysis Process for BTC and ETH News Using FinBERT}
    \label{fig:sentiment}
\end{figure}

The following steps outline the sentiment analysis workflow using FinBERT, as illustrated in Fig. \ref{fig:sentiment}:

\begin{itemize}

\item \textbf{News Data Collection:} We have collected news articles relevant to Bitcoin (BTC) and Ethereum (ETH) from various reputable sources, including finance platforms, CoinDesk, Binance, and other trusted outlets, to ensure a comprehensive dataset.

\item \textbf{Merging News for Each Date:} All news headlines or articles for a given date have been merged into a single input for sentiment analysis.

\item \textbf{Input Preparation:} Using FinBERT-specific tokenization, we have transformed the merged text into a format that retains its context while being suitable for the FinBERT model.

\item \textbf{Sentiment Classification (FinBERT):} After tokenization, we process the text using the FinBERT model, which assigns a sentiment label (positive, neutral, or negative).

    \begin{itemize}
        \item Positive sentiments indicate an optimistic outlook regarding market conditions.
        \item Neutral sentiments reflect a balanced perspective, suggesting no significant market influence.
        \item Negative sentiments represent a pessimistic view, highlighting concerns about market trends.
    \end{itemize}

\item \textbf{Sentiment Score Generation:} Along with the sentiment label, the FinBERT model produces a sentiment score that represents the overall sentiment for that day's news, with the scores mapped to a scale from -1 to 1 based on the sentiment labels. Scores greater than 0 reflect a favorable sentiment, scores less than 0 indicate an unfavorable sentiment, and a score of 0 represents a neutral stance, suggesting no strong influence on the market.

\item \textbf{Appending Sentiment Score to Dataset:} The sentiment scores generated by FinBERT are appended to the news dataset as a new column labeled ``FinBERT Score.'' These sentiment scores are then used in conjunction with the historical price data to capture the relationship between market sentiment and cryptocurrency price movements.

\end{itemize}

This sentiment analysis process was implemented using Python's \texttt{transformers} library and Hugging Face's FinBERT model. The results were integrated into the predictive model as features, enhancing its understanding of market dynamics and improving the accuracy of price predictions for BTC and ETH.

\subsection{Computational Environment}
The computational tasks, including sentiment analysis using FinBERT and the training of the Bi-LSTM model, were conducted in a Google Colab environment equipped with an NVIDIA T4 GPU. This environment provided up to 16 GB of GPU memory and up to 25 GB of available RAM, facilitating efficient processing of large volumes of financial text and historical price data.

The experiments were carried out using Python 3.10, leveraging key libraries such as \texttt{transformers} for implementing FinBERT, \texttt{pandas} for data manipulation and handling, and \texttt{nltk} for additional sentiment analysis tasks. This setup ensured that the computations were both efficient and reproducible.

\section{Results}\label{sec6}
 
This section presents the evaluation of four models—LSTM, Bi-LSTM, FinBERT-LSTM, and FinBERT-Bi-LSTM—across various cryptocurrency prediction tasks, complemented by a daily trading simulation.

\subsection{Performance Metrics}

To assess the performance of our models, we employed three key metrics: Mean Absolute Error (MAE), Mean Absolute Percentage Error (MAPE), and Accuracy. These metrics are defined as follows:

\begin{enumerate}
    \item \textbf{Mean Absolute Error (MAE)}: This metric calculates the average magnitude of errors in predictions, providing a straightforward measure of accuracy in the same units as the data.
\begin{equation}
\text{MAE} = \frac{1}{n} \sum_{i=1}^{n} |y_i - \hat{y}_i|
\end{equation}

    \item \textbf{Mean Absolute Percentage Error (MAPE)}: MAPE provides the average absolute error as a percentage, helping to interpret prediction accuracy relative to actual values.
\begin{equation}
\text{MAPE} = \frac{1}{n} \sum_{i=1}^{n} \left|\frac{y_i - \hat{y}_i}{y_i}\right| \times 100\%
\end{equation}

    \item \textbf{Accuracy}: Defined as the complement of MAPE, this metric reflects the percentage of accurate predictions as an overall measure of model performance.
    \begin{equation}
     \text{Accuracy} = 1 - \text{MAPE}
    \end{equation}

\end{enumerate}

\subsection{Results of Intra-Day Price Prediction}

In this section, we evaluate the impact of integrating current-day sentiment scores into FinBERT-LSTM and FinBERT-Bi-LSTM models, comparing their performance against plain LSTM and Bi-LSTM models for predicting the same day's closing price.

\subsubsection{Results of Bitcoin Intra-Day Prediction}

We first evaluate the training and validation loss of LSTM, FinBERT-LSTM, Bi-LSTM, and FinBERT-Bi-LSTM models over the course of 100 epochs.

\begin{figure}[H]
    \centering
    \begin{minipage}{0.40\textwidth}
        \centering
        \includegraphics[width=\textwidth]{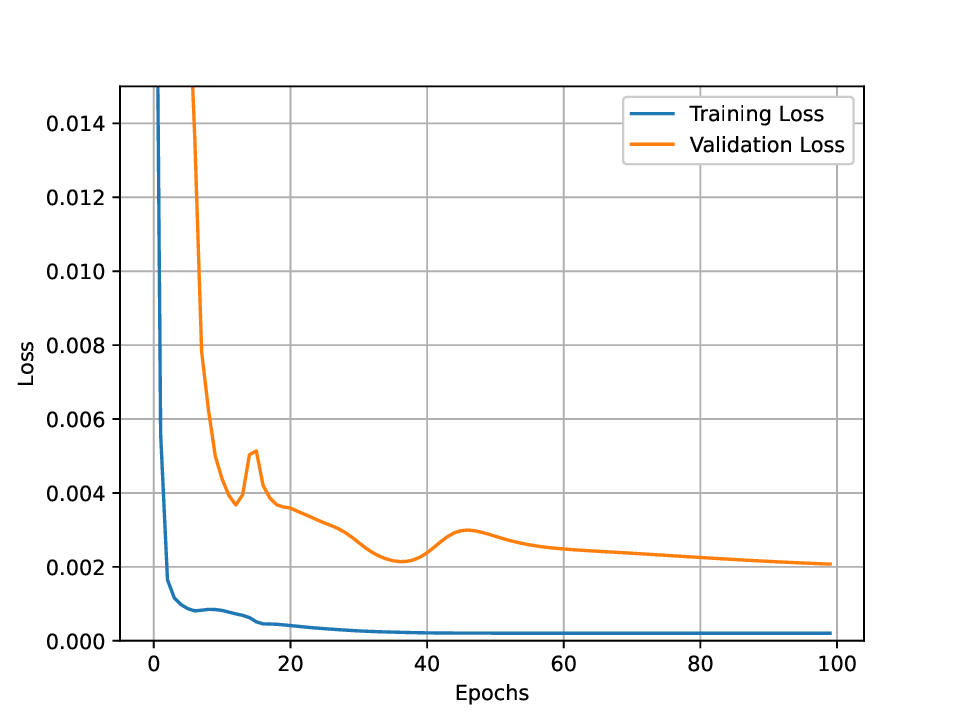}
        \caption{BTC-Training and Validation loss of LSTM Model}
        \label{fig:train-val-loss-lstm}
    \end{minipage}
    \hfill
    \begin{minipage}{0.40\textwidth}
        \centering
        \includegraphics[width=\textwidth]{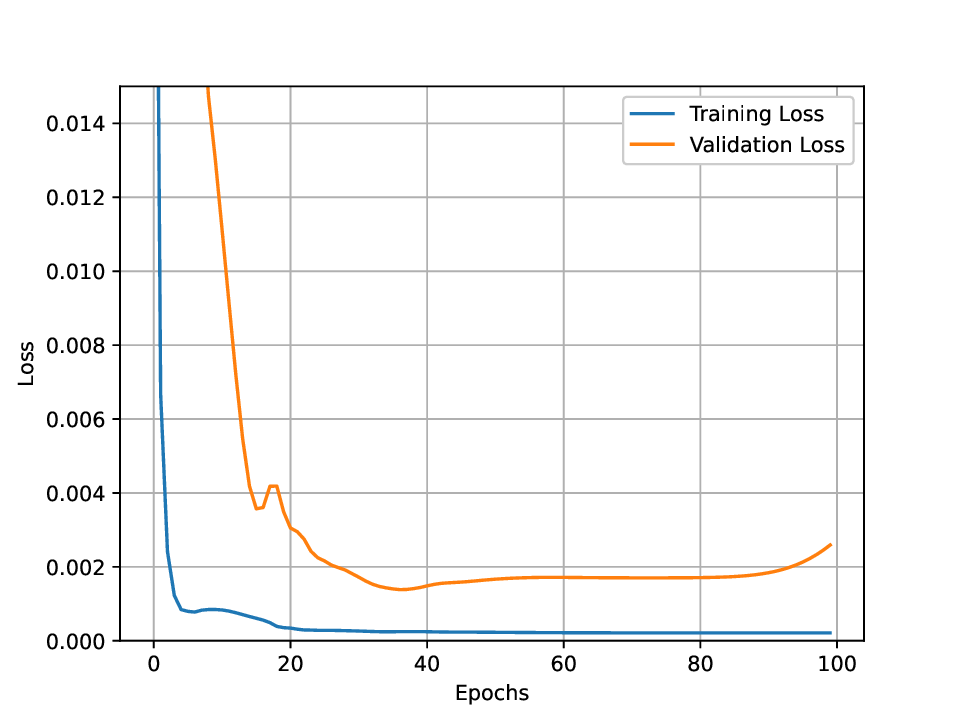}
        \caption{BTC-Training and Validation loss of FinBERT-LSTM Model}
        \label{fig:train-val-loss-finbert-lstm}
    \end{minipage}
\end{figure}

In Fig. \ref{fig:train-val-loss-lstm}, the LSTM model reaches its lowest validation loss of 0.0021 at epoch 37. Meanwhile, Fig. \ref{fig:train-val-loss-finbert-lstm} highlights the superior performance of the FinBERT-LSTM model compared to the plain LSTM model, achieving a validation loss of 0.0014 at epoch 35.

\begin{figure}[H]
    \centering
    \begin{minipage}{0.44\textwidth}
        \centering
        \includegraphics[width=\textwidth]{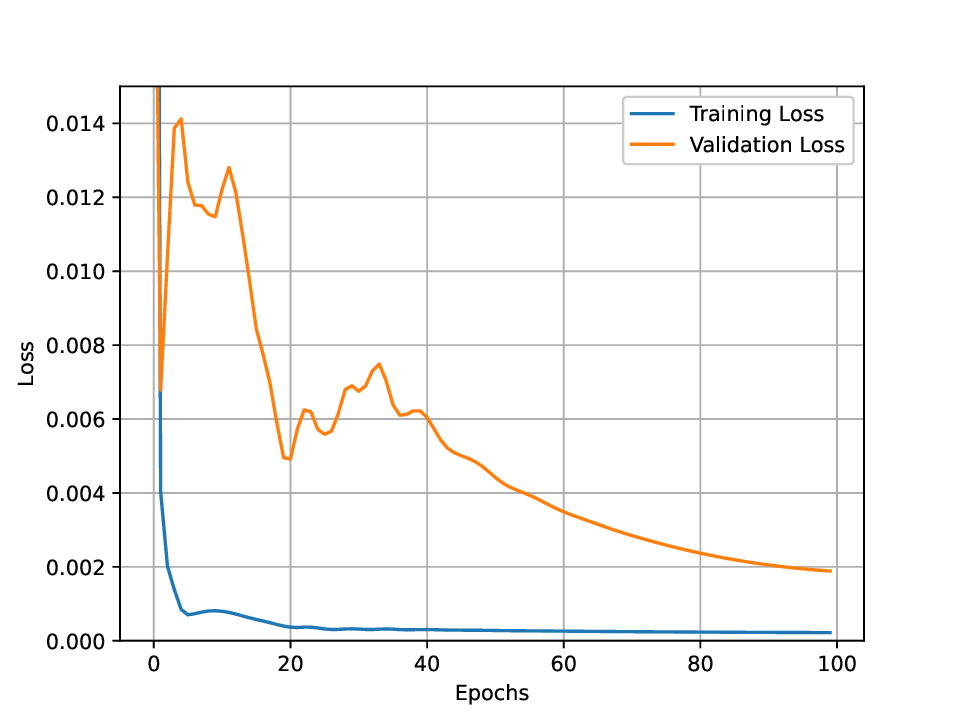}
        \caption{BTC-Training and Validation loss of Bi-LSTM Model}
        \label{fig:train-val-loss-bi-lstm}
    \end{minipage}
    \hfill
    \begin{minipage}{0.44\textwidth}
        \centering
        \includegraphics[width=\textwidth]{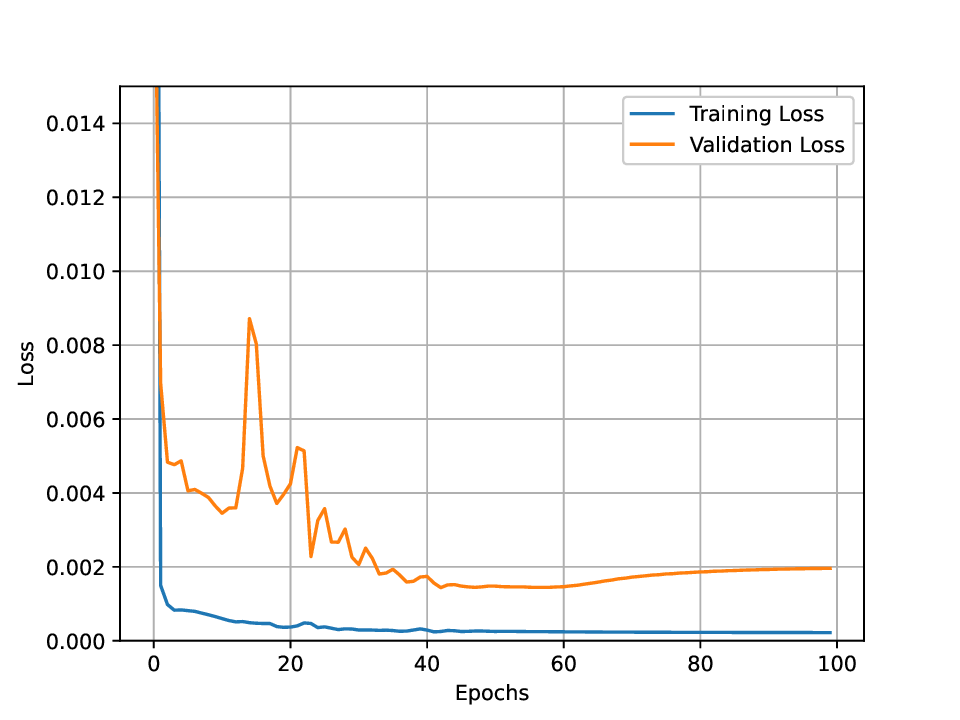}
        \caption{BTC-Training and Validation loss of FinBERT-Bi-LSTM Model}
        \label{fig:train-val-loss-finbert-bi-lstm}
    \end{minipage}
\end{figure}

Similarly, in Fig. \ref{fig:train-val-loss-bi-lstm}, the Bi-LSTM model records its lowest validation loss of 0.0019 at epoch 96.  In contrast, the FinBERT-Bi-LSTM model, as shown in Fig. \ref{fig:train-val-loss-finbert-bi-lstm}, outperforms the plain Bi-LSTM model with a validation loss of 0.0014 at epoch 43. Next, we compare the performance of all models on the test data graphically.

\begin{figure}[H]
    \centering
    \includegraphics[width=0.7\textwidth]{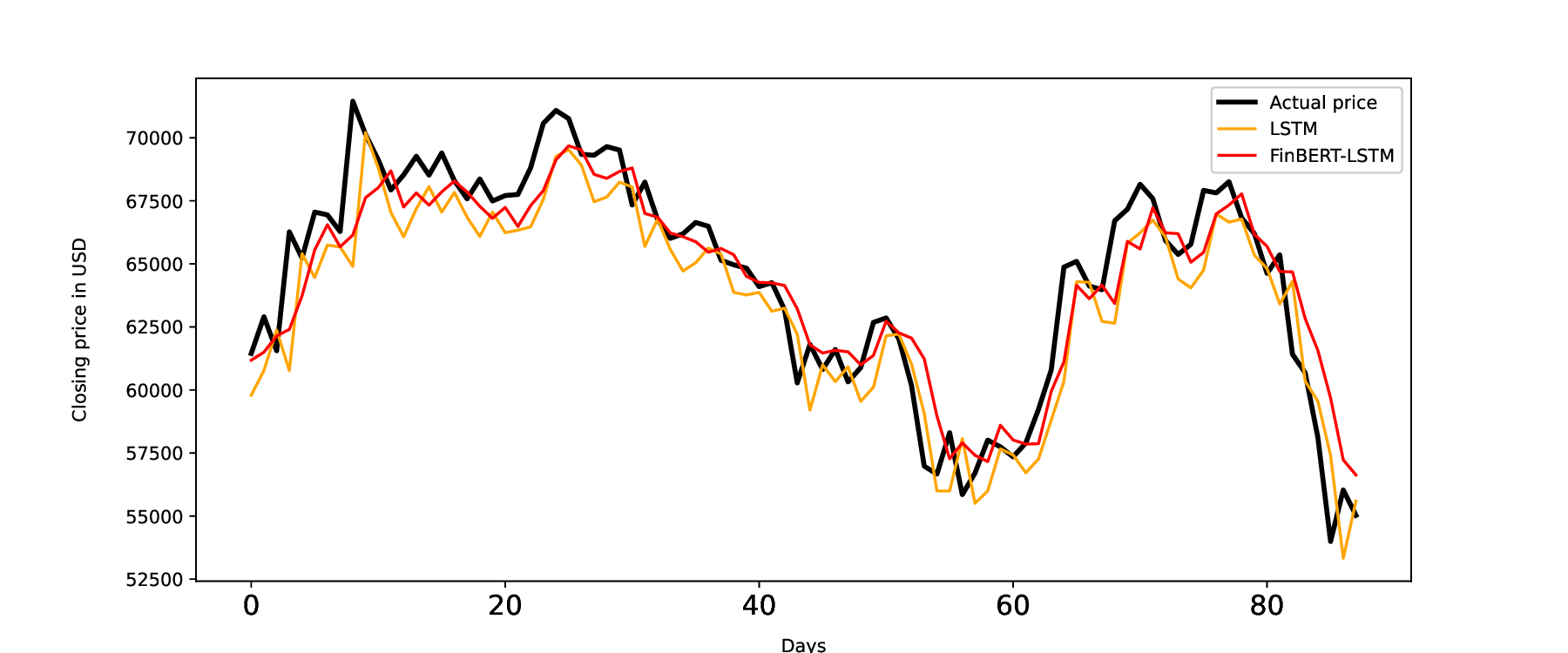}
    \caption{BTC-Actual vs Intra-Day Predicted Price Using LSTM Model and FinBERT-LSTM Model
}
    \label{fig:LSTM-vs-Finbert-LSTM}
\end{figure}

\begin{figure}[H]
    \centering
    \includegraphics[width=0.7\textwidth]{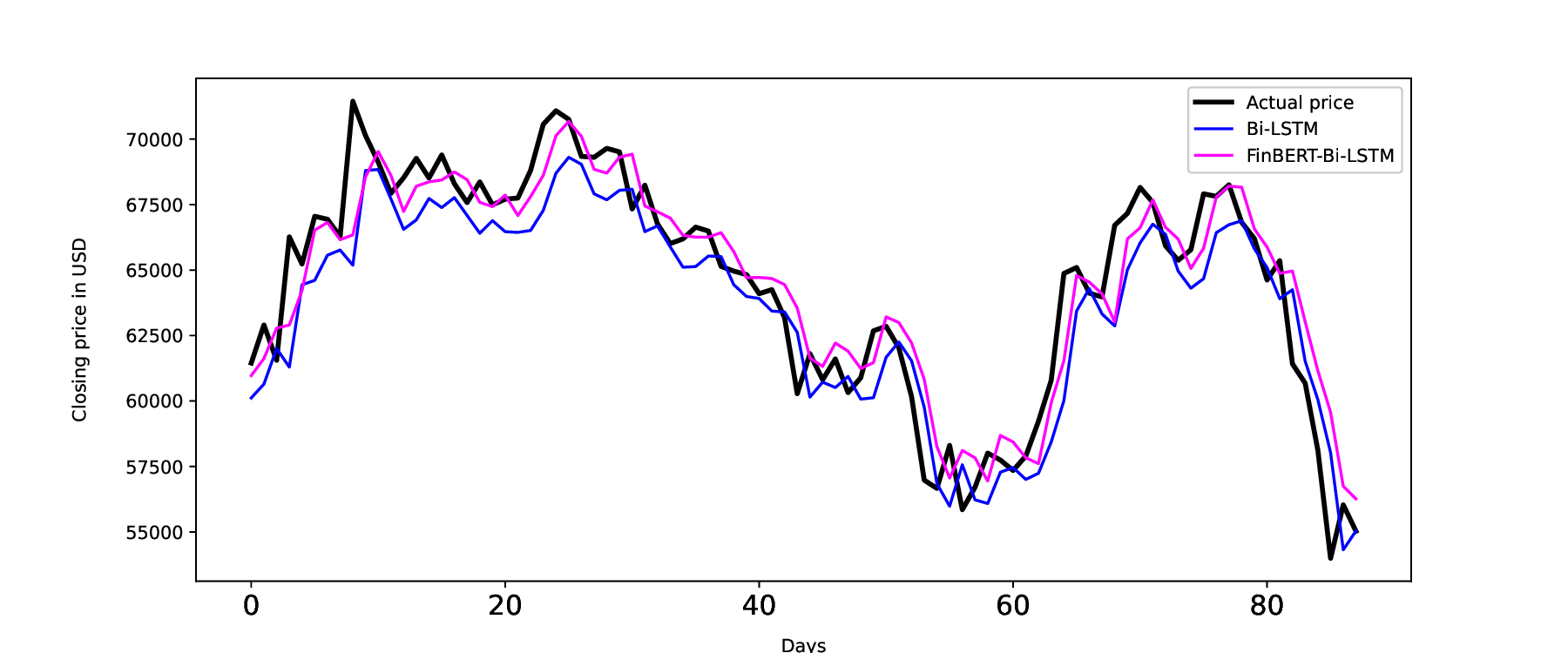}
    \caption{BTC-Actual vs Intra-Day Predicted Price Using Bi-LSTM Model and FinBERT-Bi-LSTM Model}
    \label{fig:Bi-LSTM-vs-Finbert-Bi-LSTM}
\end{figure}

In Fig. \ref{fig:LSTM-vs-Finbert-LSTM}, the actual price is depicted in black, with the LSTM predictions in yellow and the FinBERT-LSTM predictions in red. Similarly, in Fig. \ref{fig:Bi-LSTM-vs-Finbert-Bi-LSTM}, the actual price is shown in black, while the Bi-LSTM predictions are in blue, and the FinBERT-Bi-LSTM predictions in magenta.

Table \ref{Table1} highlights that both FinBERT-integrated models outperform their plain counterparts, with the FinBERT-Bi-LSTM model achieving the highest accuracy of 98.21\% for intra-day predictions supporting our hypothesis mentioned in Section~\ref{sec4}.

\begin{table}[htbp]
\centering
\caption{BTC-Performance Comparison of LSTM, FinBERT-LSTM, Bi-LSTM, and FinBERT-Bi-LSTM Models for Intra-Day Price Prediction}
\label{Table1}
\begin{tabular}{ |c|c|c|c|c| } 
\hline
\textbf{Metric} & \textbf{LSTM} & \textbf{FinBERT-LSTM} & \textbf{Bi-LSTM} & \textbf{FinBERT-Bi-LSTM} \\ 
\hline
MAE & 1453.36 & 1239.01 & 1419.26 & \textbf{1134.59}\\ 
MAPE & 0.02317\% & 0.01946\% & 0.02250\% & \textbf{0.01788\%} \\ 
Accuracy & 97.68\% & 98.05\% & 97.75\% & \textbf{98.21\%} \\ 
\hline
\end{tabular}
\end{table}

\subsubsection{Results of Ethereum Intra-Day Prediction}
Following the same process as with Bitcoin, we first evaluate the training and validation loss of the LSTM, FinBERT-LSTM, Bi-LSTM, and FinBERT-Bi-LSTM 
models over the course of 100 epochs.

\begin{figure}[H]
    \centering
    \begin{minipage}{0.44\textwidth}
        \centering
        \includegraphics[width=\textwidth]{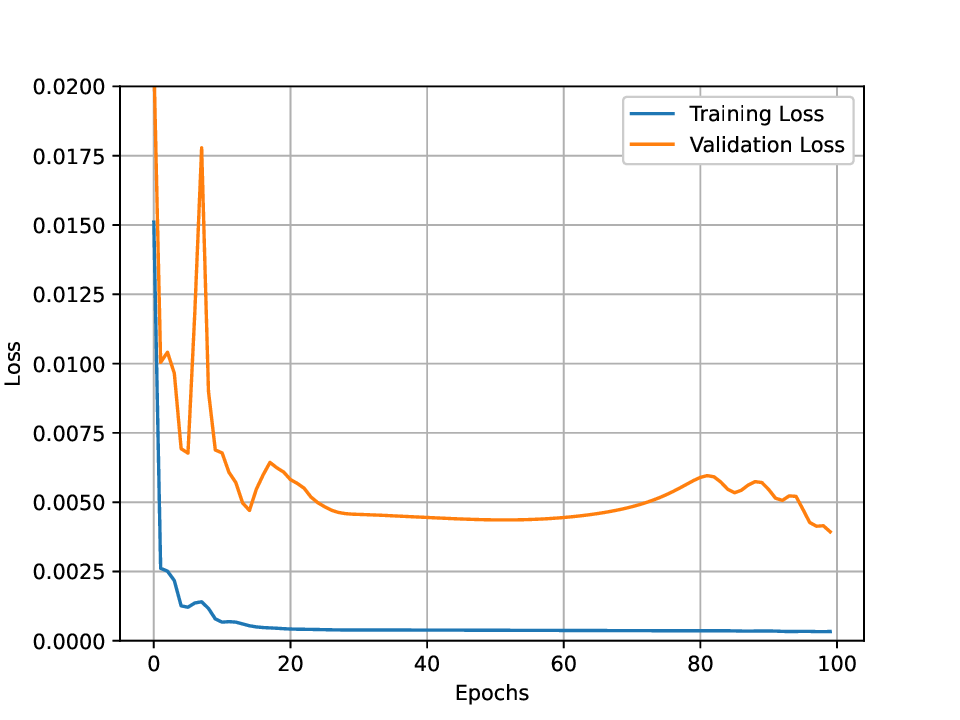}
        \caption{ETH-Training and Validation Loss of LSTM Model}
        \label{fig:26}
    \end{minipage}
    \hfill
    \begin{minipage}{0.44\textwidth}
        \centering
        \includegraphics[width=\textwidth]{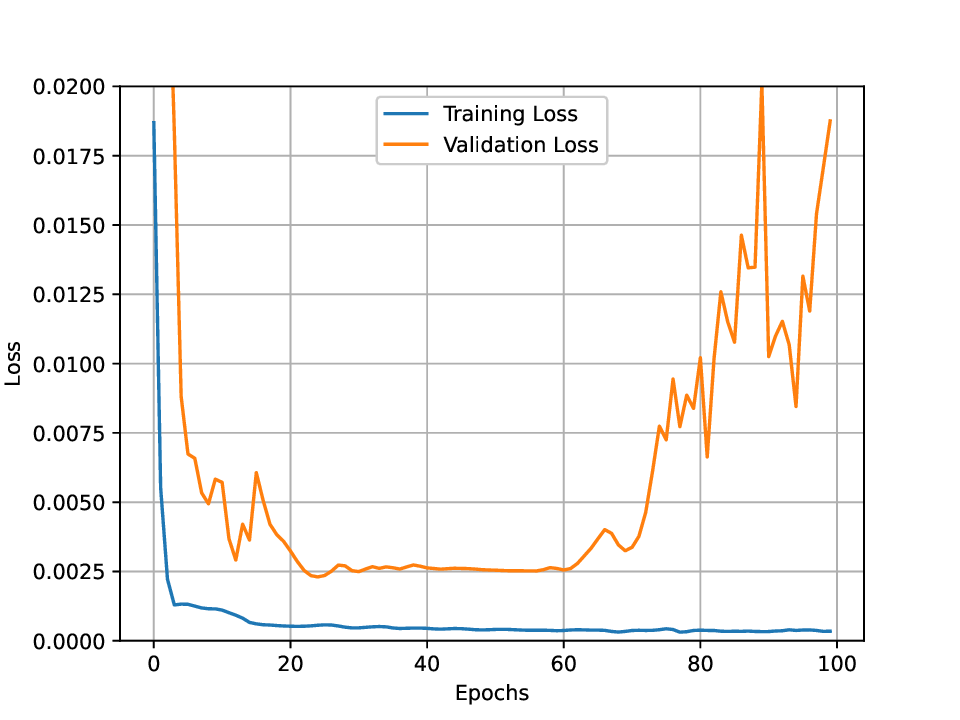}
        \caption{ETH-Training and Validation Loss of FinBERT-LSTM Model}
        \label{fig:27}
    \end{minipage}
\end{figure}

\begin{figure}[H]
    \centering
    \begin{minipage}{0.44\textwidth}
        \centering
        \includegraphics[width=\textwidth]{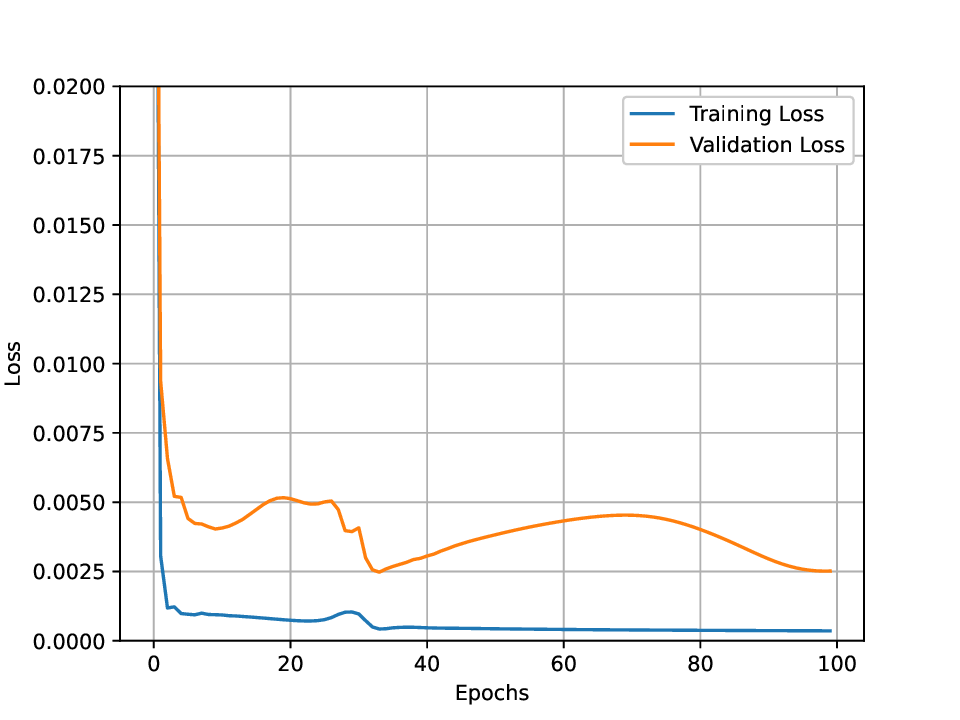}
        \caption{ETH-Training and Validation loss of Bi-LSTM Model}
        \label{fig:29}
    \end{minipage}
    \hfill
    \begin{minipage}{0.44\textwidth}
        \centering
        \includegraphics[width=\textwidth]{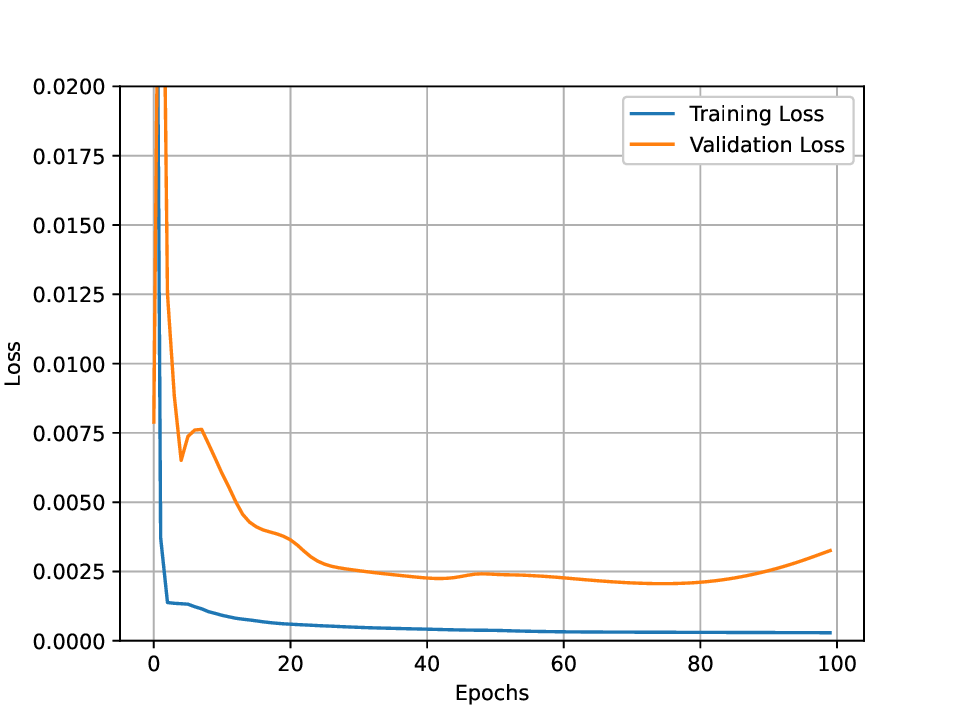}
        \caption{ETH-Training and Validation loss of FinBERT-Bi-LSTM Model}
        \label{fig:30}
    \end{minipage}
\end{figure}

In Fig. \ref{fig:26}, the lowest validation loss for the LSTM model is recorded at epoch 100, with a value of 0.0039. In Fig. \ref{fig:27}, the FinBERT-LSTM model shows superior performance, with the lowest validation loss of 0.0023, recorded at epoch 25. In Fig. \ref{fig:29}, the Bi-LSTM model achieved its lowest validation loss of 0.0025 at the 34th epoch, while in Fig. \ref{fig:30}, the FinBERT-Bi-LSTM model reached its lowest validation loss of 0.0021, first recorded at the 67th epoch.

Next, we graphically compare the performance of all the models over the test data.

\begin{figure}[H]
    \centering
    \includegraphics[width=0.7\textwidth]{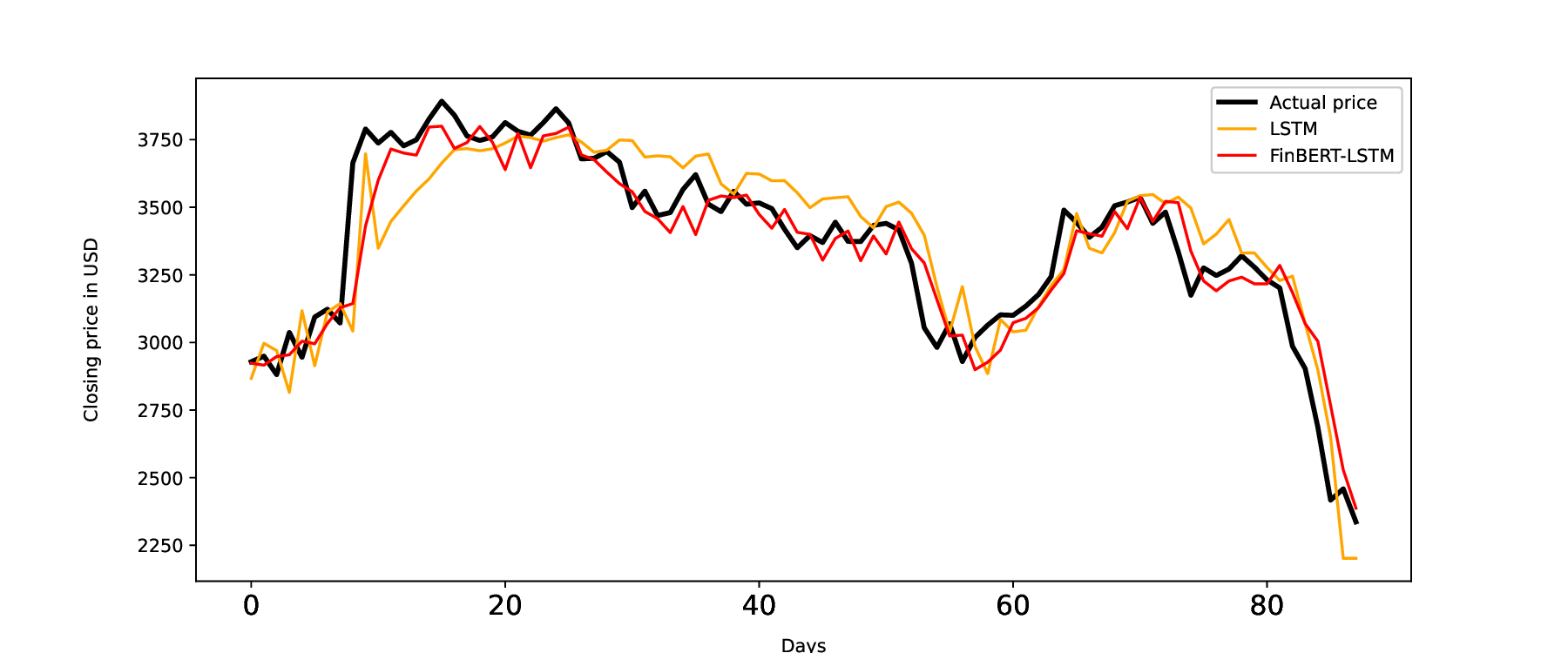}
    \caption{ETH-Actual vs Intra-Day Predicted Price Using LSTM Model and FinBERT-LSTM Model}
    \label{fig:28}
\end{figure}
\begin{figure}[H]
    \centering
    \includegraphics[width=0.7\textwidth]{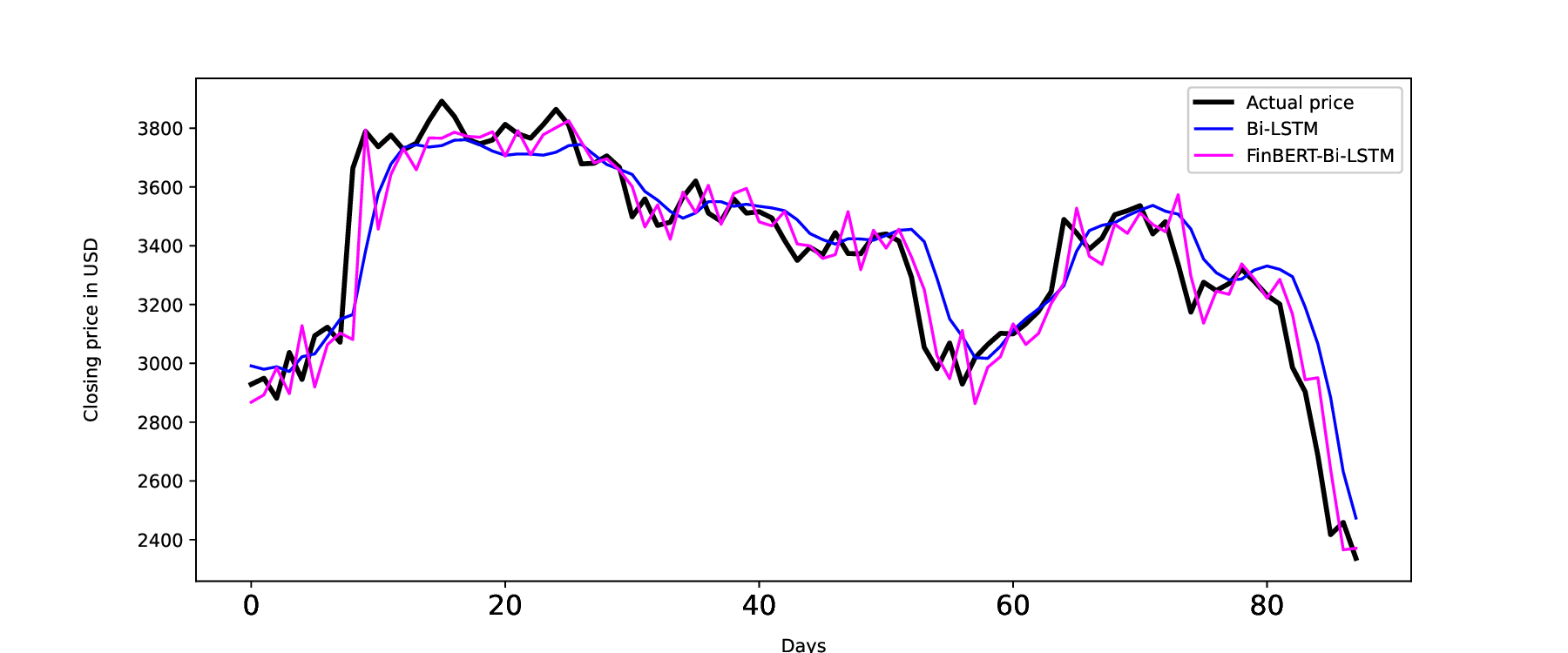}
    \caption{ETH-Actual vs Intra-Day Predicted Price Using BI-LSTM Model and FinBERT-Bi-LSTM Model}
    \label{fig:bilstm_vs_finbert_bilstm}
\end{figure}

In Fig. \ref{fig:28}, the actual price is depicted in black, with LSTM predictions in orange and FinBERT-LSTM predictions in red. Similarly, in Fig. \ref{fig:bilstm_vs_finbert_bilstm}, the actual price is again shown in black, while the Bi-LSTM predictions appear in blue and the FinBERT-Bi-LSTM predictions in magenta.

A quantitative comparison of all models for intra-day price prediction is presented in Table \ref{Table2}, where the FinBERT-LSTM model shows an improved accuracy of 97.27\% compared to 96.16\% for the LSTM model in the case of Ethereum. Moreover, the Bi-LSTM model outperforms the LSTM, with the FinBERT-Bi-LSTM model achieving the highest accuracy of 97.50\%.

\vspace{0.01cm} 
\begin{table}[htbp]
\centering
\caption{ETH-Performance Comparison of LSTM, FinBERT-LSTM, Bi-LSTM, and FinBERT-Bi-LSTM Models for Intra-Day Price Prediction}
\label{Table2}
\begin{tabular}{ |c|c|c|c|c| } 
\hline
\textbf{Metric} & \textbf{LSTM} & \textbf{FinBERT-LSTM} & \textbf{Bi-LSTM} & \textbf{FinBERT-Bi-LSTM} \\ 
\hline
MAE & 126.57 & 90.05 & 95.71 & \textbf{80.81} \\ 
MAPE & 0.03840\% & 0.02711\% & 0.02901\% & \textbf{0.02500\%} \\ 
Accuracy & 96.16\% & 97.27\% & 97.10\% & \textbf{97.50\%} \\ 
\hline
\end{tabular}
\end{table}

\subsection{Results of One-Day-Ahead Price Prediction}
In this section, we continue to use current-day sentiment data for both training and validation for the FinBERT-integrated models. Therefore, the training and validation losses for all models, both with and without sentiment integration, remain unchanged. The key difference here is that we incorporate the previous day's sentiment data during testing to evaluate the effect of lagged sentiment on prediction accuracy for the FinBERT-integrated models.
\subsubsection{Results of Bitcoin One-Day-Ahead Prediction}

First, we graphically analyze the performance of all the models over test data.
\begin{figure}[h]
    \centering
    \includegraphics[width=0.7\textwidth]{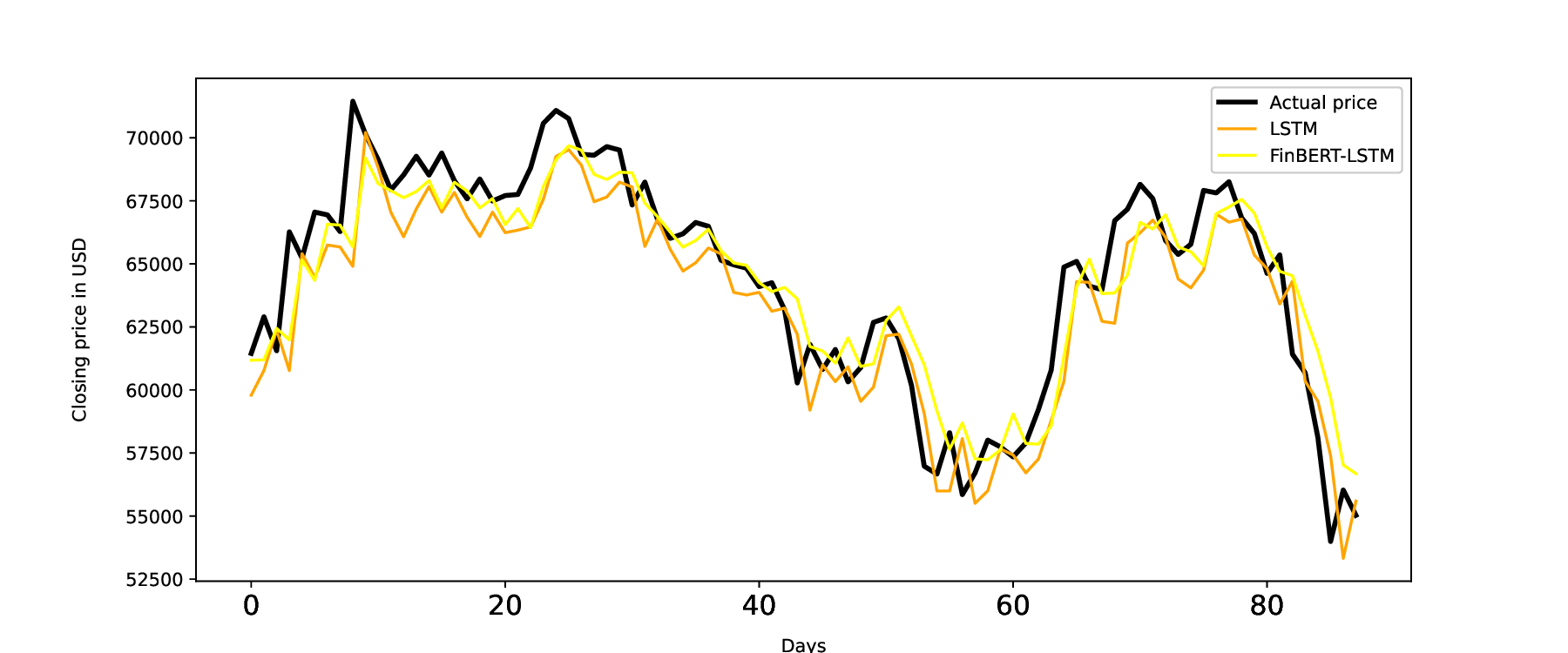}
    \caption{BTC-Actual vs One-Day-Ahead Predicted Price Using LSTM Model and FinBERT-LSTM Model}
    \label{fig10}
\end{figure}
\begin{figure}[h]
    \centering
    \includegraphics[width=0.7\textwidth]{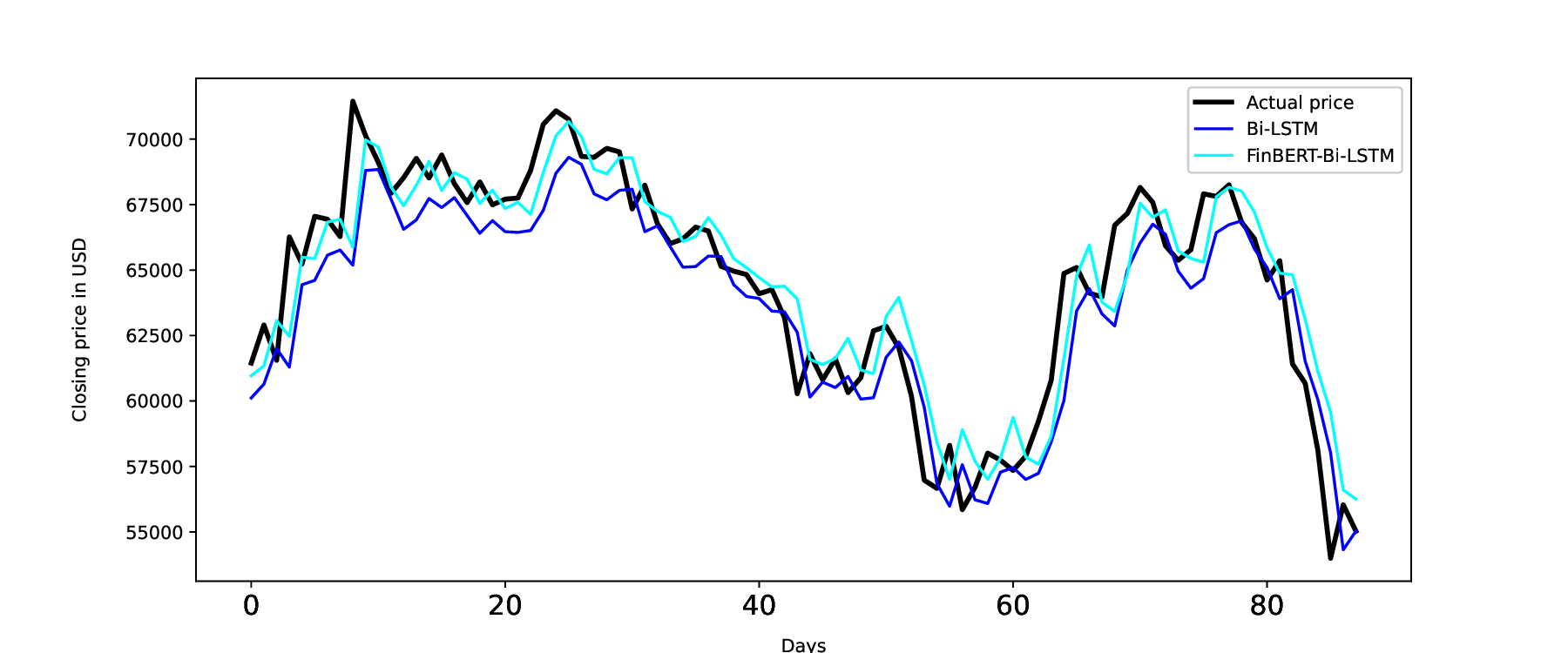}
    \caption{BTC-Actual vs One-Day-Ahead Predicted Price Using Bi-LSTM Model and FinBERT-Bi-LSTM Model}
    \label{fig11}
\end{figure}

In Fig. \ref{fig10}, the actual price is shown in black, with LSTM predictions in orange and FinBERT-LSTM predictions now in yellow. Similarly, in Fig. \ref{fig11}, the actual price is depicted in black, with Bi-LSTM predictions in blue and FinBERT-Bi-LSTM predictions now in sky blue. Notice that the LSTM and Bi-LSTM predictions remain unchanged, as reflected by the same colors used for the intra-day price prediction. We now move on to the quantitative comparison of the models.

\begin{table}[htbp]
\centering
\caption{BTC-Performance Comparison of LSTM, FinBERT-LSTM, Bi-LSTM, and FinBERT-Bi-LSTM Models for One-Day-Ahead Price Prediction}
\label{Table3}
\begin{tabular}{ |c|c|c|c|c| } 
\hline
\textbf{Metric} & \textbf{LSTM} & \textbf{FinBERT-LSTM} & \textbf{Bi-LSTM} & \textbf{FinBERT-Bi-LSTM} \\ 
\hline
MAE & 1453.36 & 1274.51 & 1419.26 & \textbf{1223.49} \\ 
MAPE & 0.02317\% & 0.02009\% & 0.02250\% & \textbf{0.01928\%} \\ 
Accuracy & 97.68\% & 97.99\% & 97.75\% & \textbf{98.07\%} \\ 
\hline
\end{tabular}
\end{table}

Table \ref{Table3} shows that both FinBERT-integrated models outperform their plain counterparts, even when using the previous day’s sentiment data for testing, with the FinBERT-Bi-LSTM model achieving the highest accuracy of 98.07\% for one-day-ahead predictions aligning with our hypothesis mentioned in Section~\ref{sec4}.

To further explore, instead of using current-day sentiment data for training and validation, we used the previous day’s sentiment also for both training and validation, along with the test data, in the FinBERT-integrated models. The corresponding results are presented in Table \ref{Table4}.

\begin{table}[htbp]
\centering
\caption{BTC-Performance Comparison of LSTM, FinBERT-LSTM, Bi-LSTM, and FinBERT-Bi-LSTM Models Using Previous Day Sentiment in All Phases for One-Day-Ahead Price Prediction}
\label{Table4}
\begin{tabular}{ |c|c|c|c|c| } 
\hline
\textbf{Metric} & \textbf{LSTM} & \textbf{FinBERT-LSTM} & \textbf{Bi-LSTM} & \textbf{FinBERT-Bi-LSTM} \\ 
\hline
MAE & 1453.36 & 1450.66 & 1419.26 & 1532.75 \\ 
MAPE & 0.02317\% & 0.02280\% & 0.02250\% & 0.02415\% \\ 
Accuracy & 97.68\% & 97.72\% & 97.75\% & 97.58\% \\ 
\hline
\end{tabular}
\end{table}

As shown in Table \ref{Table4}, the performance of the FinBERT-integrated models declines when using previous-day sentiment for training, validation, and testing, compared to the earlier case in Table \ref{Table3}.
 Interestingly, the FinBERT-Bi-LSTM model even performs worse than the
 plain Bi-LSTM. This performance decline occurs because the models better capture the relationship between market sentiment and the current day’s closing price when trained on current-day sentiment. However, for one-day-ahead predictions, the previous day’s sentiment is the last available data for testing, as current-day sentiment would not yet be accessible in real-world scenarios. Therefore, the better approach for one-day-ahead predictions is to train and validate the models using current-day sentiment, while employing previous-day sentiment during testing.

\subsubsection{Results of Ethereum One-Day-Ahead Prediction}

Following the same approach as for Bitcoin, we begin by presenting the graphical comparison of all the models over the test data.

\begin{figure}[h]
    \centering
    \includegraphics[width=0.7\textwidth]{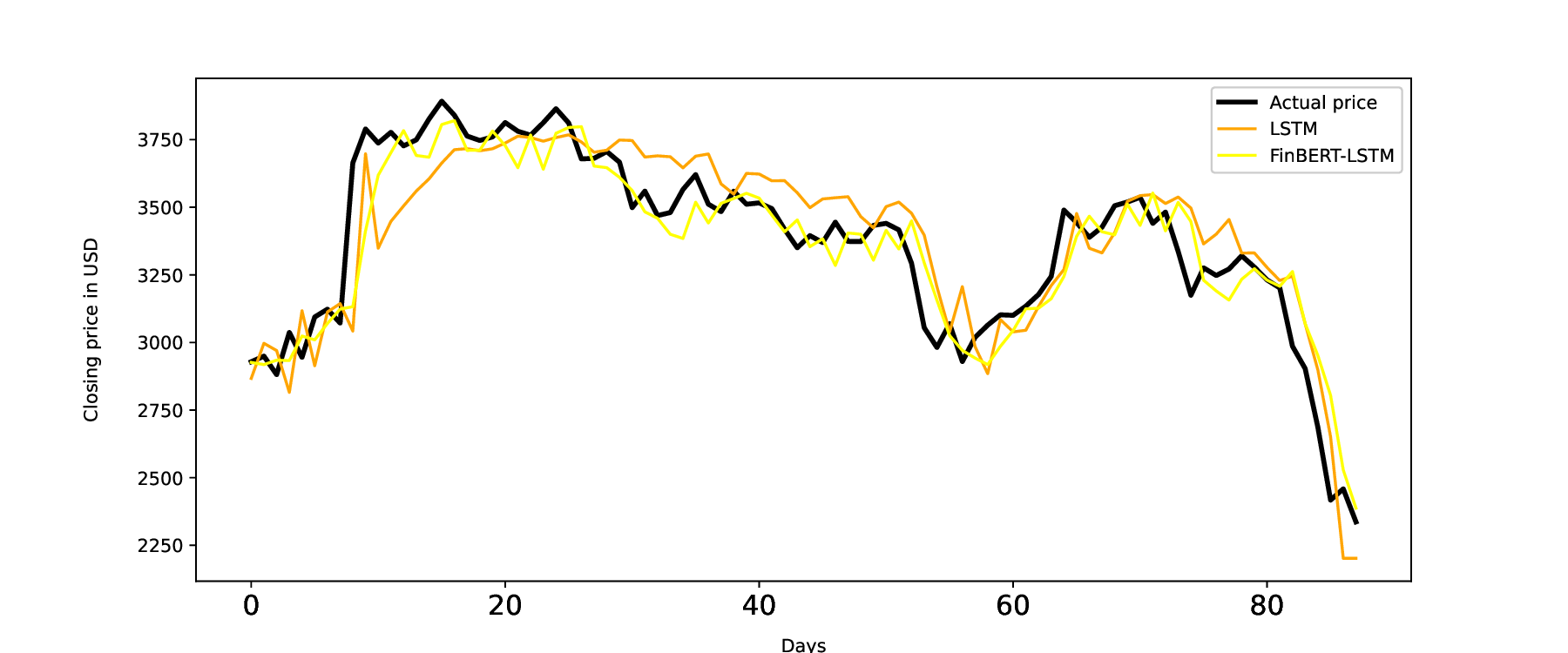}
    \caption{ETH-Actual vs One-Day-Ahead Predicted Price Using LSTM Model and FinBERT-LSTM Model}
    \label{fig:32}
\end{figure}

\begin{figure}[h]
    \centering
    \includegraphics[width=0.7\textwidth]{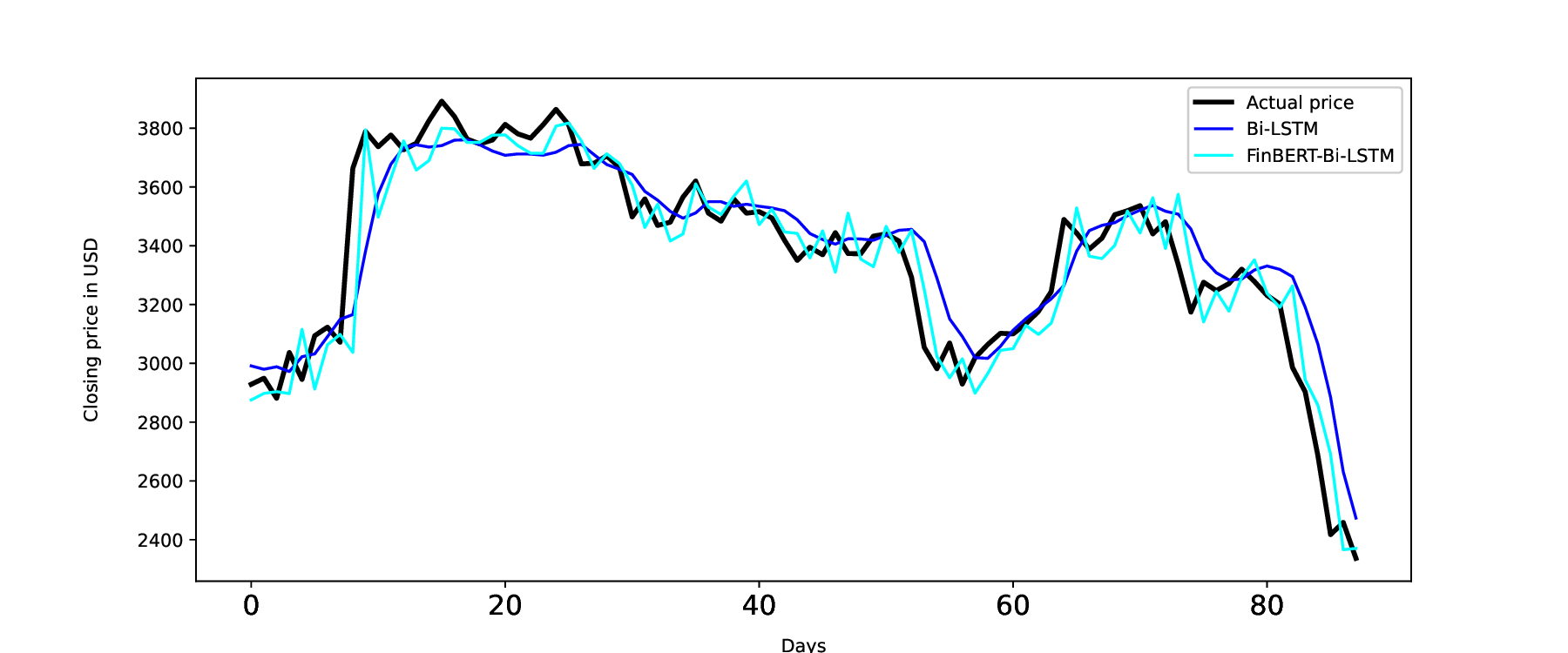}
    \caption{ETH-Actual vs One-Day-Ahead Predicted Price Using Bi-LSTM Model and FinBERT-Bi-LSTM Model}
    \label{fig:33}
\end{figure}

In Fig. \ref{fig:32}, the actual price is shown in black, with LSTM predictions remaining in orange and FinBERT-LSTM predictions now in yellow. Similarly, in Fig. \ref{fig:33}, the actual price is depicted in black, with Bi-LSTM predictions in blue and FinBERT-Bi-LSTM predictions now in sky blue.


\vspace{0.01cm} 
\begin{table}[htbp]
\centering
\caption{ETH-Performance Comparison of LSTM, FinBERT-LSTM, Bi-LSTM, and FinBERT-Bi-LSTM Models for One-Day-Ahead Price Prediction}
\label{Table5}
\begin{tabular}{ |c|c|c|c|c| } 
\hline
\textbf{Metric} & \textbf{LSTM} & \textbf{FinBERT-LSTM} & \textbf{Bi-LSTM} & \textbf{FinBERT-Bi-LSTM} \\ 
\hline
MAE & 126.57 & 92.09 & 95.71 & \textbf{85.66} \\ 
MAPE & 0.03840\% & 0.02808\% & 0.02901\% & \textbf{0.02639\%} \\ 
Accuracy & 96.16\% & 97.19\% & 97.10\% & \textbf{97.36\%} \\ 
\hline
\end{tabular}
\end{table}

A quantitative comparison of all models for one-day-ahead price prediction is presented in Table \ref{Table5}, where the FinBERT-integrated models outperform their plain counterparts, even when using previous-day sentiment during testing, demonstrating the effectiveness of sentiment integration. Notably, the FinBERT-Bi-LSTM model achieves the highest accuracy, also in Ethereum, at 97.36\% for one-day-ahead predictions. Next, we demonstrate the same phenomenon of using the previous day's sentiment data across all phases: training, validation, and testing, as shown in Table \ref{Table6}.

\vspace{0.01cm} 
\begin{table}[htbp]
\centering
\caption{ETH-Performance Comparison of LSTM, FinBERT-LSTM, Bi-LSTM, and FinBERT-Bi-LSTM Models Using Previous Day Sentiment in All Phases for One-Day-Ahead Price Prediction}
\label{Table6}
\begin{tabular}{ |c|c|c|c|c| } 
\hline
\textbf{Metric} & \textbf{LSTM} & \textbf{FinBERT-LSTM} & \textbf{Bi-LSTM} & \textbf{FinBERT-Bi-LSTM} \\ 
\hline
MAE & 126.57 & 104.13 & 95.71 & 135.28 \\ 
MAPE & 0.03840\% & 0.03174\% & 0.02901\% & 0.04098\% \\ 
Accuracy & 96.16\% & 96.83\% & 97.10\% & 95.90\% \\ 
\hline
\end{tabular}
\end{table}

From Table \ref{Table6}, we observe that using previous-day sentiment across training, validation, and testing results in worse performance for the FinBERT-integrated models compared to their plain counterparts, with the FinBERT-Bi-LSTM model performing particularly poorly. This follows the same trend as in the BTC case, where using current-day sentiment for training and validation while applying previous-day sentiment for one-day-ahead testing, provides better predictive accuracy for the FinBERT-integrated models.

\subsection{Results of One-Day-Ahead Trading Strategy}
\label{sec:one_day_ahead_trading}
To assess the potential profitability of our one-day-ahead predictions, we implemented a straightforward trading strategy. As outlined earlier, the logic was simple: buy when the predicted closing price for the next day exceeds the previous day's actual closing price by a specified threshold, sell when it falls below a certain threshold, and hold if the price change stays within these thresholds.

\subsubsection{Results of Bitcoin One-Day-Ahead Trading Strategy}

\textbf{Profitability Assessment with Perfect Price Predictions:} We chose an initial capital of \$100,000 for BTC and examined price trends across the data prior to the testing set. Over this period, prices increased by an average of 0.0029 and the standard deviation was 0.0254. The buy threshold was set at a level of 0.030, which is close to one standard deviation above the average, indicating that we only enter the market when there is a significant price rise, reflecting our conservative approach to buying. On the other hand, the sell threshold was established at -0.010, approximately half a standard deviation below the average, allowing for quicker sales in the event of a price drop. This strategy ensures that we capitalize on substantial upward movements while minimizing losses by acting swiftly when prices decline.

To validate the strategy, we assumed perfect price prediction (i.e., knowing the actual prices in advance) and applied it to the data prior to the testing set, as well as to the testing dataset. With this assumption, the strategy yielded a significant profit of \$2,908,232,886.22 over 497 days of data prior to the testing set, while the testing phase generated a profit of \$41,233.62 over 88 days. A substantial profit from the data prior to the testing phase is expected, as the thresholds were optimized based on this information. Although the thresholds become less optimal in the testing phase due to market volatility, the straightforward strategy still manages to deliver a reasonable profit despite the challenging conditions.

Now we apply the same strategy over the test data using one-day-ahead price predictions generated by all of our models. The LSTM and Bi-LSTM models made no trades during the test period. In contrast, both the FinBERT-LSTM and FinBERT-Bi-LSTM models executed trades based on their predicted prices.
Below, we present the plots showing the buy and sell signals generated by both the FinBERT-LSTM and FinBERT-Bi-LSTM models.

\begin{figure}[H]
    \centering
    \includegraphics[width=0.7\textwidth]{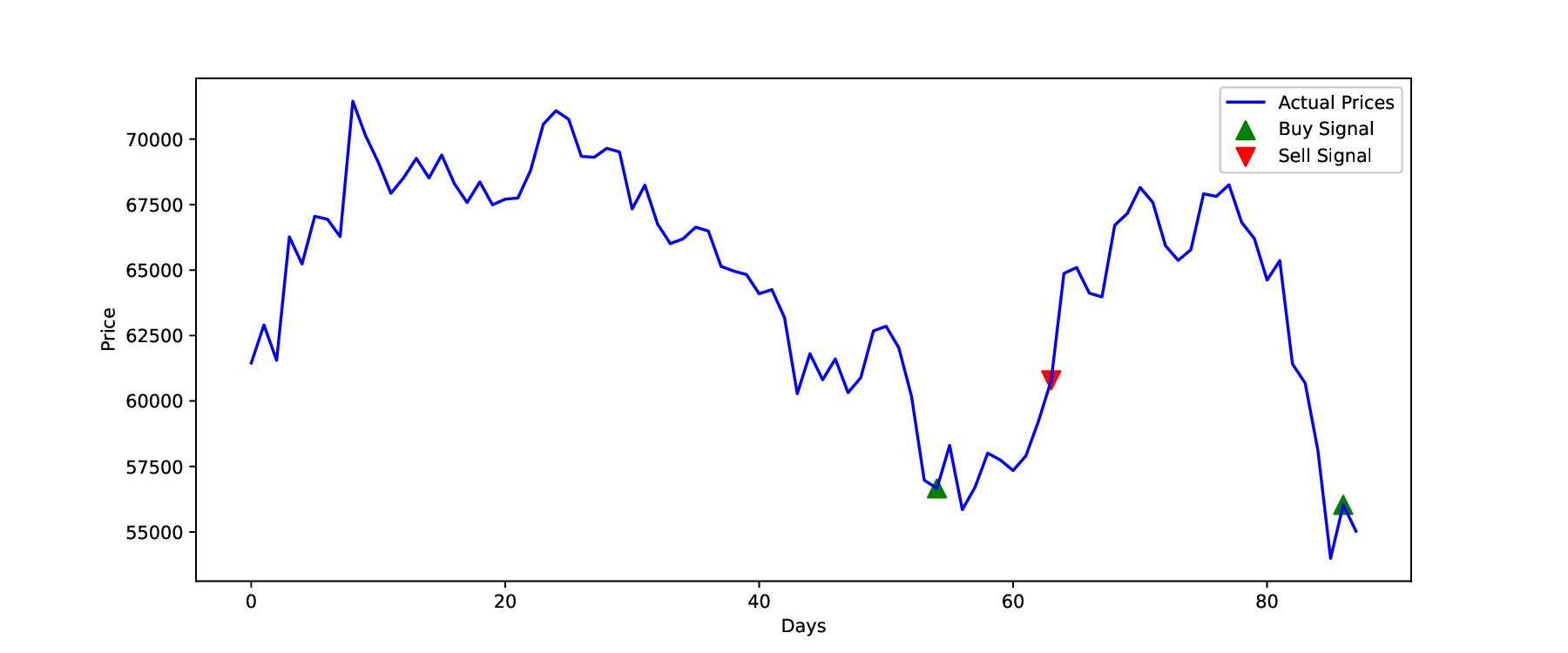}
    \caption{BTC-Trading Signals from FinBERT-LSTM Predictions}
    \label{fig:12}
\end{figure}
\begin{figure}[H]
    \centering
    \includegraphics[width=0.7\textwidth]{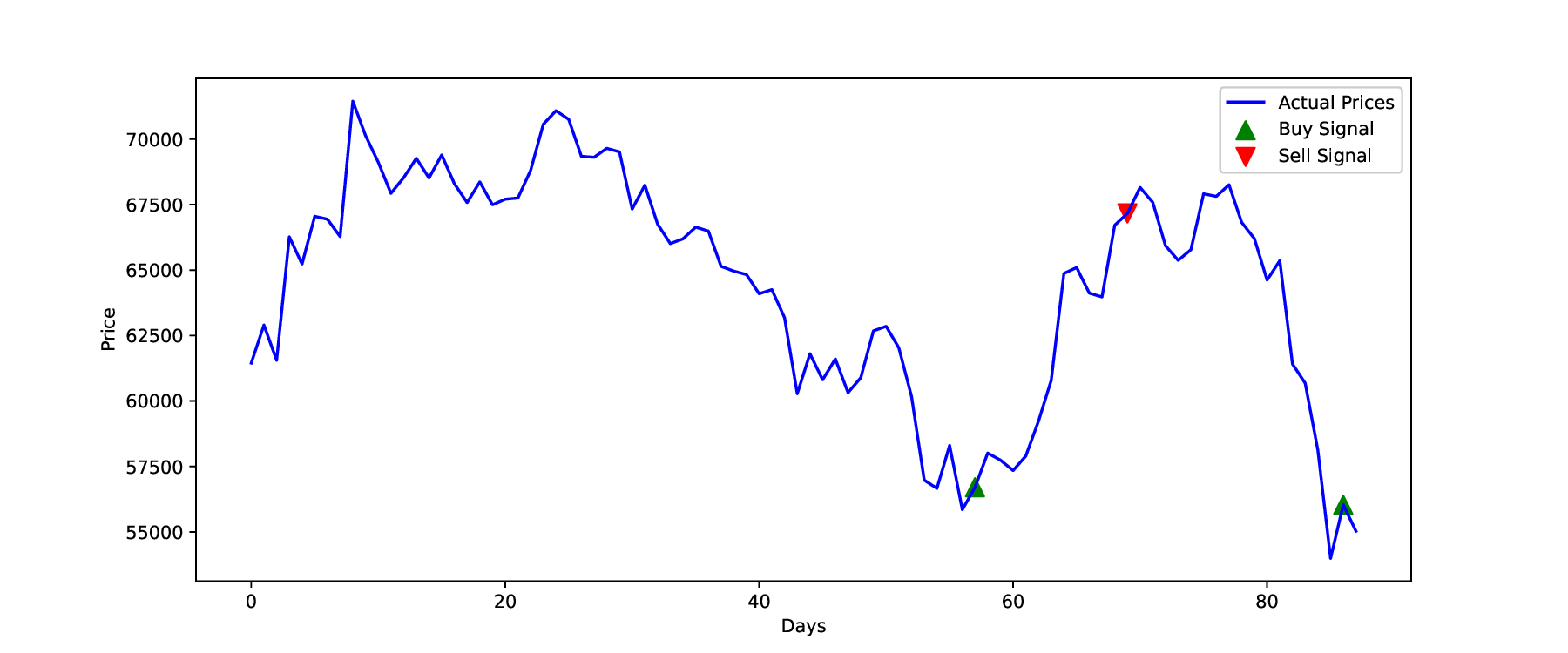}
    \caption{BTC-Trading Signals from FinBERT-Bi-LSTM Predictions}
    \label{fig:13}
\end{figure}
In Fig. \ref{fig:12} and Fig. \ref{fig:13}, green signals indicate buy actions and red signals indicate sell actions. In Fig. \ref{fig:12}, the FinBERT-LSTM model achieved a profit of \$5,807.56 through one-day-ahead predictions, while in Fig. \ref{fig:13}, the FinBERT-Bi-LSTM model generated a profit of \$21,579.95, which is more than half of what could have been achieved with perfect price prediction over the test data.

\subsubsection{Results of Ethereum One-Day-Ahead Trading Strategy}
\label{sec:one-day-ahead-eth}

\textbf{Profitability Assessment with Perfect Price Predictions:} For ETH, we again started with an initial capital of \$100,000 and analyzed price trends across the data prior to the testing set. Over this period, prices increased by an average of 0.0021 and the standard deviation was 0.0269. The buy threshold for Ethereum (ETH) was set at a level of 0.025, which is close to 0.85 standard deviations above the average, indicating that we are less conservative in our buying approach compared to Bitcoin (BTC). Similarly, the sell threshold was established at -0.015, approximately 0.65 standard deviations below the average, reflecting a less aggressive stance on selling. Despite these adjustments, our underlying motive remains the same: to capitalize on significant upward movements while 
minimizing losses by acting promptly when prices decline. Applying this strategy with perfect price predictions resulted in a profit of \$20,060,355,418.60 over 497 days of data prior to the testing set, and \$68,451.78 over 88 days of testing, thereby validating our straightforward trading strategy. The significant profit during the data prior to the testing set and the lower, yet still substantial, profit in testing is consistent with the findings from the Bitcoin (BTC) case.

Now using the same strategy, we simulated trading based on the one-day-ahead price predictions generated by all the models.

\begin{figure}[h]
    \centering
    \includegraphics[width=0.7\textwidth]{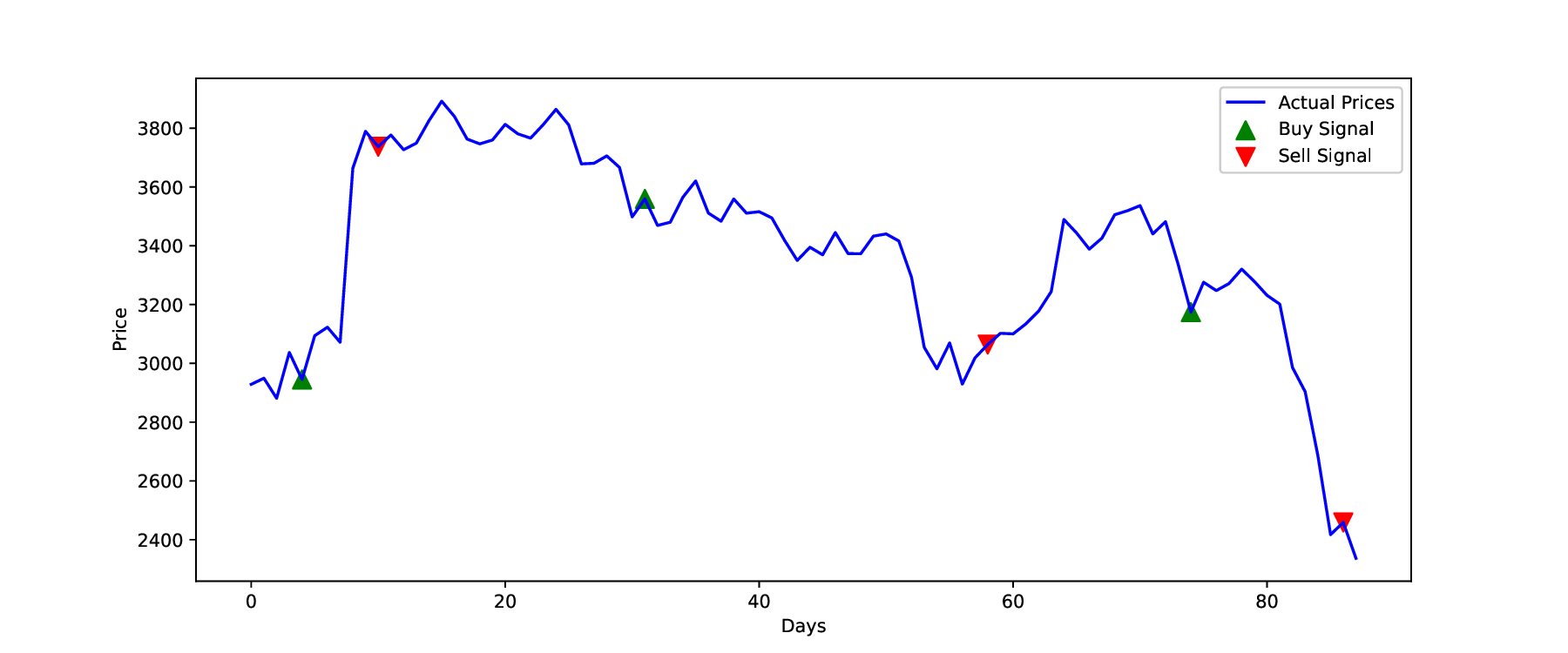}
    \caption{ETH-Trading Signals from LSTM Predictions}
    \label{fig:34}
\end{figure}
\begin{figure}[h]
    \centering
    \includegraphics[width=0.7\textwidth]{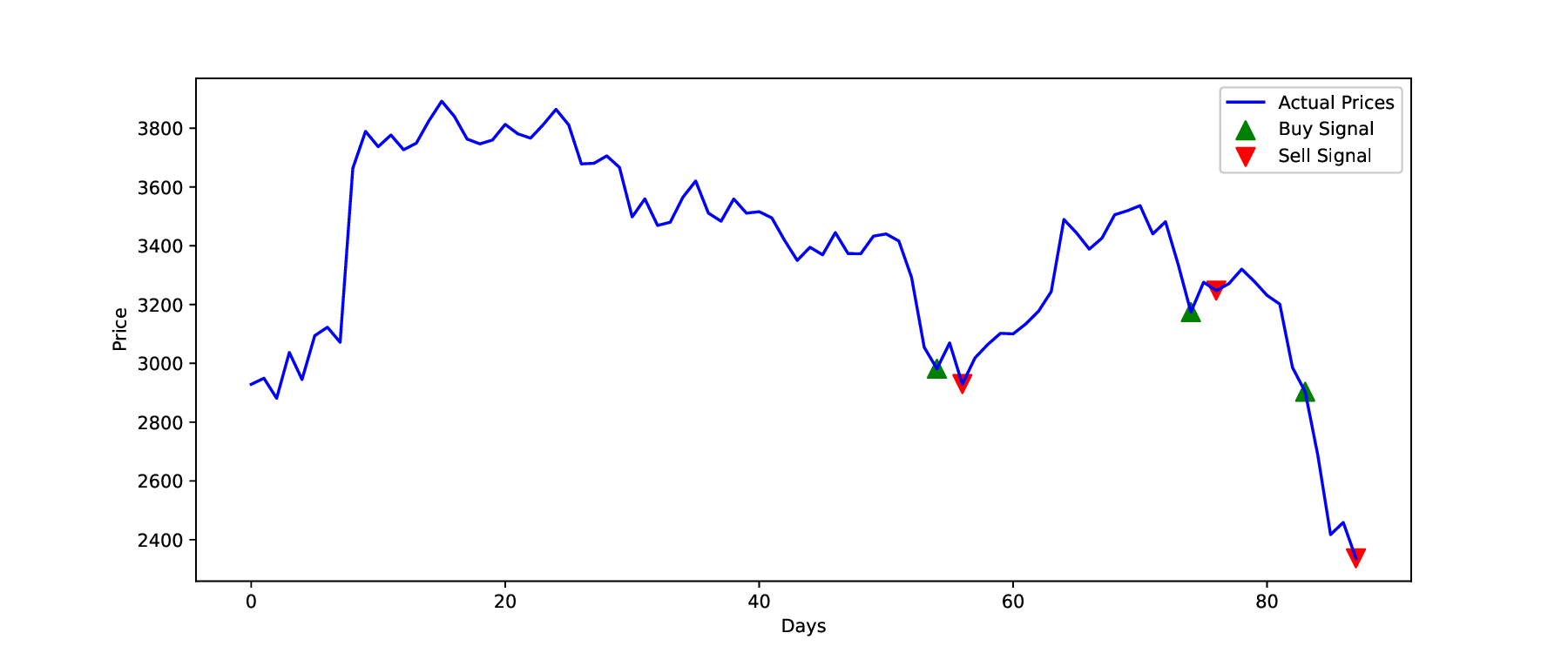}
    \caption{ETH-Trading Signals from FinBERT-LSTM Predictions}
    \label{fig:35}
\end{figure}

\begin{figure}[H]
    \centering
    \includegraphics[width=0.7\textwidth]{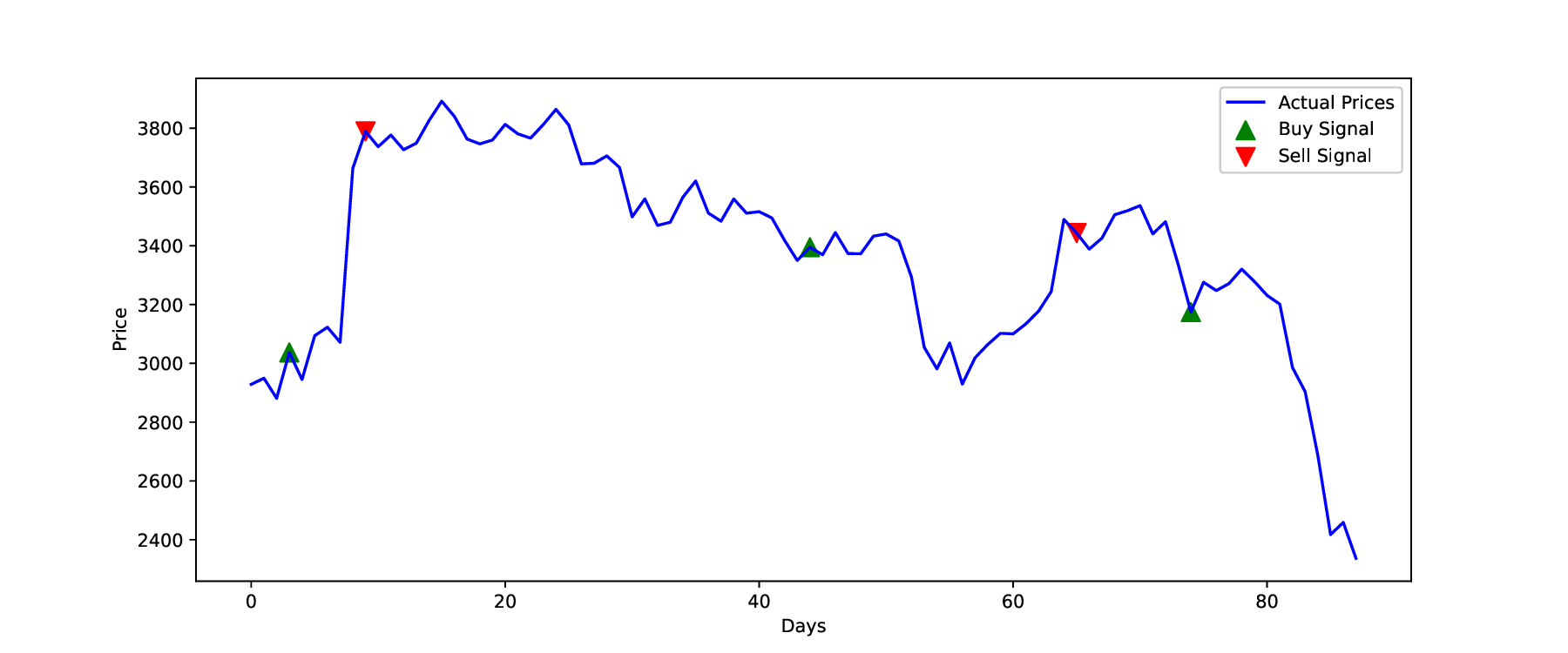}
    \caption{ETH-Trading Signals from Bi-LSTM Predictions}
    \label{fig:36}
\end{figure}
\begin{figure}[H]
    \centering
    \includegraphics[width=0.7\textwidth]{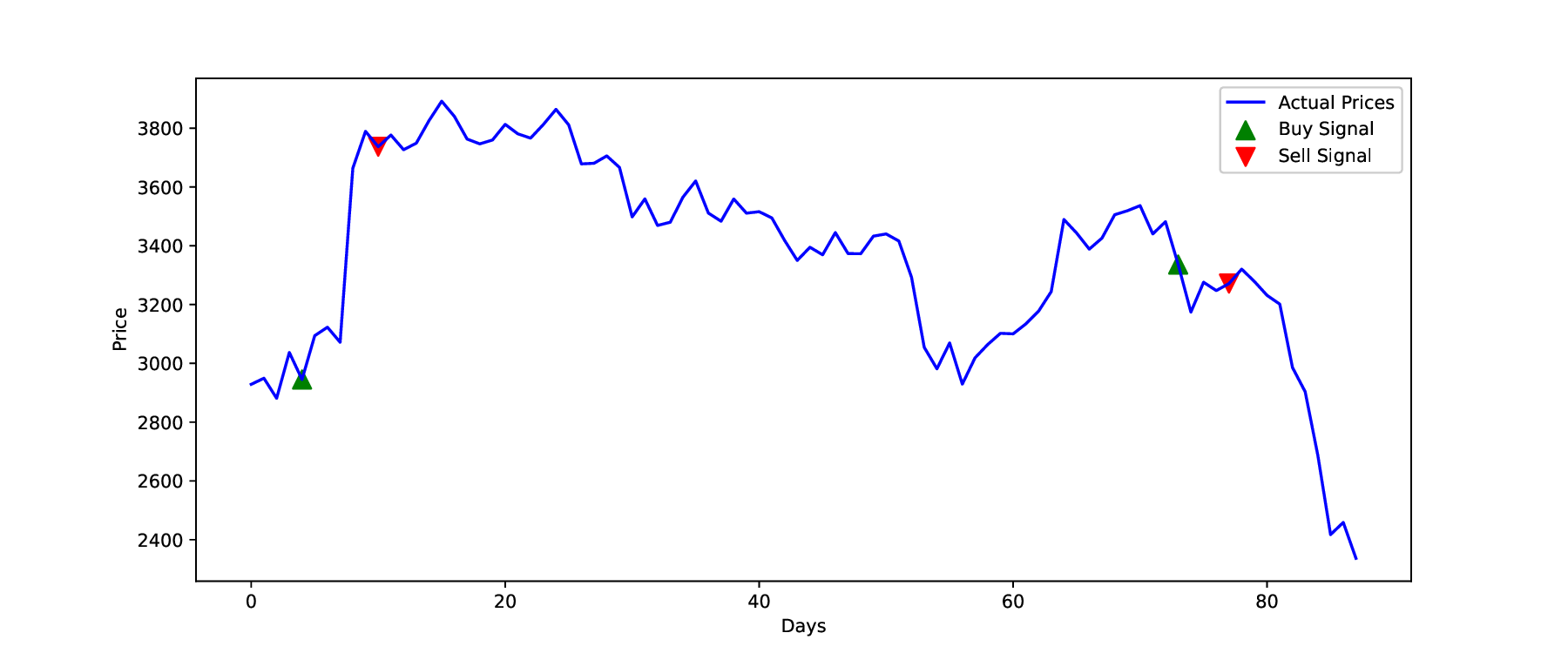}
    \caption{ETH-Trading Signals from FinBERT-Bi-LSTM Predictions}
    \label{fig:37}
\end{figure}

In Fig. \ref{fig:34}, Fig. \ref{fig:35}, Fig. \ref{fig:36}, and Fig. \ref{fig:37}, green signals represent buy actions, and red signals indicate sell actions for all the models: LSTM, FinBERT-LSTM, Bi-LSTM, and FinBERT-Bi-LSTM. In Fig. \ref{fig:34}, Fig. \ref{fig:35}, and Fig. \ref{fig:36}, we encountered losses for the LSTM, FinBERT-LSTM, and Bi-LSTM models, with the LSTM model showing a loss of \$22,003.73. The FinBERT-LSTM model exhibited a loss of \$20,207.27, and the Bi-LSTM model demonstrated a loss of \$7,241.03. Finally, in Fig. \ref{fig:37}, the FinBERT-Bi-LSTM model achieved a profit of \$16,365.85.

\subsection{Results of Future 30-Day Price Prediction using MDT
}

In the Maximum Data Training(MDT) approach for future prediction, the models were trained on data available until the start of the testing period and then used to make predictions for the following 30 days. The FinBERT-integrated models incorporated sentiment data during training, while the non-integrated models relied solely on price data for their predictions.

\label{sec:30day1BTC}
\subsubsection{Results of Bitcoin Future 30-Day Price Prediction using MDT
}

\begin{figure}[h]
    \centering
    \includegraphics[width=0.68\textwidth]{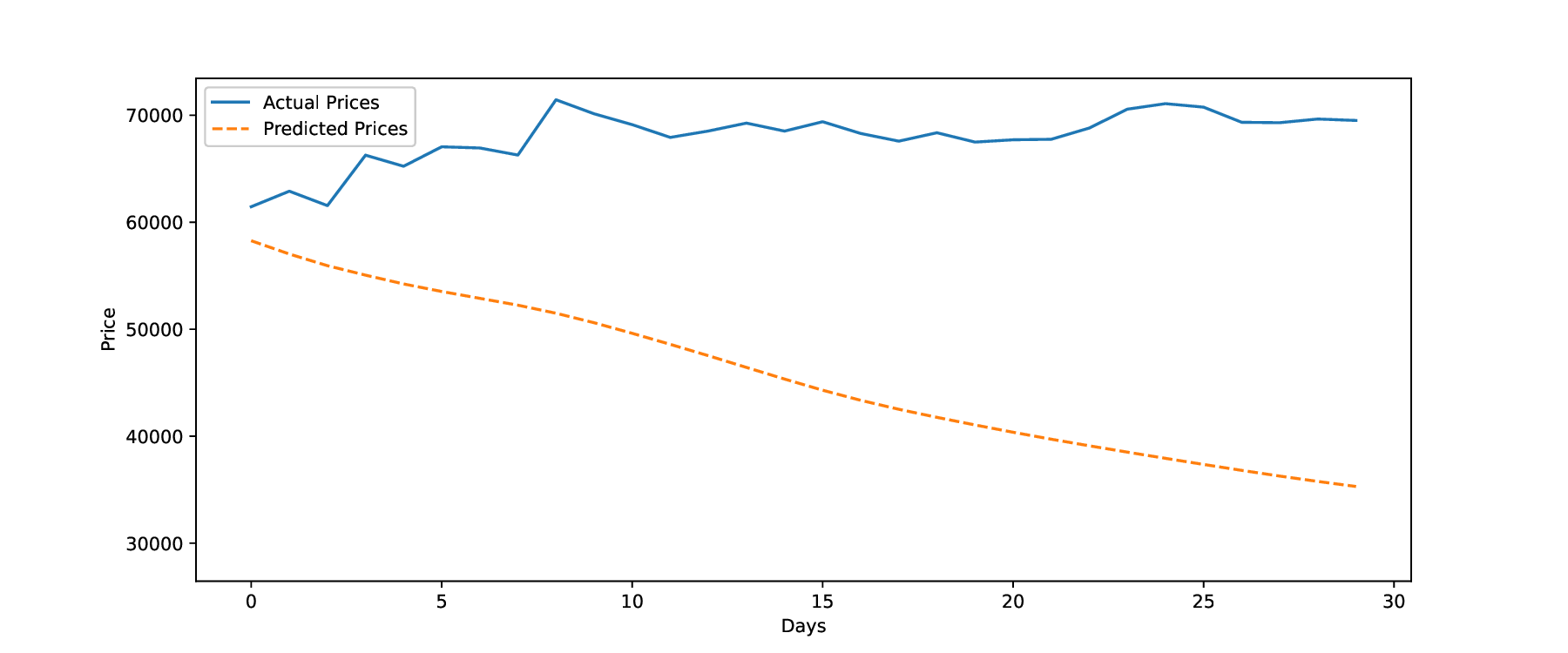}
    \caption{BTC-Actual 30-Day Prices vs Future 30-Day Price Prediction by LSTM using MDT}
    \label{fig:14}
\end{figure}

\begin{figure}[h]
    \centering
    \includegraphics[width=0.68\textwidth]{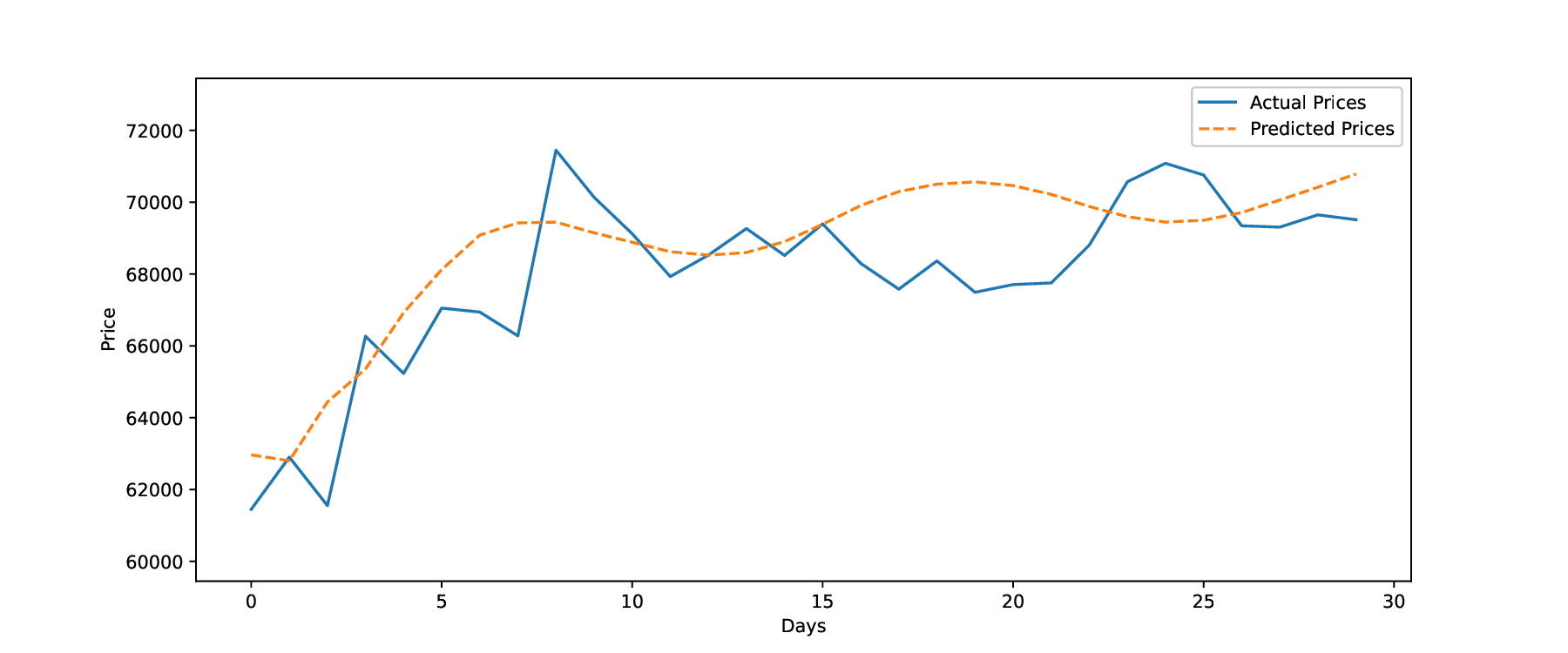}
    \caption{BTC-Actual 30-Day Prices vs Future 30-Day Price Prediction by FinBERT-LSTM using MDT}
    \label{fig:15}
\end{figure}

\begin{figure}[H]
    \centering
    \includegraphics[width=0.68\textwidth]{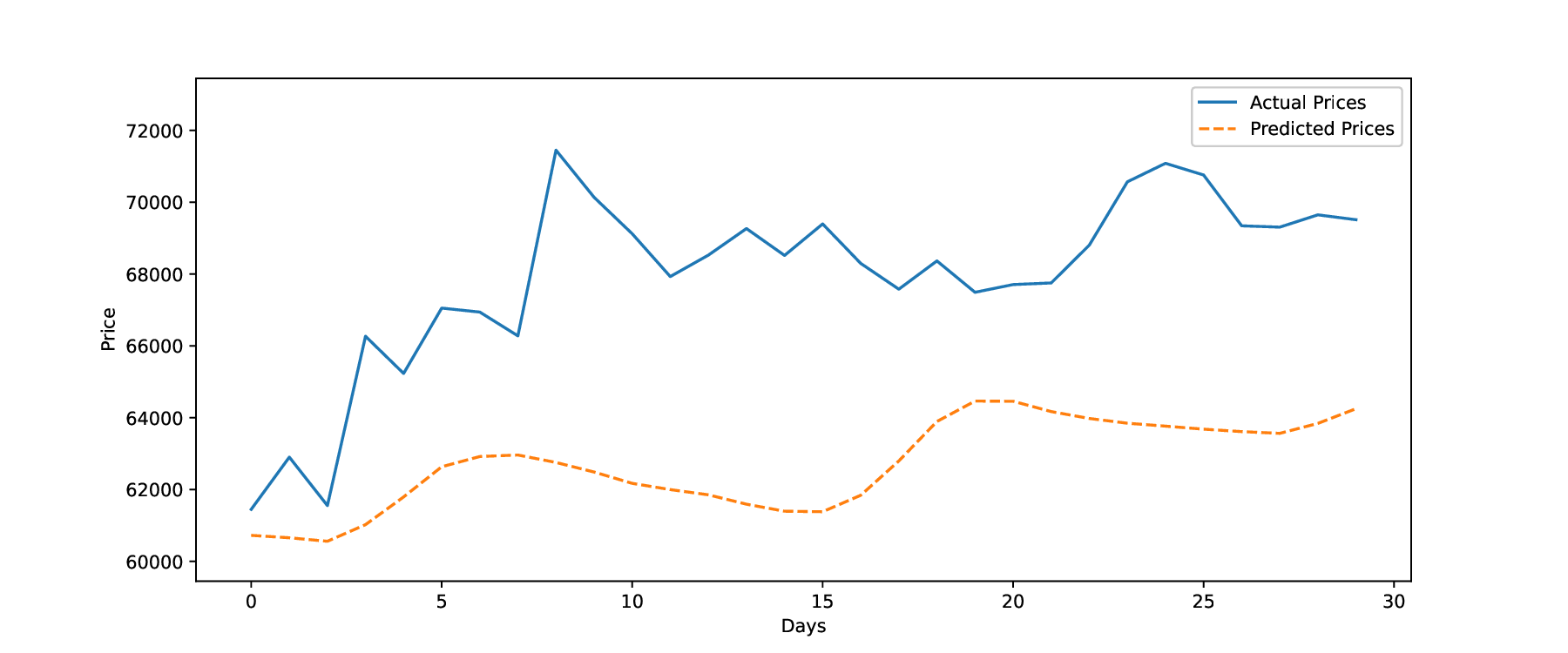}
    \caption{BTC-Actual 30-Day Prices vs Future 30-Day Price Prediction by Bi-LSTM using MDT}
    \label{fig:16}
\end{figure}

\begin{figure}[H]
    \centering
    \includegraphics[width=0.68\textwidth]{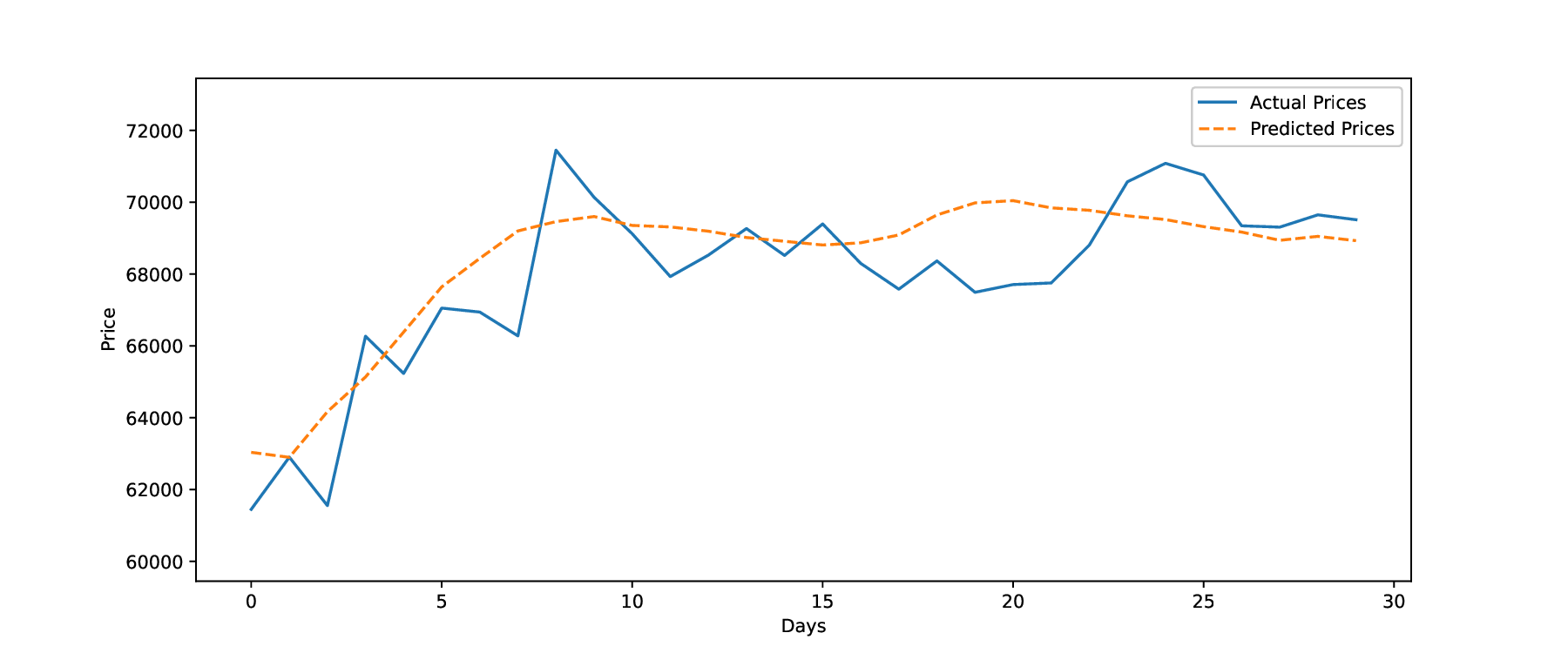}
    \caption{BTC-Actual Prices and 30-Day Future Price Predictions by FinBERT-Bi-LSTM using MDT}
    \label{fig:17}
\end{figure}

We first present the graphs comparing the predicted prices for 30 days with the actual prices. In Fig. \ref{fig:14}, \ref{fig:15}, \ref{fig:16}, and \ref{fig:17}, the blue line represents the actual future 30-day price, while the orange dots give the price predictions from the LSTM, FinBERT-LSTM, Bi-LSTM, and FinBERT-Bi-LSTM models, respectively.


Table \ref{Table7} shows that the FinBERT-LSTM model significantly outperforms the LSTM model in 30-day future price predictions, achieving much lower error rates and a higher accuracy of 97.96\% compared to the LSTM's 67.53\%. Additionally, the Bi-LSTM model achieves a significantly higher accuracy than the LSTM, but the FinBERT-Bi-LSTM model further improves upon the plain Bi-LSTM, reaching the highest accuracy of 98.30\% for 30-day future predictions showing that our hypothesis mentioned in Section \ref{sec4} works for future multi-day forecasting in this approach.

\vspace{0.01cm} 
\begin{table}[htbp]
\centering
\caption{BTC-Performance Comparison of LSTM, FinBERT-LSTM, Bi-LSTM, and FinBERT-Bi-LSTM Models for 30-Day Future Prediction using MDT}
\label{Table7}
\begin{tabular}{ |c|c|c|c|c| } 
\hline
\textbf{Metric} & \textbf{LSTM} & \textbf{FinBERT-LSTM} & \textbf{Bi-LSTM} & \textbf{FinBERT-Bi-LSTM} \\ 
\hline
MAE & 22311.38 & 1377.01 & 2905.08 & \textbf{1148.52}\\ 
MAPE & 32.47\% & 0.02043\% & 4.22\% & \textbf{0.01705\%} \\ 
Accuracy & 67.53\% & 97.96\% & 95.78\% & \textbf{98.30\%} \\ 
\hline
\end{tabular}
\end{table}

\subsubsection{Results of Ethereum  Future 30-Day Price Prediction using MDT}

Following the same process as for BTC, we first present the graphs comparing the predicted prices for 30 days with the actual prices for ETH.

\begin{figure}[H]
    \centering
    \includegraphics[width=0.7\textwidth]{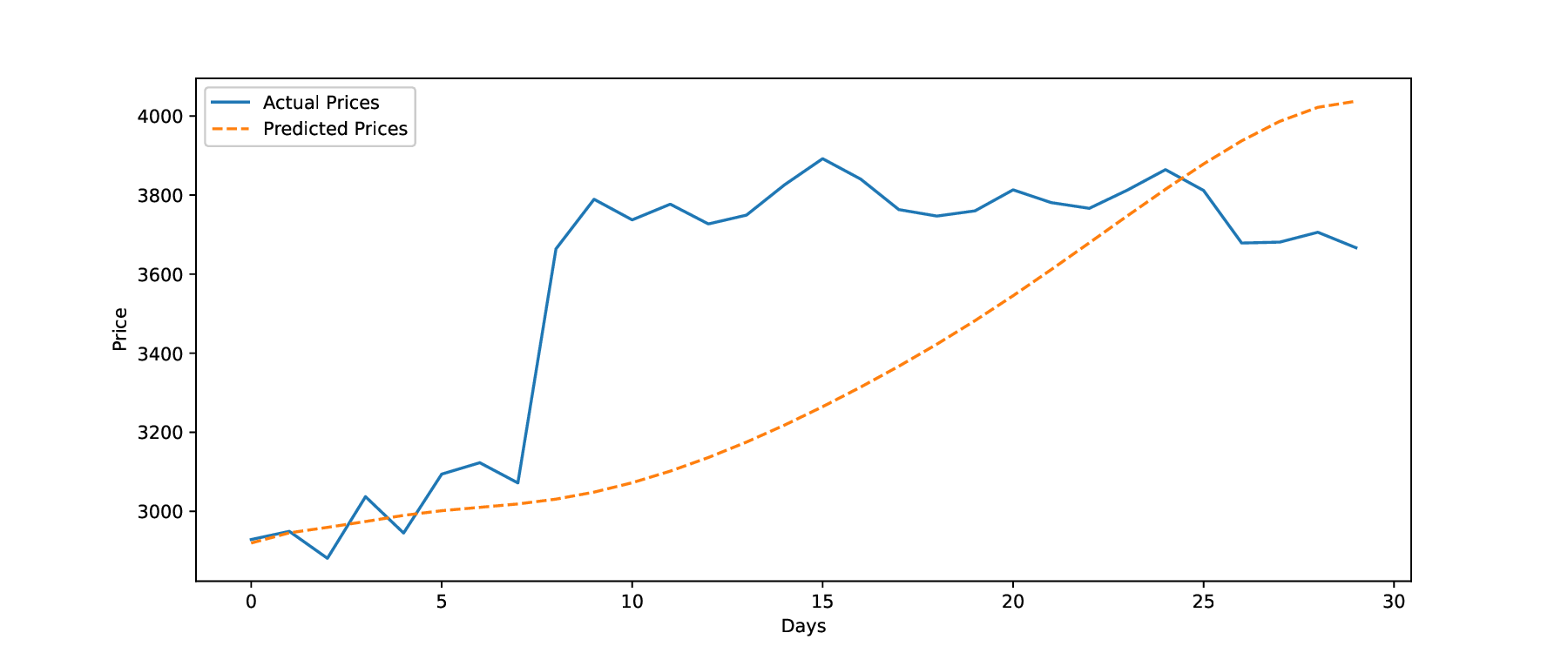}
    \caption{ETH-Actual 30-Day Prices vs Future 30-Day Price Prediction by LSTM using MDT}
    
    \label{fig:38}
\end{figure}

\begin{figure}[H]
    \centering
    \includegraphics[width=0.7\textwidth]{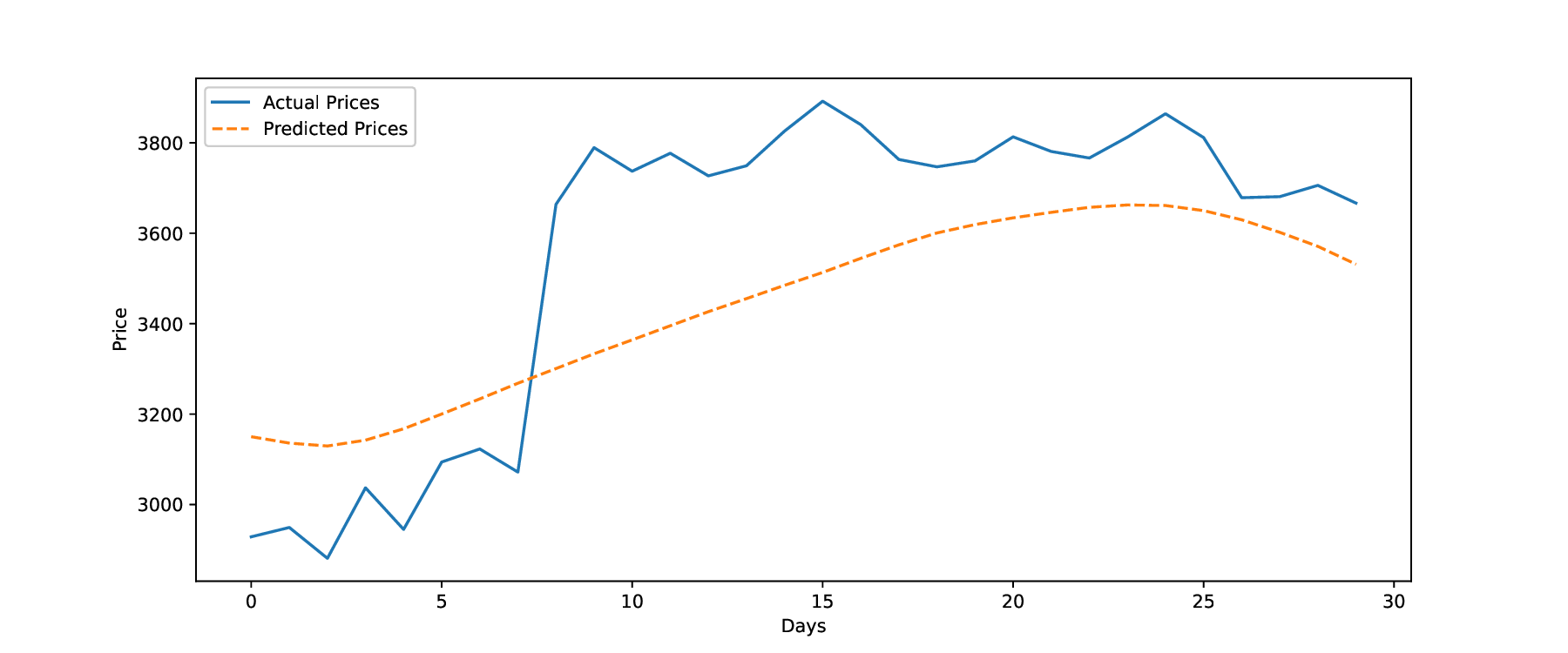}
    \caption{ETH-Actual 30-Day Prices vs Future 30-Day Price Prediction by FinBERT-LSTM using MDT}
    \label{fig:39}
\end{figure}

\begin{figure}[H]
    \centering
    \includegraphics[width=0.7\textwidth]{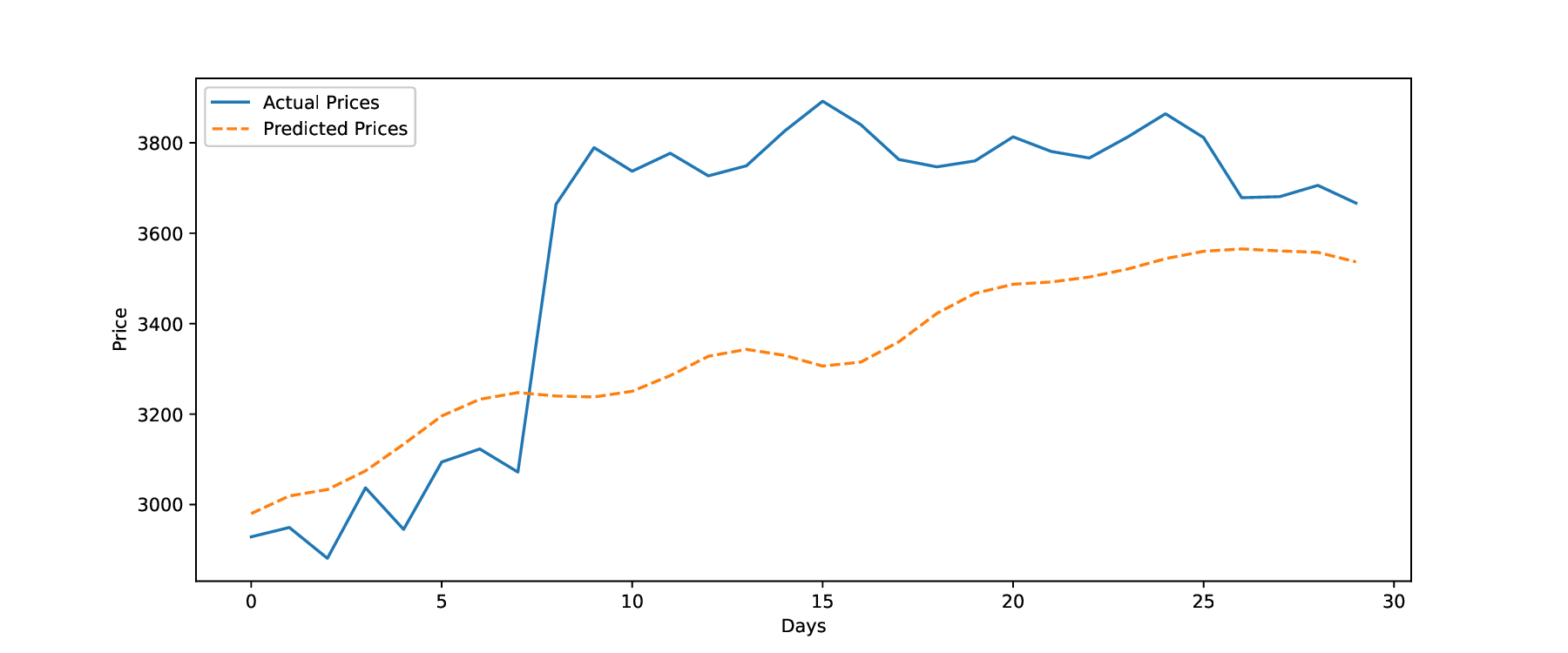}
    \caption{ETH-Actual 30-Day Prices vs Future 30-Day Price Prediction by Bi-LSTM using MDT}
    \label{fig:40}
\end{figure}
\begin{figure}[H]
    \centering
    \includegraphics[width=0.7\textwidth]{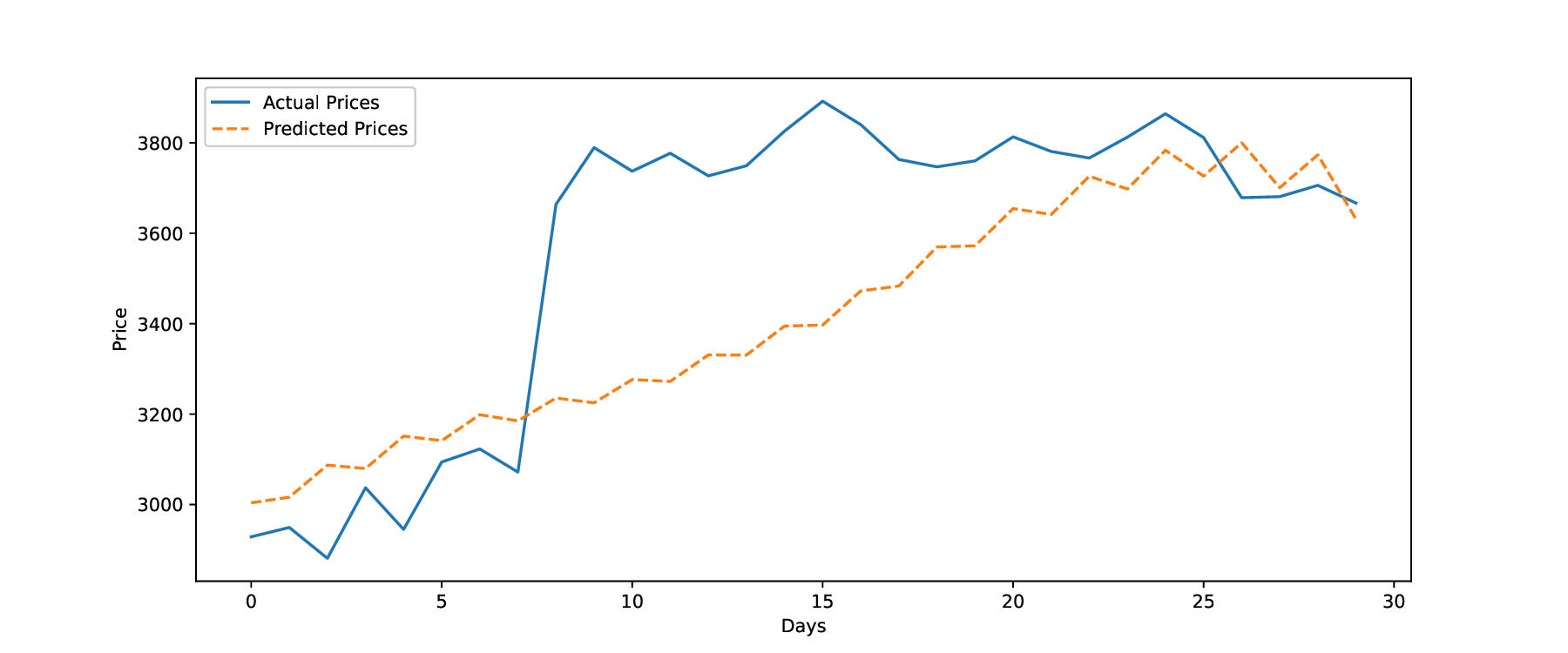}
    \caption{ETH-Actual 30-Day Prices vs Future 30-Day Price Prediction by FinBERT-Bi-LSTM using MDT}
    \label{fig:41}
\end{figure}

In Fig. \ref{fig:38}, Fig. \ref{fig:39}, Fig. \ref{fig:40}, and Fig. \ref{fig:41}, the actual prices are represented in blue, while the orange dots represent the predicted prices from the LSTM, FinBERT-LSTM, Bi-LSTM, and FinBERT-Bi-LSTM models, respectively. As shown in Fig. \ref{fig:38} to Fig. \ref{fig:41}, a significant price spike starting around the 7th day and continuing until just before the 10th day made accurate predictions challenging. However, from Table \ref{Table8}, we see that the FinBERT-LSTM model demonstrates better performance than the LSTM model in 30-day future predictions, achieving a higher accuracy of 94.03\% compared to 91.88\%, along with lower error metrics. Additionally, the FinBERT-Bi-LSTM model enhances prediction accuracy, slightly outperforming the FinBERT-LSTM with an accuracy of 94.14\%.

\vspace{0.01cm} 
\begin{table}[htbp]
\centering
\caption{ETH-Performance Comparison of LSTM, FinBERT-LSTM, Bi-LSTM, and FinBERT-Bi-LSTM Models for 30-Day Future Prediction using MDT}
\label{Table8}
\begin{tabular}{ |c|c|c|c|c| } 
\hline
\textbf{Metric} & \textbf{LSTM} & \textbf{FinBERT-LSTM} & \textbf{Bi-LSTM} & \textbf{FinBERT-Bi-LSTM} \\ 
\hline
MAE & 301.72 & 213.05 & 284.16 & \textbf{213.59} \\ 
MAPE & 8.12\% & 0.05973\% & 7.72\% & \textbf{0.05857\%} \\ 
Accuracy & 91.88\% & 94.03\% & 92.28\% & \textbf{94.14\%} \\ 
\hline
\end{tabular}
\end{table}

\subsection{Results of Future 30-Day Price Prediction using VET
}
\label{sec:30day2BTC}
In the Validation-Enhanced Training(VET) approach, a validation set was introduced between the training and testing data to improve model selection for future 30-day predictions over the test data, though it reduced the amount of training data available immediately before testing. A key difference between the non-integrated and FinBERT-integrated models is that the latter incorporates sentiment data during training, similar to the previous approach.

\subsubsection{Results of Bitcoin Future 30-Day Price Prediction using VET}

We first present the training loss and validation MAE for all the models.

\begin{figure}[H]
    \centering
    \begin{minipage}{0.44\textwidth}
        \centering
        \includegraphics[width=\textwidth]{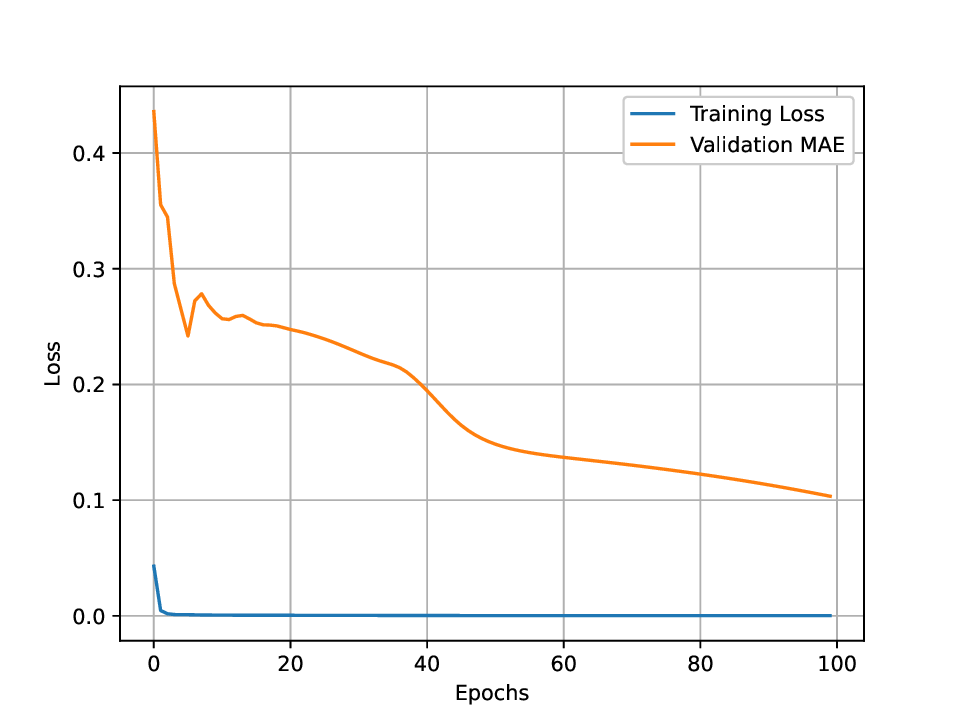}
        \caption{BTC-Training Loss and Validation MAE in Future Price Prediction for LSTM}
        \label{fig:18}
    \end{minipage}
    \hfill
    \begin{minipage}{0.44\textwidth}
        \centering
        \includegraphics[width=\textwidth]{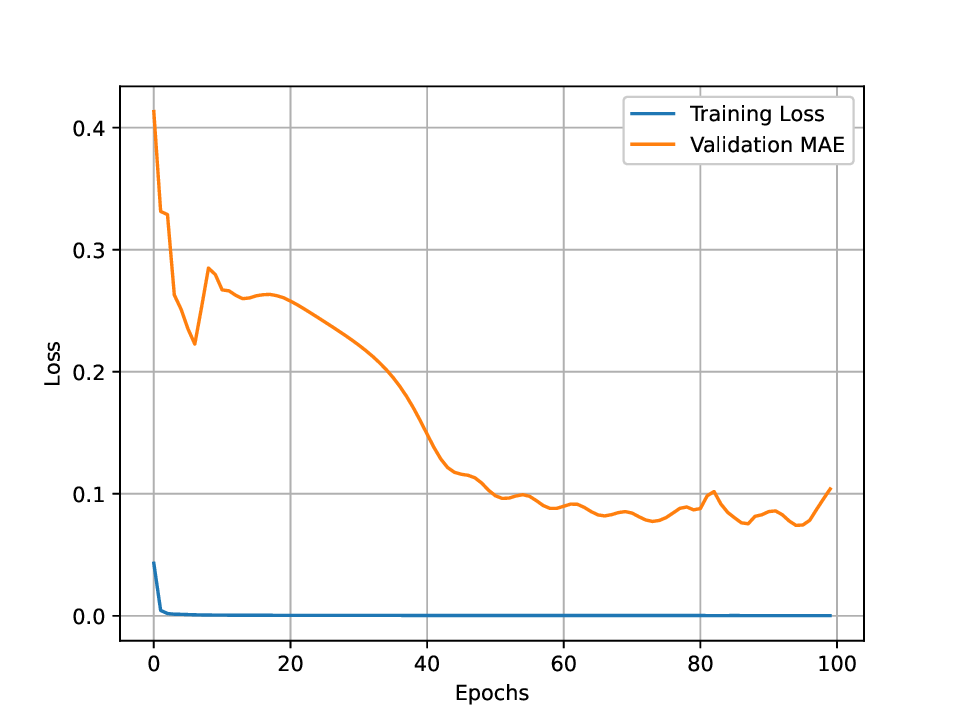}
        \caption{BTC-Training Loss and Validation MAE in Future Price Prediction for FinBERT-LSTM}
        \label{fig:19}
    \end{minipage}
\end{figure}

\begin{figure}[H]
    \centering
    \begin{minipage}{0.44\textwidth}
        \centering
        \includegraphics[width=\textwidth]{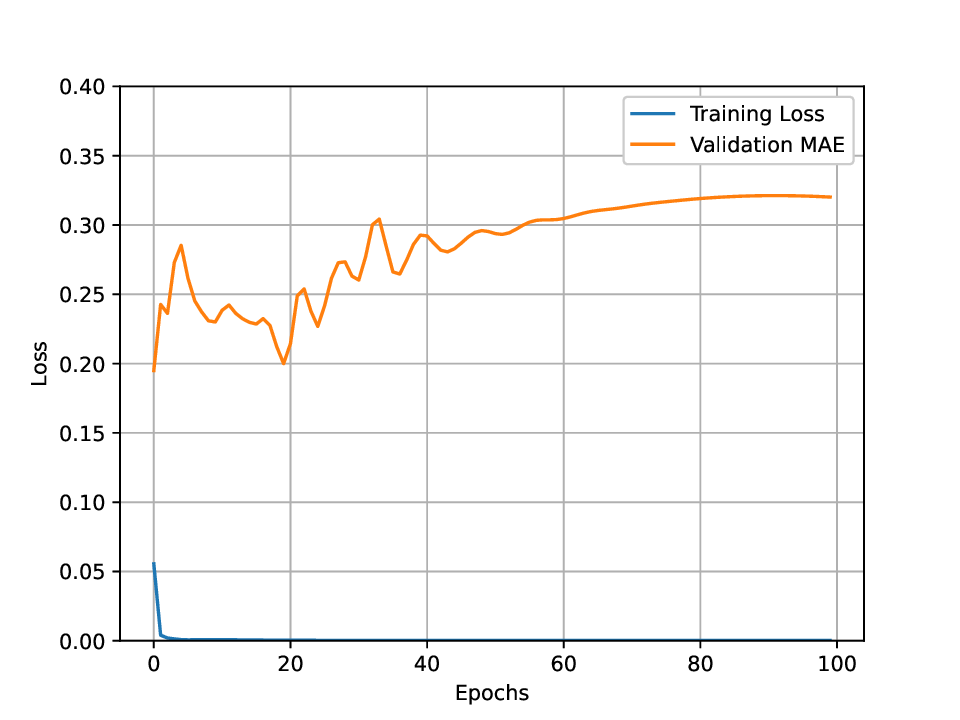}
        \caption{BTC-Training Loss and Validation MAE in Future Price Prediction for Bi-LSTM}
        \label{fig:22}
    \end{minipage}
    \hfill
    \begin{minipage}{0.44\textwidth}
        \centering
        \includegraphics[width=\textwidth]{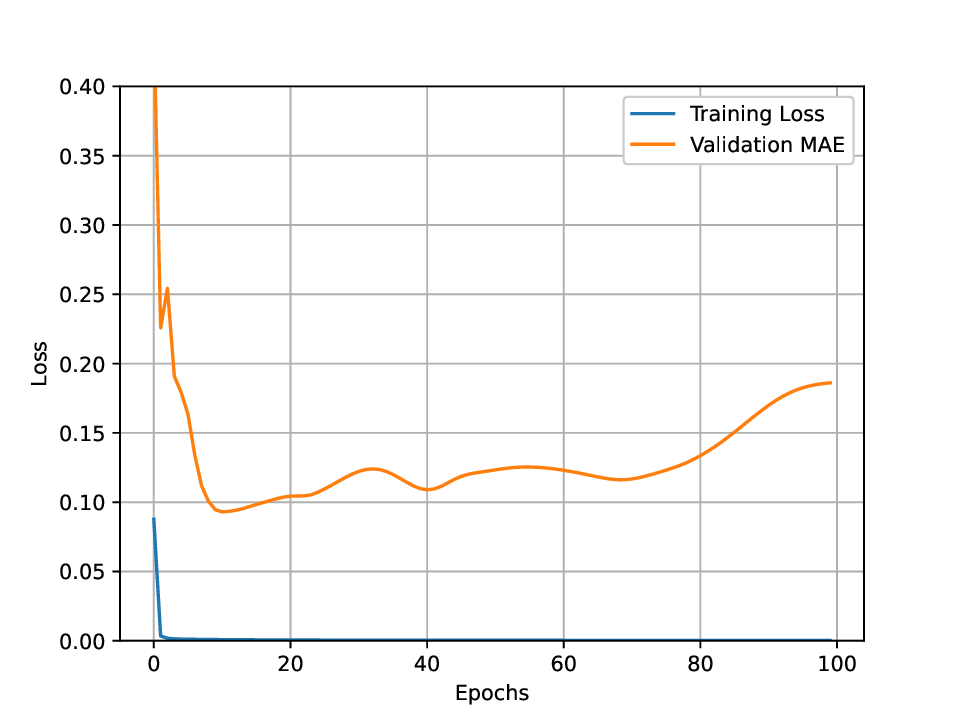}
        \caption{BTC-Training Loss and Validation MAE in Future Price Prediction for Finbert-Bi-LSTM}
        \label{fig:23}
    \end{minipage}
\end{figure}

In Fig. \ref{fig:18}, the LSTM model achieved its lowest validation MAE of 0.103 at the 100th epoch, while in Fig \ref{fig:19}, the FinBERT-LSTM model reached a lower validation MAE of 0.0744 at the 95th epoch. Similarly, in Fig. \ref{fig:22}, the Bi-LSTM model achieved its lowest validation MAE of 0.2 at the 20th epoch, while in Fig. \ref{fig:23}, the FinBERT-Bi-LSTM model demonstrated superior performance with the lowest validation MAE of 0.093 at the 11th epoch. After selecting the best models, we proceed with predicting the future 30 days and present the visual comparison of predicted prices and actual prices for all the models.

\begin{figure}[H]
    \centering
    \includegraphics[width=0.60\textwidth]{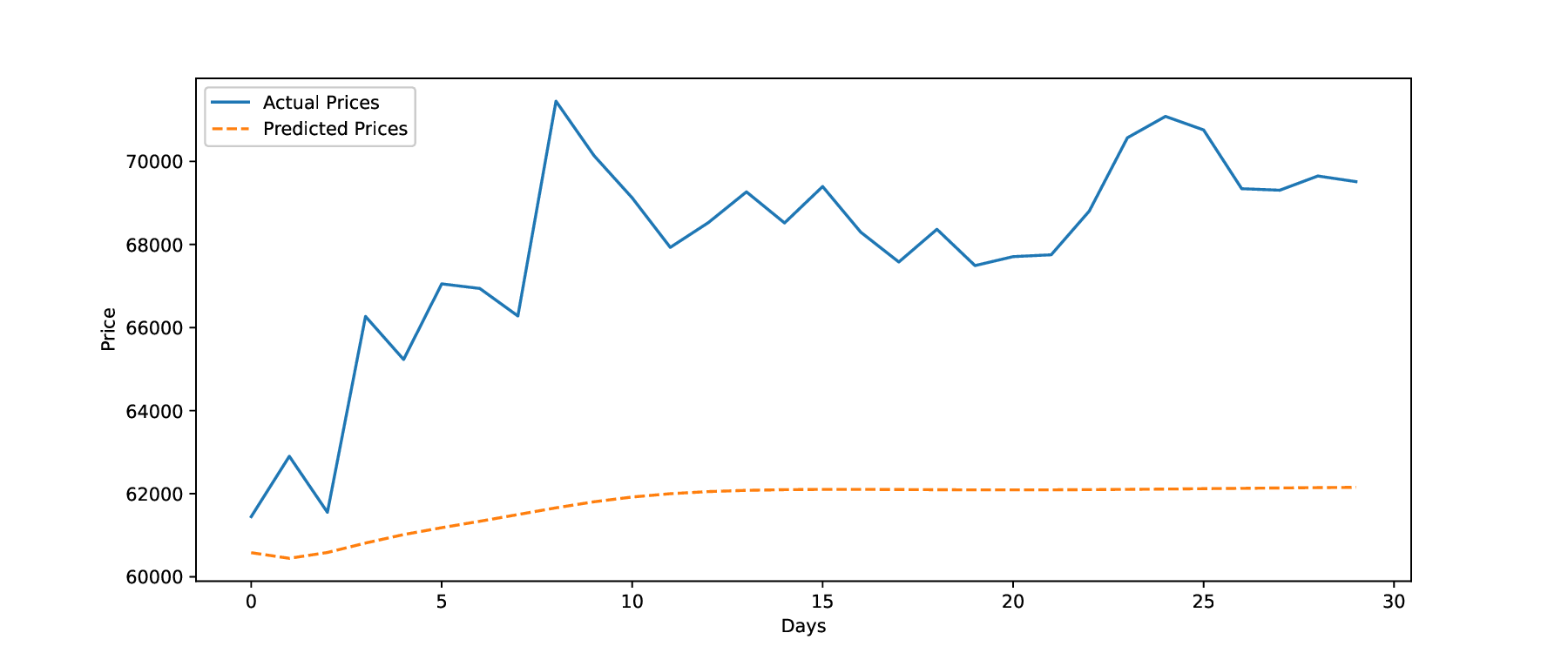}
    \caption{BTC-Actual 30-Day Prices vs Future 30-Day Price Prediction by LSTM using VET}
    \label{fig:20}
\end{figure}

\begin{figure}[H]
    \centering
    \includegraphics[width=0.60\textwidth]{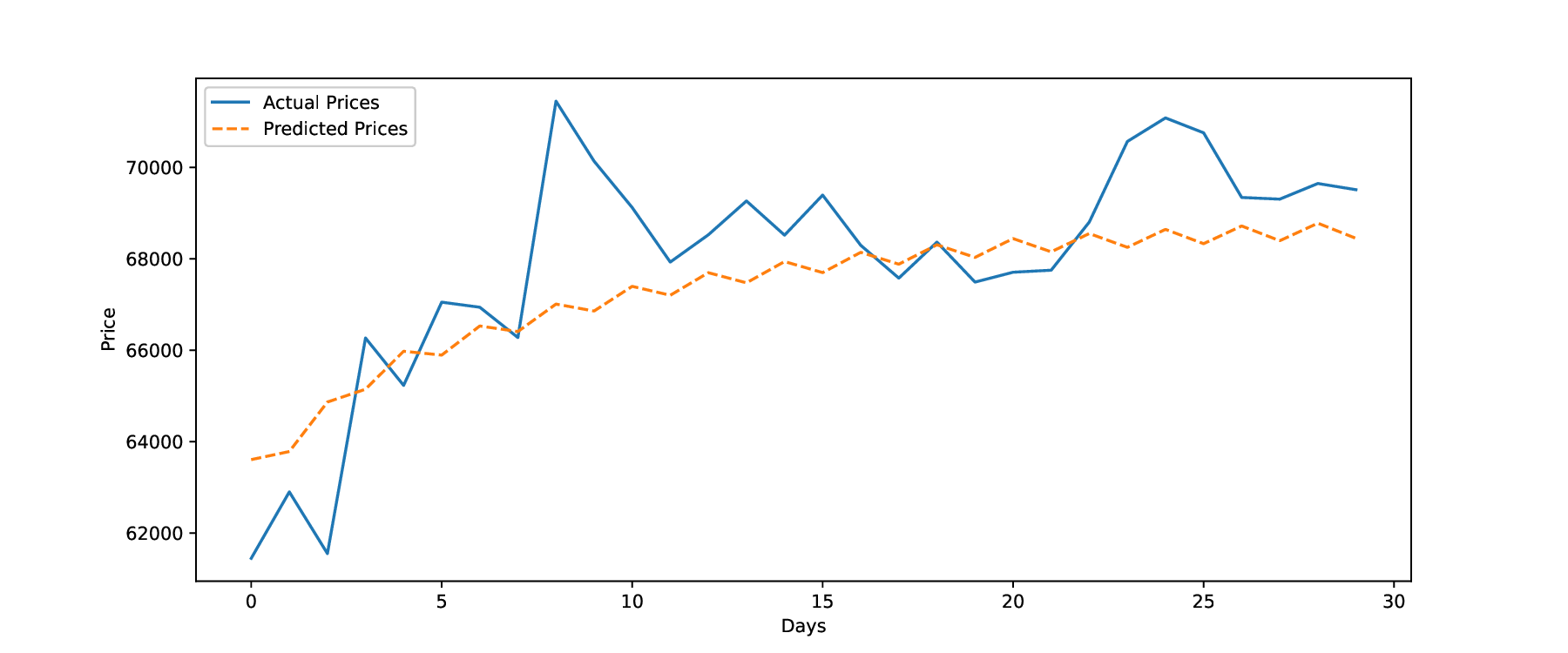}
    \caption{BTC-Actual 30-Day Prices vs Future 30-Day Price Prediction by FinBERT-LSTM using VET}
    \label{fig:21}
\end{figure}

\begin{figure}[H]
    \centering
    \includegraphics[width=0.60\textwidth]{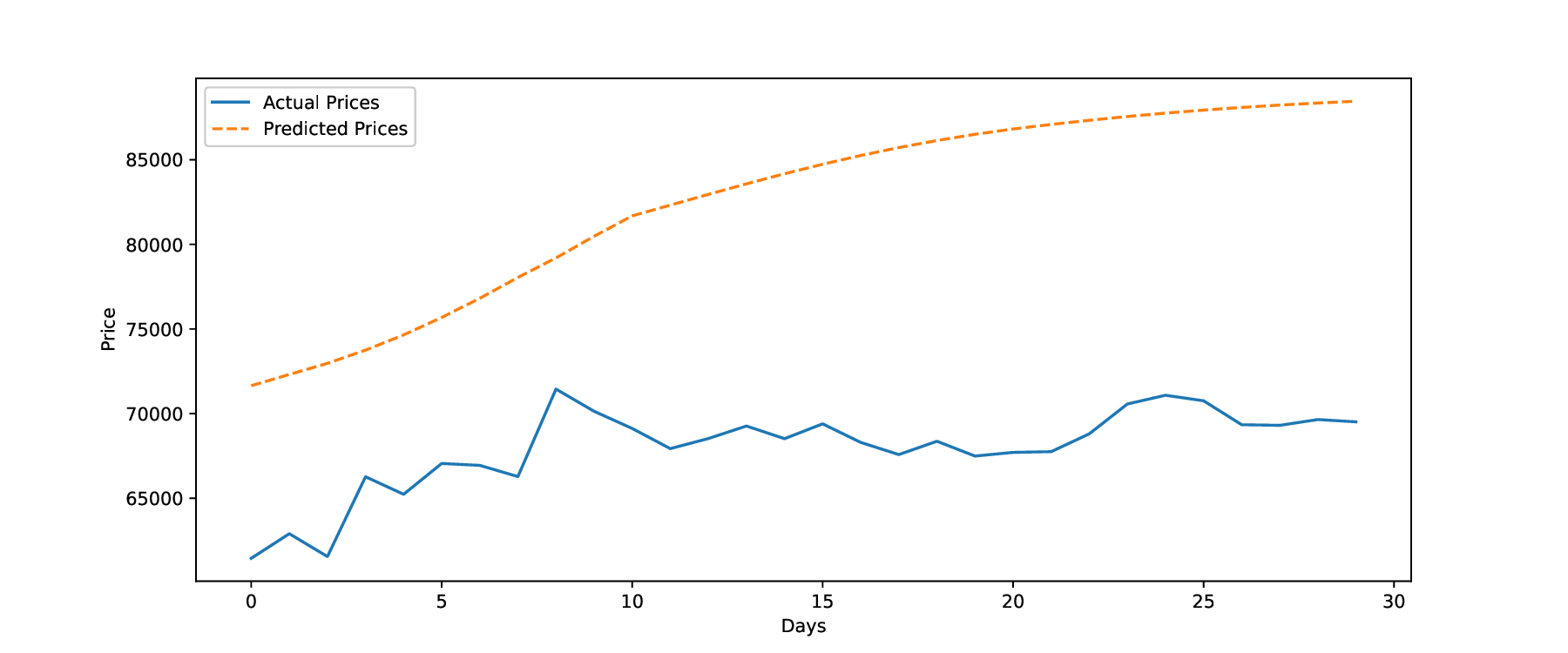}
    \caption{BTC-Comparison of Actual 30-Day Prices vs Future 30-Day Price Prediction by Bi-LSTM using VET
}
    \label{fig:24}
\end{figure}

\begin{figure}[H]
    \centering
    \includegraphics[width=0.60\textwidth]{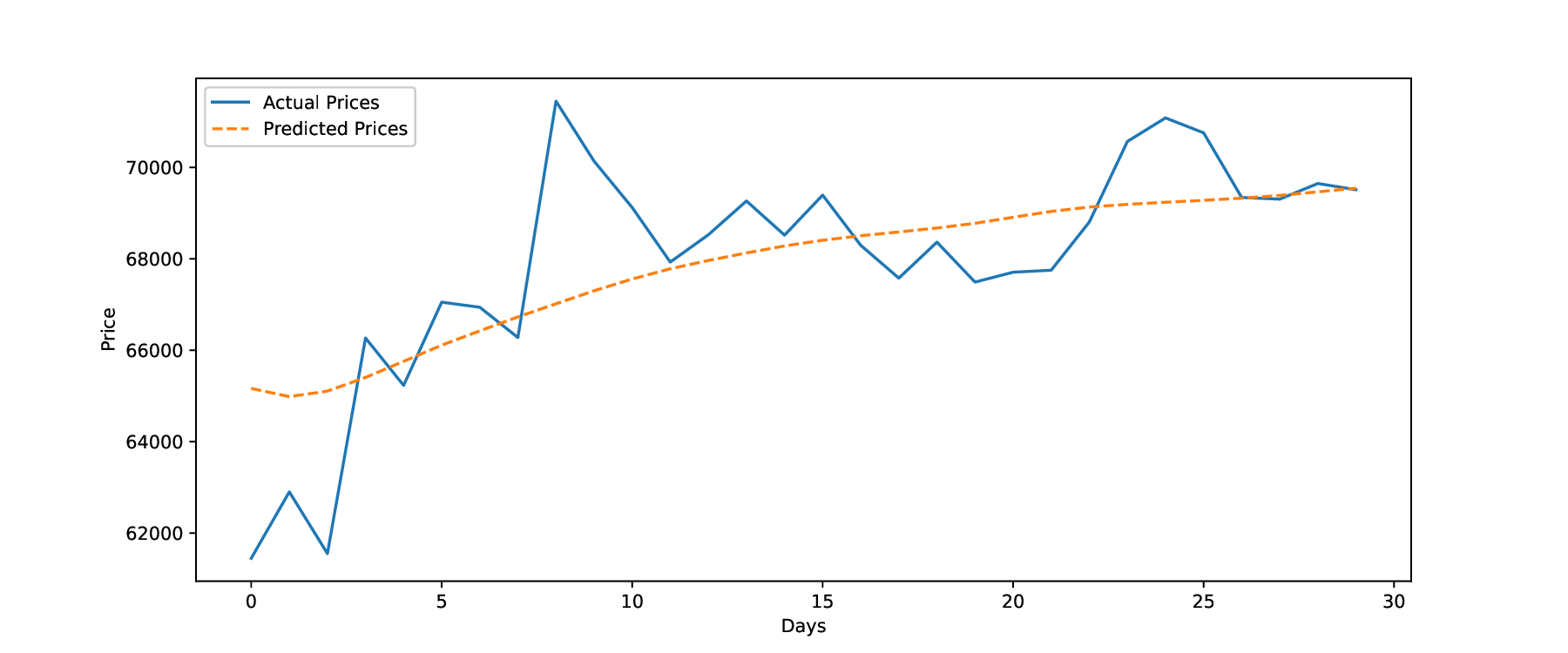}
    \caption{BTC-Comparison of Actual 30-Day Prices vs Future 30-Day Price Prediction by Finbert-Bi-LSTM using VET
}
    \label{fig:25}
\end{figure}

In Fig. \ref{fig:20}, Fig. \ref{fig:21}, Fig. \ref{fig:24}, and Fig. \ref{fig:25}, the actual prices are consistently shown in blue, while the orange dots represent the predicted prices for the LSTM, FinBERT-LSTM, Bi-LSTM, and FinBERT-Bi-LSTM models, respectively.
 
A quantitative comparison in Table \ref{Table9} shows that the FinBERT-LSTM model significantly outperforms the LSTM model in 30-day future predictions, achieving a higher accuracy of 98.14\% with lower error metrics. Interestingly, the table also reveals that while the Bi-LSTM model performs worse than the LSTM model, the FinBERT-Bi-LSTM model achieves the highest accuracy of 98.25\%, slightly surpassing the FinBERT-LSTM's 98.14\%
showing that our hypothesis mentioned in Section \ref{sec4}  works for future multi-day forecasting with this approach also.

\vspace{0.01cm} 
\begin{table}[htbp]
\centering
\caption{BTC-Performance Comparison of LSTM, FinBERT-LSTM, Bi-LSTM, and FinBERT-Bi-LSTM Models for 30-Day Future Prediction using VET}
\label{Table9}
\begin{tabular}{ |c|c|c|c|c| } 
\hline
\textbf{Metric} & \textbf{LSTM} & \textbf{FinBERT-LSTM} & \textbf{Bi-LSTM} & \textbf{FinBERT-Bi-LSTM} \\ 
\hline
MAE & 6181.94 & 1269.13 & 14598.75 & \textbf{1173.13}\\ 
MAPE & 0.08999\% & 0.01865\% & 0.21419\% & \textbf{0.01748\%} \\ 
Accuracy & 91.00\% & 98.14\% & 78.58\% & \textbf{98.25\%} \\ 
\hline
\end{tabular}
\end{table}

\subsubsection{Results of Ethereum Future 30-Day Price Prediction using VET}

First, we present the training loss and validation MAE for all the models.

\begin{figure}[H]
    \centering
    \begin{minipage}{0.40\textwidth}
        \centering
        \includegraphics[width=\textwidth]{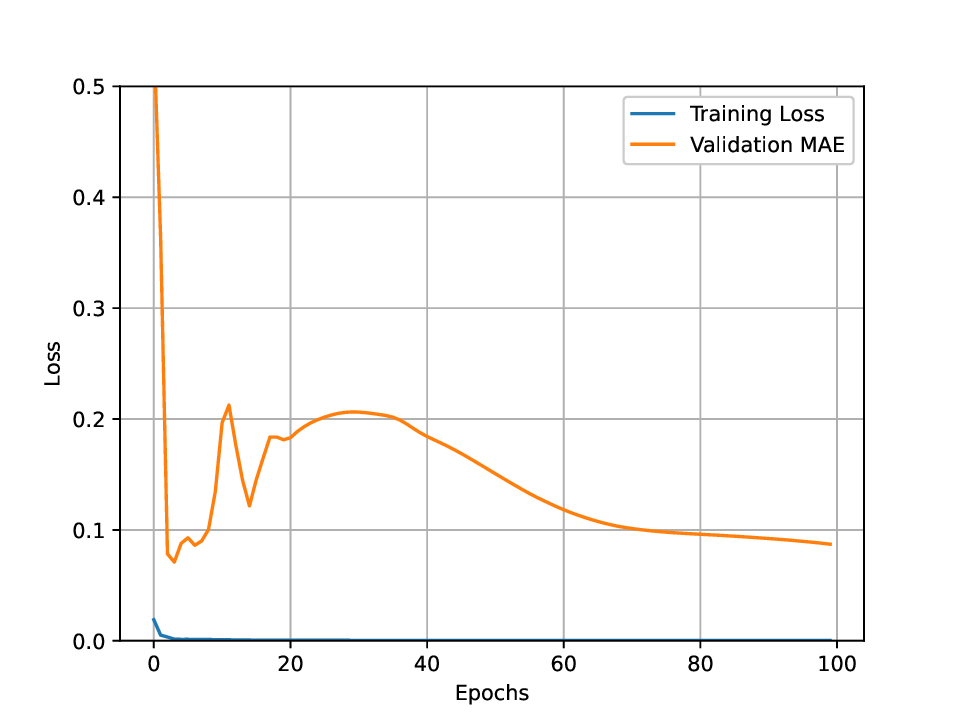}
        \caption{ETH-Training Loss and Validation
MAE in Future Price Prediction for LSTM}
        \label{fig:42}
    \end{minipage}
    \hfill
    \begin{minipage}{0.40\textwidth}
        \centering
        \includegraphics[width=\textwidth]{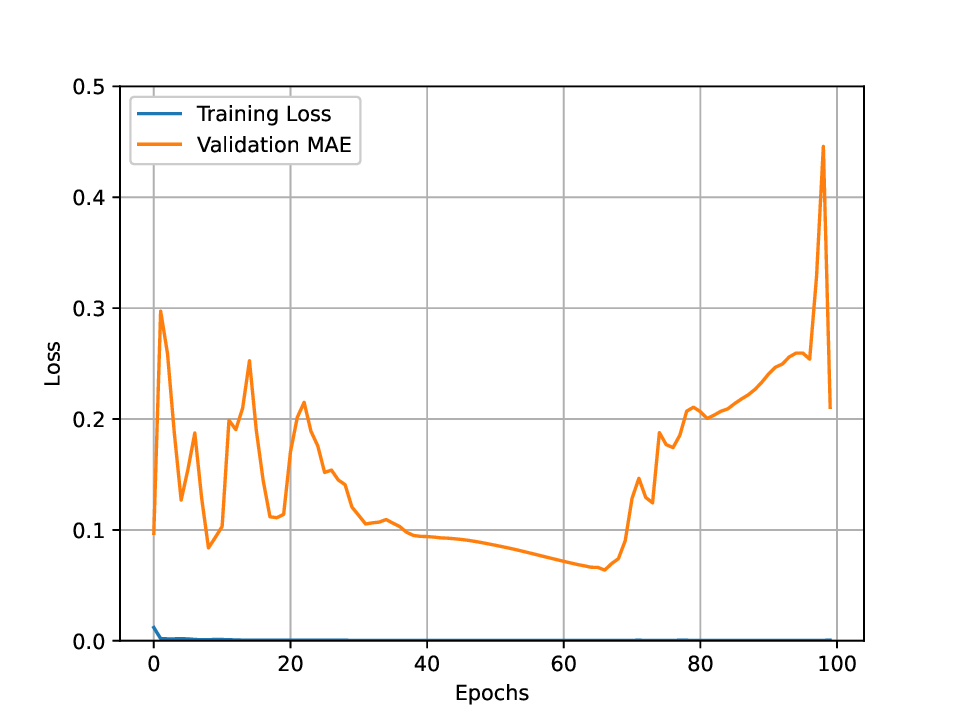}
        \caption{ETH-Training Loss and Validation
MAE in Future Price Prediction for FinBERT-LSTM}
        \label{fig:43}
    \end{minipage}
\end{figure}

\begin{figure}[H]
    \centering
    \begin{minipage}{0.40\textwidth}
        \centering
        \includegraphics[width=\textwidth]{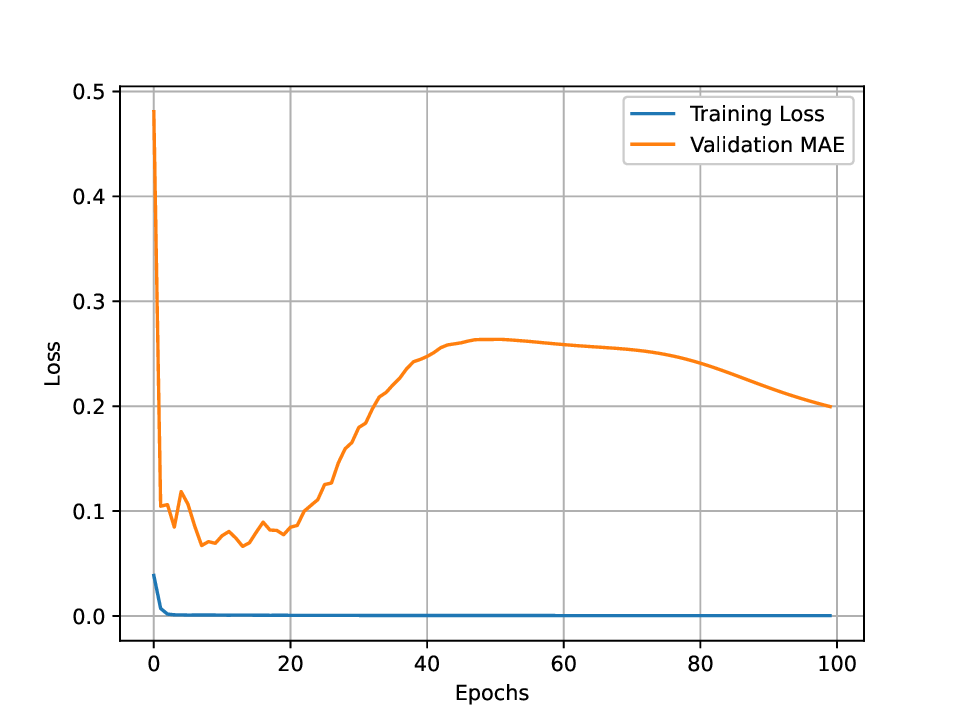}
        \caption{ETH-Training Loss and Validation
MAE in Future Price Prediction for Bi-LSTM}
        \label{fig:46}
    \end{minipage}
    \hfill
    \begin{minipage}{0.40\textwidth}
        \centering
        \includegraphics[width=\textwidth]{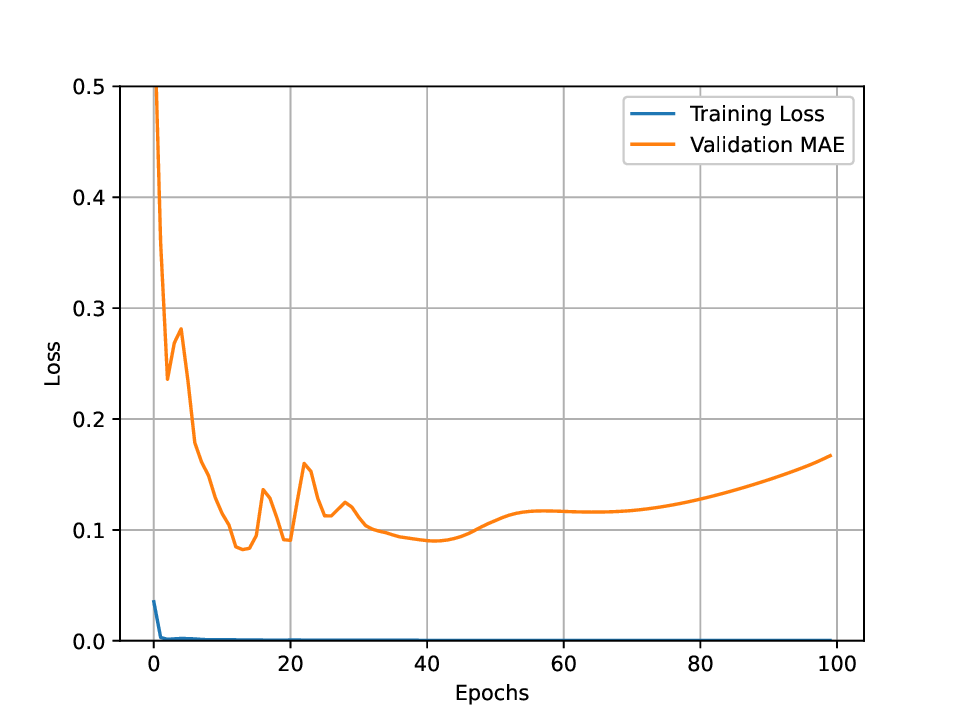}
        \caption{ETH-Training Loss and Validation
MAE in Future Price Prediction for FinBERT-Bi-LSTM}
        \label{fig:47}
    \end{minipage}
\end{figure}

For the LSTM model in Fig. \ref{fig:42}, the lowest validation MAE of 0.0709 is recorded at the 4th epoch, while the FinBERT-LSTM model in Fig. \ref{fig:43} achieves a lower validation MAE of 0.0637 at the 67th epoch. Similarly, for the Bi-LSTM model in Fig. \ref{fig:46}, the lowest validation MAE of 0.0663 is reached at the 14th epoch, while the FinBERT-Bi-LSTM model in Fig. \ref{fig:47} achieves its lowest validation MAE of 0.0822, also at the 14th epoch. Interestingly, in this case, the plain Bi-LSTM model slightly outperformed the FinBERT-Bi-LSTM model in terms of validation MAE.

Next, we present the graphs comparing the predicted prices for 30 days with the actual prices.

\begin{figure}[H]
    \centering
    \includegraphics[width=0.60\textwidth]{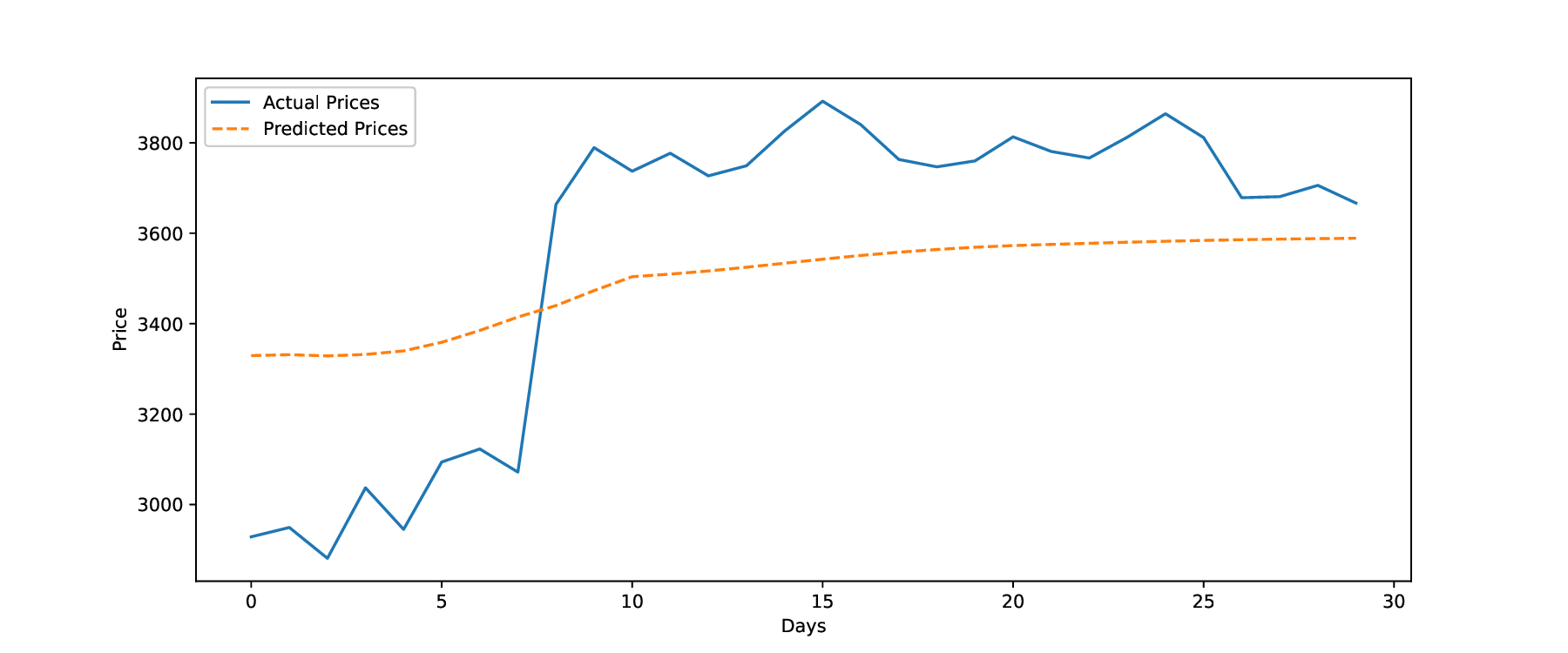}
    \caption{ETH-Actual 30-Day Prices vs Future 30-Day Price Prediction by LSTM using VET}
    \label{fig:44}
\end{figure}
\begin{figure}[H]
    \centering
    \includegraphics[width=0.60\textwidth]{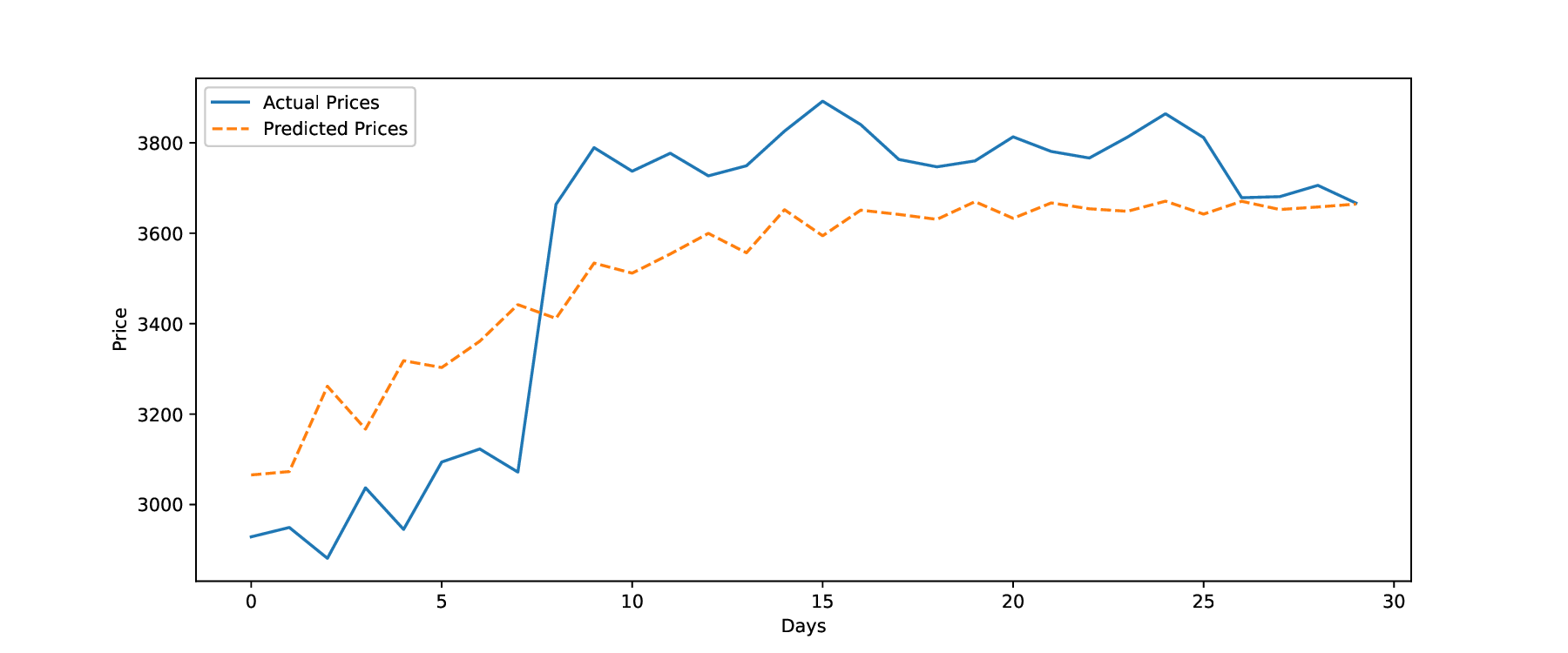}
    \caption{ETH-Actual 30-Day Prices vs Future 30-Day Price Prediction by FinBERT-LSTM using VET}
    \label{fig:45}
\end{figure}

\begin{figure}[H]
    \centering
    \includegraphics[width=0.60\textwidth]{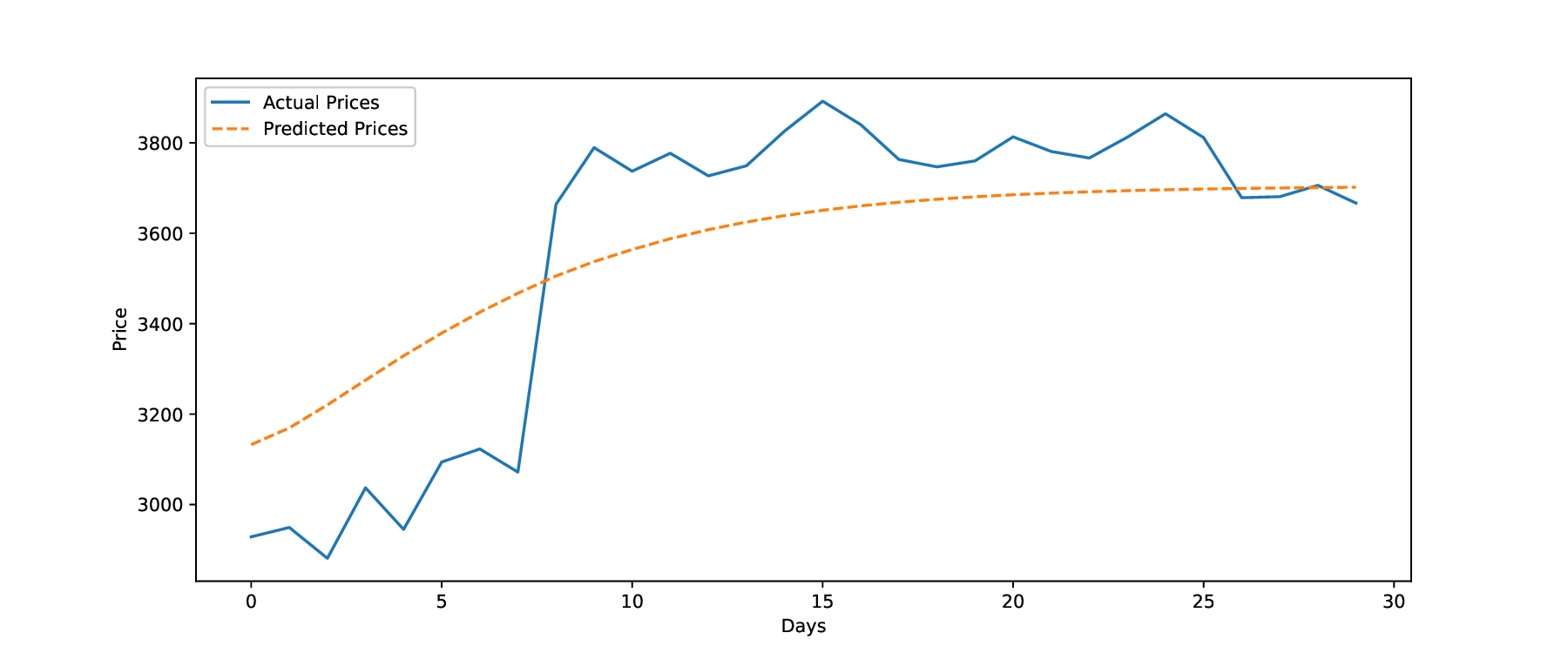}
    \caption{ETH-Actual 30-Day Prices vs Future 30-Day Price Prediction by Bi-LSTM using VET}
    \label{fig:48}
\end{figure}

\begin{figure}[H]
    \centering
    \includegraphics[width=0.60\textwidth]{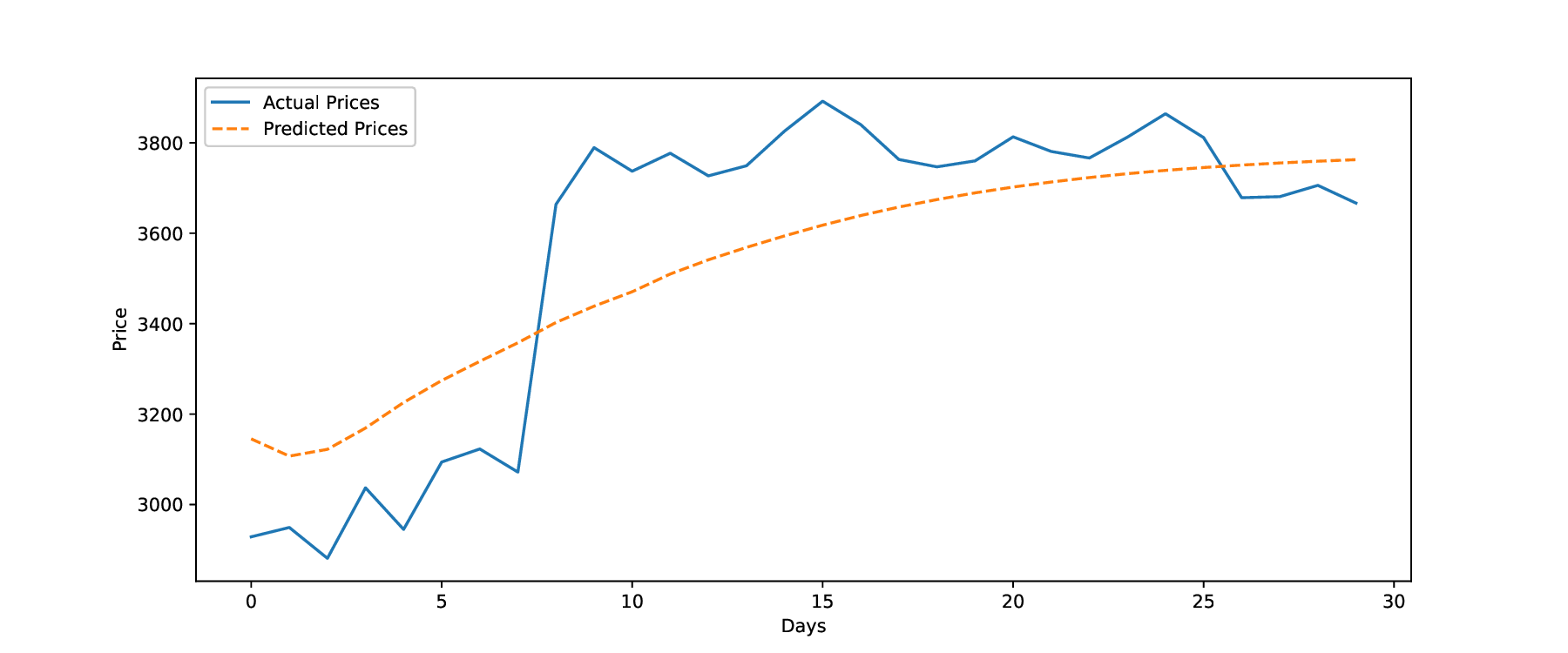}
    \caption{ETH-Actual 30-Day Prices vs Future 30-Day Price Prediction by FinBERT-Bi-LSTM using VET }
    \label{fig:49}
\end{figure}

In Fig. \ref{fig:44}, Fig. \ref{fig:45}, Fig. \ref{fig:48}, and Fig. \ref{fig:49}, the actual prices are shown in blue, while the predicted prices are represented by orange dots for the LSTM, FinBERT-LSTM, Bi-LSTM, and FinBERT-Bi-LSTM models, respectively. This approach, although trained on fewer data, is more generalized, and interestingly all the models predict a price rise quite early, even before the significant price increase starting around the 7th day and continuing until just before the 10th day. In contrast, the earlier approach predicted the price rise more gradually after the major price increase.

Now, we present the quantitative performance measures for all the models in Table \ref{Table10}. The table shows that the FinBERT-LSTM model outperforms the LSTM model in the 30-day future prediction for Ethereum, achieving a higher accuracy of 94.93\% compared to 92.71\%, along with lower error metrics. Notably, while the plain Bi-LSTM performs even better than the FinBERT-LSTM model, the FinBERT-Bi-LSTM model slightly surpasses the Bi-LSTM, achieving the highest accuracy of 95.25\% for 30-day future prediction.

\vspace{0.01cm} 
\begin{table}[htbp]
\centering
\caption{ETH-Performance Comparison of LSTM, FinBERT-LSTM, Bi-LSTM, and FinBERT-Bi-LSTM Models for 30-Day Future Prediction using VET}
\label{Table10}
\begin{tabular}{ |c|c|c|c|c| } 
\hline
\textbf{Metric} & \textbf{LSTM} & \textbf{FinBERT-LSTM} & \textbf{Bi-LSTM} & \textbf{FinBERT-Bi-LSTM} \\ 
\hline
MAE & 251.11 & 174.81 & 167.09 & \textbf{164.84} \\ 
MAPE & 0.07294\% & 0.05073\% & 0.04955\% & \textbf{0.04755\% }\\ 
Accuracy & 92.71\% & 94.93\% & 95.04\% & \textbf{95.25\%} \\ 
\hline
\end{tabular}
\end{table}

\subsection{Findings}

Our experimental analysis compared the performance of various models, including traditional LSTM, Bi-LSTM, FinBERT-LSTM, and FinBERT-Bi-LSTM. We present the following key findings from our analysis:

  \begin{itemize}

    \item \textbf{Impact of Sentiment Integration on Intra-Day Predictions:} Integrating news sentiment into predictive models consistently improved the accuracy of intra-day price predictions for both LSTM and Bi-LSTM models. Notably, the FinBERT-Bi-LSTM model achieved the highest accuracy, benefiting from the rich context provided by sentiment data, which captures market behaviors beyond price movements alone.

    \item \textbf{Previous-Day Sentiment Impact on One-Day-Ahead Predictions:} Incorporating the previous day’s sentiment data as the last available sentiment information in the FinBERT-integrated models still outperformed the plain models in one-day-ahead predictions, with FinBERT-Bi-LSTM achieving the highest performance. However, when previous-day sentiment was used during both training and validation along with testing, the models showed reduced accuracy, particularly for FinBERT-Bi-LSTM, where performance even fell below that of the plain Bi-LSTM model. Training and validating with current-day sentiment data while using previous-day sentiment for testing proved to be a more effective approach for FinBERT-integrated models, yielding better accuracy across models.

    \item \textbf{Profitability in Real Trading:} We assessed the profitability of relying solely on the model’s one-day-ahead predictions within a simple daily trading strategy. For Bitcoin (BTC), the LSTM and Bi-LSTM models did not execute any trades, while FinBERT-LSTM and FinBERT-Bi-LSTM generated profits, with FinBERT-Bi-LSTM achieving nearly half of the profit that could be obtained with perfect price predictions. In the case of Ethereum (ETH), all models resulted in losses, except for FinBERT-Bi-LSTM, which produced a substantial profit, highlighting its strength in trading performance.

   \item \textbf{Long-Term Forecasting Performance:} Predicting long-term cryptocurrency prices is especially challenging due to high market volatility. Our FinBERT-enhanced models showed superior accuracy in these predictions across two approaches:
\begin{itemize}
    \item Training until the start of the test period without using validation data, which provided more training data and yielded better results for Bitcoin.
    \item Using validation data between the training and testing phases. Although this approach left less training data immediately preceding the test data, it produced more generalized results and performed better for Ethereum, particularly in managing sudden, unexpected price spikes.

      \end{itemize}
\end{itemize}

In both approaches, the FinBERT-Bi-LSTM model achieved the highest accuracy, demonstrating its robustness in long-term forecasting for volatile cryptocurrency markets.

\section{Conclusions \& Future works}\label{sec7}

This research introduces the FinBERT-Bi-LSTM model as an effective tool for predicting prices in volatile cryptocurrency markets like Bitcoin (BTC) and Ethereum (ETH). The integration of news sentiment into machine learning models significantly enhances both short-term and long-term forecasting accuracy. Our approach utilizes a unique dataset combining financial news with FinBERT-generated sentiment scores, providing deeper insights into sentiment-driven market behaviors. Notably, the FinBERT-Bi-LSTM model achieves an accuracy of 98.21\% for intra-day predictions and 98.07\% for one-day-ahead predictions on Bitcoin (BTC), resulting in a remarkable profit from the one-day-ahead strategy. On Ethereum (ETH), it achieves an accuracy of 97.5\% for intra-day predictions and 97.36\% for one-day-ahead predictions, generating a significant profit. Furthermore, in long-term price predictions, the Maximum Data Training approach yields an impressive accuracy of 98.3\% for BTC, while the Validation-Enhanced Training approach performs better for ETH with an accuracy of 95.25\%.

Looking ahead, our findings pave the way for several promising avenues of exploration. We can refine daily trading strategies by applying more complex approaches and expanding the models themselves to integrate features such as technical indicators and macroeconomic variables. Although we have utilized reliable sources in this study, in the future, we may consider incorporating sentiment from social media, where false news is more prevalent. In that case, it is essential to detect and filter out this fake news to enhance the reliability of sentiment-driven models. Ultimately, this study enhances the field of financial analytics and sets the stage for future research into more advanced sentiment-aware forecasting tools. These tools have the potential to improve investment strategies and support informed decision-making across various financial markets.

\section*{Declarations}

\begin{itemize}[label=$\bullet$]
\item \textbf{Funding:} No funding was received for conducting this study.

\item \textbf{Competing Interests:} The authors have no competing interests to declare that are relevant to the content of this article.

\item \textbf{Data Availability:} The dataset will be made available upon request.
\item \textbf{Code Availability:} The source code will be made available upon request.
\item \textbf{Author Contributions:} All authors contributed to the conception and design of the study. All authors approved the final manuscript. The authors confirm their contribution to the paper as follows:
\begin{itemize}[label= – ]
        \item \textbf{Mabsur Fatin Bin Hossain:} conceptualization, methodology, writing - original draft, experimentation.
        \item \textbf{Lubna Zahan Lamia:} conceptualization, methodology, writing - original draft, dataset preparation.
        \item \textbf{Md Mahmudur Rahman:} editing the draft, investigation and analysis

        \item \textbf{Md Mosaddek Khan:} conceptualization, supervision, finalizing the draft

    \end{itemize}
\end{itemize}

\bibliography{sn-bibliography}


\begin{thebibliography}{29}
\ifx \bisbn   \undefined \def \bisbn  #1{ISBN #1}\fi
\ifx \binits  \undefined \def \binits#1{#1}\fi
\ifx \bauthor  \undefined \def \bauthor#1{#1}\fi
\ifx \batitle  \undefined \def \batitle#1{#1}\fi
\ifx \bjtitle  \undefined \def \bjtitle#1{#1}\fi
\ifx \bvolume  \undefined \def \bvolume#1{\textbf{#1}}\fi
\ifx \byear  \undefined \def \byear#1{#1}\fi
\ifx \bissue  \undefined \def \bissue#1{#1}\fi
\ifx \bfpage  \undefined \def \bfpage#1{#1}\fi
\ifx \blpage  \undefined \def \blpage #1{#1}\fi
\ifx \burl  \undefined \def \burl#1{\textsf{#1}}\fi
\ifx \doiurl  \undefined \def \doiurl#1{\url{https://doi.org/#1}}\fi
\ifx \betal  \undefined \def \betal{\textit{et al.}}\fi
\ifx \binstitute  \undefined \def \binstitute#1{#1}\fi
\ifx \binstitutionaled  \undefined \def \binstitutionaled#1{#1}\fi
\ifx \bctitle  \undefined \def \bctitle#1{#1}\fi
\ifx \beditor  \undefined \def \beditor#1{#1}\fi
\ifx \bpublisher  \undefined \def \bpublisher#1{#1}\fi
\ifx \bbtitle  \undefined \def \bbtitle#1{#1}\fi
\ifx \bedition  \undefined \def \bedition#1{#1}\fi
\ifx \bseriesno  \undefined \def \bseriesno#1{#1}\fi
\ifx \blocation  \undefined \def \blocation#1{#1}\fi
\ifx \bsertitle  \undefined \def \bsertitle#1{#1}\fi
\ifx \bsnm \undefined \def \bsnm#1{#1}\fi
\ifx \bsuffix \undefined \def \bsuffix#1{#1}\fi
\ifx \bparticle \undefined \def \bparticle#1{#1}\fi
\ifx \barticle \undefined \def \barticle#1{#1}\fi
\bibcommenthead
\ifx \bconfdate \undefined \def \bconfdate #1{#1}\fi
\ifx \botherref \undefined \def \botherref #1{#1}\fi
\ifx \url \undefined \def \url#1{\textsf{#1}}\fi
\ifx \bchapter \undefined \def \bchapter#1{#1}\fi
\ifx \bbook \undefined \def \bbook#1{#1}\fi
\ifx \bcomment \undefined \def \bcomment#1{#1}\fi
\ifx \oauthor \undefined \def \oauthor#1{#1}\fi
\ifx \citeauthoryear \undefined \def \citeauthoryear#1{#1}\fi
\ifx \endbibitem  \undefined \def \endbibitem {}\fi
\ifx \bconflocation  \undefined \def \bconflocation#1{#1}\fi
\ifx \arxivurl  \undefined \def \arxivurl#1{\textsf{#1}}\fi
\csname PreBibitemsHook\endcsname

\bibitem[\protect\citeauthoryear{Dhingra et~al.}{2024}]{dhingra}
\begin{barticle}
\bauthor{\bsnm{Dhingra}, \binits{B.}},
\bauthor{\bsnm{Batra}, \binits{S.}},
\bauthor{\bsnm{Aggarwal}, \binits{V.}},
\bauthor{\bsnm{Yadav}, \binits{M.}},
\bauthor{\bsnm{Kumar}, \binits{P.}}:
\batitle{Stock market volatility: a systematic review}.
\bjtitle{Journal of Modelling in Management}
\bvolume{19}(\bissue{3}),
\bfpage{925}--\blpage{952}
(\byear{2024})
\doiurl{10.1108/JM2-04-2023-0080}
\end{barticle}
\endbibitem

\bibitem[\protect\citeauthoryear{Alameer et~al.}{2019}]{Alameer2019}
\begin{barticle}
\bauthor{\bsnm{Alameer}, \binits{Z.}},
\bauthor{\bsnm{Ewees}, \binits{A.}},
\bauthor{\bsnm{Elsayed Abd~Elaziz}, \binits{M.}},
\bauthor{\bsnm{Ye}, \binits{H.}}:
\batitle{Forecasting gold price fluctuations using improved multilayer perceptron neural network and whale optimization algorithm}.
\bjtitle{Resources Policy}
\bvolume{61},
\bfpage{250}--\blpage{260}
(\byear{2019})
\doiurl{10.1016/j.resourpol.2019.02.014}
\end{barticle}
\endbibitem

\bibitem[\protect\citeauthoryear{Mittal and Goel}{2020}]{mittal2020cryptocurrency}
\begin{barticle}
\bauthor{\bsnm{Mittal}, \binits{H.}},
\bauthor{\bsnm{Goel}, \binits{S.}}:
\batitle{Economic, legal and financial perspectives on cryptocurrencies: A review on cryptocurrency growth, opportunities and future prospects}.
\bjtitle{World Review of Entrepreneurship Management and Sustainable Development}
\bvolume{16}(\bissue{6}),
\bfpage{611}--\blpage{623}
(\byear{2020})
\doiurl{10.1504/WREMSD.2020.10033440}
\end{barticle}
\endbibitem

\bibitem[\protect\citeauthoryear{Nakamoto}{2008}]{nakamoto2008bitcoin}
\begin{botherref}
\oauthor{\bsnm{Nakamoto}, \binits{S.}}:
Bitcoin: A Peer-to-Peer Electronic Cash System.
Published online at \url{https://bitcoin.org/bitcoin.pdf}
(2008)
\end{botherref}
\endbibitem

\bibitem[\protect\citeauthoryear{Hu et~al.}{2021}]{hu2021transaction}
\begin{barticle}
\bauthor{\bsnm{Hu}, \binits{T.}},
\bauthor{\bsnm{Liu}, \binits{X.}},
\bauthor{\bsnm{Chen}, \binits{T.}},
\bauthor{\bsnm{Zhang}, \binits{X.}},
\bauthor{\bsnm{Huang}, \binits{X.}},
\bauthor{\bsnm{Niu}, \binits{W.}},
\bauthor{\bsnm{Lu}, \binits{J.}},
\bauthor{\bsnm{Zhou}, \binits{K.}},
\bauthor{\bsnm{Liu}, \binits{Y.}}:
\batitle{Transaction-based classification and detection approach for ethereum smart contract}.
\bjtitle{Information Processing and Management}
\bvolume{58}(\bissue{2}),
\bfpage{102462}
(\byear{2021})
\doiurl{10.1016/j.ipm.2020.102462}
\end{barticle}
\endbibitem

\bibitem[\protect\citeauthoryear{Sovbetov}{2018}]{sovbetov2018factors}
\begin{barticle}
\bauthor{\bsnm{Sovbetov}, \binits{Y.}}:
\batitle{Factors influencing cryptocurrency prices: Evidence from bitcoin, ethereum, dash, litecoin, and monero}.
\bjtitle{Journal of Economics and Financial Analysis}
\bvolume{2}(\bissue{2}),
\bfpage{1}--\blpage{27}
(\byear{2018})
\end{barticle}
\endbibitem

\bibitem[\protect\citeauthoryear{Cheah and Fry}{2015}]{cheah2015speculative}
\begin{barticle}
\bauthor{\bsnm{Cheah}, \binits{E.-T.}},
\bauthor{\bsnm{Fry}, \binits{J.}}:
\batitle{Speculative bubbles in bitcoin markets? an empirical investigation into the fundamental value of bitcoin}.
\bjtitle{Economics Letters}
\bvolume{130},
\bfpage{32}--\blpage{36}
(\byear{2015})
\doiurl{10.1016/j.econlet.2015.02.029}
\end{barticle}
\endbibitem

\bibitem[\protect\citeauthoryear{Corbet et~al.}{2018}]{Corbet2018}
\begin{barticle}
\bauthor{\bsnm{Corbet}, \binits{S.}},
\bauthor{\bsnm{Meegan}, \binits{A.}},
\bauthor{\bsnm{Larkin}, \binits{C.}},
\bauthor{\bsnm{Lucey}, \binits{B.}},
\bauthor{\bsnm{Yarovaya}, \binits{L.}}:
\batitle{Exploring the dynamic relationships between cryptocurrencies and other financial assets}.
\bjtitle{Economics Letters}
\bvolume{165},
\bfpage{28}--\blpage{34}
(\byear{2018})
\doiurl{10.1016/j.econlet.2018.01.004}
\end{barticle}
\endbibitem

\bibitem[\protect\citeauthoryear{Baur et~al.}{2018}]{baur2018bitcoin}
\begin{barticle}
\bauthor{\bsnm{Baur}, \binits{D.G.}},
\bauthor{\bsnm{Hong}, \binits{K.}},
\bauthor{\bsnm{Lee}, \binits{A.D.}}:
\batitle{Bitcoin: Medium of exchange or speculative assets?}
\bjtitle{Journal of International Financial Markets, Institutions and Money}
\bvolume{54},
\bfpage{177}--\blpage{189}
(\byear{2018})
\doiurl{10.1016/j.intfin.2017.12.004}
\end{barticle}
\endbibitem

\bibitem[\protect\citeauthoryear{Urquhart}{2016}]{urquhart2016inefficiency}
\begin{barticle}
\bauthor{\bsnm{Urquhart}, \binits{A.}}:
\batitle{The inefficiency of bitcoin}.
\bjtitle{Economics Letters}
\bvolume{148},
\bfpage{80}--\blpage{82}
(\byear{2016})
\doiurl{10.1016/j.econlet.2016.09.019}
\end{barticle}
\endbibitem

\bibitem[\protect\citeauthoryear{Katsiampa}{2017}]{katsiampa2017volatility}
\begin{barticle}
\bauthor{\bsnm{Katsiampa}, \binits{P.}}:
\batitle{Volatility estimation for bitcoin: A comparison of garch models}.
\bjtitle{Economics Letters}
\bvolume{158},
\bfpage{3}--\blpage{6}
(\byear{2017})
\doiurl{10.1016/j.econlet.2017.06.023}
\end{barticle}
\endbibitem

\bibitem[\protect\citeauthoryear{Bouri et~al.}{2017}]{bouri2017returnvolatility}
\begin{barticle}
\bauthor{\bsnm{Bouri}, \binits{E.}},
\bauthor{\bsnm{Azzi}, \binits{G.}},
\bauthor{\bsnm{Dyhrberg}, \binits{A.H.}}:
\batitle{On the return-volatility relationship in the bitcoin market around the price crash of 2013}.
\bjtitle{Economics: The Open-Access, Open-Assessment E-Journal}
\bvolume{11},
\bfpage{1}--\blpage{16}
(\byear{2017})
\doiurl{10.5018/economics-ejournal.ja.2017-2}
\end{barticle}
\endbibitem

\bibitem[\protect\citeauthoryear{Rizwan et~al.}{2019}]{rizwan2019bitcoin}
\begin{botherref}
\oauthor{\bsnm{Rizwan}, \binits{M.}},
\oauthor{\bsnm{Narejo}, \binits{S.}},
\oauthor{\bsnm{Javed}, \binits{M.}}:
Bitcoin price prediction using Deep Learning Algorithm.
Paper presented at the 13th International Conference on Mathematics, Actuarial Science, Computer Science and Statistics (MACS)
(2019).
\doiurl{10.1109/MACS48846.2019.9024772}
\end{botherref}
\endbibitem

\bibitem[\protect\citeauthoryear{Seabe et~al.}{2023}]{seabe2023forecasting}
\begin{barticle}
\bauthor{\bsnm{Seabe}, \binits{P.L.}},
\bauthor{\bsnm{Moutsinga}, \binits{C.}},
\bauthor{\bsnm{Pindza}, \binits{E.}}:
\batitle{Forecasting cryptocurrency prices using lstm, gru, and bi-directional lstm: A deep learning approach}.
\bjtitle{Fractal and Fractional}
\bvolume{7}(\bissue{2}),
\bfpage{203}
(\byear{2023})
\doiurl{10.3390/fractalfract7020203}
\end{barticle}
\endbibitem

\bibitem[\protect\citeauthoryear{Karalevicius et~al.}{2018}]{karalevicius2018sentiment}
\begin{barticle}
\bauthor{\bsnm{Karalevicius}, \binits{V.}},
\bauthor{\bsnm{Degrande}, \binits{N.}},
\bauthor{\bsnm{Weerdt}, \binits{J.D.}}:
\batitle{Using sentiment analysis to predict interday bitcoin price movements}.
\bjtitle{Journal of Risk Finance}
\bvolume{19}(\bissue{1}),
\bfpage{56}--\blpage{75}
(\byear{2018})
\doiurl{10.1108/JRF-06-2017-0092}
\end{barticle}
\endbibitem

\bibitem[\protect\citeauthoryear{Valencia et~al.}{2019}]{valencia2019price}
\begin{barticle}
\bauthor{\bsnm{Valencia}, \binits{F.}},
\bauthor{\bsnm{Gómez-Espinosa}, \binits{A.}},
\bauthor{\bsnm{Valdés-Aguirre}, \binits{B.}}:
\batitle{Price movement prediction of cryptocurrencies using sentiment analysis and machine learning}.
\bjtitle{Entropy}
\bvolume{21}(\bissue{6}),
\bfpage{589}
(\byear{2019})
\doiurl{10.3390/e21060589}
\end{barticle}
\endbibitem

\bibitem[\protect\citeauthoryear{Hartmann et~al.}{2019}]{hartmann2019comparing}
\begin{barticle}
\bauthor{\bsnm{Hartmann}, \binits{J.}},
\bauthor{\bsnm{Huppertz}, \binits{J.}},
\bauthor{\bsnm{Schamp}, \binits{C.}},
\bauthor{\bsnm{Heitmann}, \binits{M.}}:
\batitle{Comparing automated text classification methods}.
\bjtitle{International Journal of Research in Marketing}
\bvolume{36}(\bissue{1}),
\bfpage{20}--\blpage{38}
(\byear{2019})
\doiurl{10.1016/j.ijresmar.2018.09.009}
\end{barticle}
\endbibitem

\bibitem[\protect\citeauthoryear{Sousa et~al.}{2019}]{sousabert}
\begin{botherref}
\oauthor{\bsnm{Sousa}, \binits{M.G.}},
\oauthor{\bsnm{Sakiyama}, \binits{K.}},
\oauthor{\bsnm{Rodrigues}, \binits{L.d.S.}},
\oauthor{\bsnm{Moraes}, \binits{P.H.}},
\oauthor{\bsnm{Fernandes}, \binits{E.R.}},
\oauthor{\bsnm{Matsubara}, \binits{E.T.}}:
BERT for Stock Market Sentiment Analysis.
Paper presented at the 31st International Conference on Tools with Artificial Intelligence (ICTAI), Portland, OR, USA, 2019
(2019)
\end{botherref}
\endbibitem

\bibitem[\protect\citeauthoryear{Araci}{2019}]{araci2019finbert}
\begin{botherref}
\oauthor{\bsnm{Araci}, \binits{D.T.}}:
FinBERT: Financial Sentiment Analysis with Pre-trained Language Models.
Preprint at \url{https://arxiv.org/abs/1908.10063v1}
(2019)
\end{botherref}
\endbibitem

\bibitem[\protect\citeauthoryear{Das et~al.}{2024}]{arijit2024}
\begin{botherref}
\oauthor{\bsnm{Das}, \binits{A.}},
\oauthor{\bsnm{Nandi}, \binits{T.}},
\oauthor{\bsnm{Saha}, \binits{P.}},
\oauthor{\bsnm{Das}, \binits{S.}},
\oauthor{\bsnm{Mukherjee}, \binits{S.}},
\oauthor{\bsnm{Naskar}, \binits{S.K.}},
\oauthor{\bsnm{Saha}, \binits{D.}}:
Effect of Leader’s Voice on Financial Market: An Empirical Deep Learning Expedition on NASDAQ, NSE, and Beyond.
Preprint at \url{https://arxiv.org/abs/2403.12161v1}
(2024)
\end{botherref}
\endbibitem

\bibitem[\protect\citeauthoryear{Halder}{2022}]{halder2022}
\begin{botherref}
\oauthor{\bsnm{Halder}, \binits{S.}}:
FinBERT-LSTM: Deep Learning-based Stock Price Prediction Using News Sentiment Analysis.
Preprint at \url{https://arxiv.org/abs/2211.07392}
(2022)
\end{botherref}
\endbibitem

\bibitem[\protect\citeauthoryear{Gu et~al.}{2024}]{gu2024}
\begin{botherref}
\oauthor{\bsnm{Gu}, \binits{W.}},
\oauthor{\bsnm{Zhong}, \binits{Y.}},
\oauthor{\bsnm{Li}, \binits{S.}},
\oauthor{\bsnm{Wei}, \binits{C.}},
\oauthor{\bsnm{Dong}, \binits{L.}},
\oauthor{\bsnm{Wang}, \binits{Z.}},
\oauthor{\bsnm{Yan}, \binits{C.}}:
Predicting Stock Prices with FinBERT-LSTM: Integrating News Sentiment Analysis.
Preprint at \url{https://www.arxiv.org/abs/2407.16150}
(2024)
\end{botherref}
\endbibitem

\bibitem[\protect\citeauthoryear{Jiang and Zeng}{2023}]{jiang2023financial}
\begin{botherref}
\oauthor{\bsnm{Jiang}, \binits{T.}},
\oauthor{\bsnm{Zeng}, \binits{A.}}:
Financial sentiment analysis using FinBERT with application in predicting stock movement.
Preprint at \url{https://arxiv.org/abs/2306.02136}
(2023)
\end{botherref}
\endbibitem

\bibitem[\protect\citeauthoryear{Girsang and Stanley}{2023}]{girsang2023hybrid}
\begin{barticle}
\bauthor{\bsnm{Girsang}, \binits{A.S.}},
\bauthor{\bsnm{Stanley}}:
\batitle{Hybrid lstm and gru for cryptocurrency price forecasting based on social network sentiment analysis using finbert}.
\bjtitle{IEEE Access}
\bvolume{11},
\bfpage{120530}--\blpage{120540}
(\byear{2023})
\doiurl{10.1109/ACCESS.2023.3324535}
\end{barticle}
\endbibitem

\bibitem[\protect\citeauthoryear{Chevalier}{2018}]{chevalier2018larnn}
\begin{botherref}
\oauthor{\bsnm{Chevalier}, \binits{G.}}:
LARNN: Linear Attention Recurrent Neural Network.
arXiv preprint arXiv:1808.05578.
Preprint at \url{https://arxiv.org/abs/1808.05578}
(2018)
\end{botherref}
\endbibitem

\bibitem[\protect\citeauthoryear{Hochreiter and Schmidhuber}{1997}]{hochreiter1997long}
\begin{barticle}
\bauthor{\bsnm{Hochreiter}, \binits{S.}},
\bauthor{\bsnm{Schmidhuber}, \binits{J.}}:
\batitle{Long short-term memory}.
\bjtitle{Neural Computation}
\bvolume{9},
\bfpage{1735}--\blpage{1780}
(\byear{1997})
\doiurl{10.1162/neco.1997.9.8.1735}
\end{barticle}
\endbibitem

\bibitem[\protect\citeauthoryear{Graves and Schmidhuber}{2005}]{graves2005framewise}
\begin{botherref}
\oauthor{\bsnm{Graves}, \binits{A.}},
\oauthor{\bsnm{Schmidhuber}, \binits{J.}}:
Framewise phoneme classification with bidirectional LSTM networks.
Paper presented at the IEEE International Joint Conference on Neural Networks (IJCNN), 31 July--4 August 2005
(2005)
\end{botherref}
\endbibitem

\bibitem[\protect\citeauthoryear{Schuster and Paliwal}{1997}]{schuster1997bidirectional}
\begin{barticle}
\bauthor{\bsnm{Schuster}, \binits{M.}},
\bauthor{\bsnm{Paliwal}, \binits{K.K.}}:
\batitle{Bidirectional recurrent neural networks}.
\bjtitle{IEEE Trans. Signal Process.}
\bvolume{45},
\bfpage{2673}--\blpage{2681}
(\byear{1997})
\doiurl{10.1109/78.650093}
\end{barticle}
\endbibitem

\bibitem[\protect\citeauthoryear{Ben~Said et~al.}{2021}]{ben2021predicting}
\begin{barticle}
\bauthor{\bsnm{Ben~Said}, \binits{A.}},
\bauthor{\bsnm{Erradi}, \binits{A.}},
\bauthor{\bsnm{Aly}, \binits{H.}},
\bauthor{\bsnm{Mohamed}, \binits{A.}}:
\batitle{Predicting covid-19 cases using bidirectional lstm on multivariate time series}.
\bjtitle{Environmental Science and Pollution Research}
\bvolume{28},
\bfpage{56043}--\blpage{56052}
(\byear{2021})
\doiurl{10.1007/s11356-021-14286-7}
\end{barticle}
\endbibitem

\end{thebibliography}

\end{document}